\documentclass[prb,a4paper,showpacs,twocolumn,superscriptaddress,longbibliography]{revtex4-2}

\usepackage{textcomp}
\usepackage{amssymb}
\usepackage{amsmath}
\usepackage{amsfonts}
\usepackage{graphicx}
\graphicspath{{Figures/}}
\usepackage{bm}
\usepackage{xcolor}
\usepackage{multirow}
\usepackage{natbib}
\usepackage{hyperref}
\usepackage{mathrsfs}

\begin{document}

\title{Interaction effects on Andreev states from an electromagnetic environment}

\author{E. S. Samuelsen}
\email{e.s.s.samuelsen@tudelft.nl}
\affiliation{Kavli Institute of Nanoscience, Delft University of Technology, 2628 CJ Delft, The Netherlands}

\author{Y. V. Nazarov}

\affiliation{Kavli Institute of Nanoscience, Delft University of Technology, 2628 CJ Delft, The Netherlands}

\begin{abstract}
Here we present a detailed theoretical investigation of the environment-induced interaction correction to the ground states and addition energies in a short superconducting junction. We derive general formulas for these corrections linking them to the frequency-dependent admittance of the junction and the environment impedance. We specify two environmental models: that of an external impedance and the Mattis-Bardeen model. In both cases we assume $G_Q Z \ll 1$, where $G_Q$ is the conductance quantum and $Z$ the typical environmental resistance.

While an expectation is that the typical scale of the relative correction is of the order $G_Q Z$, we have found that in many cases the dimensionless scale $L G_Q \Delta$ provides a better estimation, where $\Delta$ is the superconducting energy gap and $L$ an external inductance. We discuss the logarithmic renormalizations of the energies by low- and high-frequency phase fluctuations and single out a non-superconducting contribution related to the renormalization of transmission eigenvalues. We investigate in detail the peculiarities of the corrections at: i. small phases, where the interaction causes a current jump, ii. Andreev bound state energies close to the gap edge, iii. Andreev bound state energies close to Fermi level. In all these cases the correction may become comparable with the unperturbed energy, even for $G_Q Z \ll 1$, and we briefly sketch non-perturbative models relevant for the situations.

\end{abstract}

\maketitle

\section{Introduction}
\label{sec:intro}
Hybrid superconducting devices remain in focus of attention of the condensed matter community for several decades. There are numerous applications of these devices in quantum sensing~\cite{jeanneretApplicationJosephsonEffect2009}, and they are used as platforms  to realize superconducting~\cite{kjaergaardSuperconductingQubitsCurrent2020} qubits of several kinds as well as topologically protected Majorana~\cite{beenakkerSearchMajoranaFermions2013} qubits.
Disorder-robust superconducting state~\cite{bardeenTheorySuperconductivity1957, andersonTheoryDirtySuperconductors1959} is a source of macroscopic coherence. The superconducting devices are naturally integrated into electric circuits that simplifies their use and also allows for efficient readout and manipulation of superconducting qubits, for instance, in the framework of circuit quantum electrodynamics~\cite{blaisCircuitQuantumElectrodynamics2021}.

A generic superconducting device is a junction connecting two superconductors. Andreev reflection~\cite{andreevThermalConductivityIntermediate1964} takes place in the junction and the interference of Andreev-reflected electrons and holes results in a set of discrete localized current-carrying Andreev Bound States (ABSs). If interaction can be neglected, the energies of these states are determined from the scattering matrix characterizing the junction~\cite{beenakkerUniversalLimitCriticalcurrent1991}. A particularly illuminating situation is that of a "short" junction, a junction that is small in comparison with the superconducting coherence length. In this case, the energy dependence of the scattering matrix can be neglected at the energy scale of the superconducting energy gap $\Delta$. The two-terminal scattering matrix can be decomposed into a set of independent transmission channels and each channel labeled by $n$ houses a spin-degenerate ABS with energy~\cite{beenakkerUniversalLimitCriticalcurrent1991, NazarovBlanterQuantumTransport}
\begin{equation}
E_A^{(n)} = \Delta \sqrt{ 1 - T_n \sin^2(\phi/2)},
\end{equation}
$T_n$ being the transmission coefficient in the corresponding transport channel and $\phi$ the superconducting phase difference between the terminals.
The corresponding structure of many-body states in the junction~\cite{chtchelkatchevAndreevQuantumDots2003} becomes especially simple if there is only one relevant channel. In this case, the phase-dependent part of the ground state energy is given by the ABS energy, $E_g = -E_A$. This state has an even number of particles and is a spin singlet.  Adding a quasiparticle of either spin costs energy $E_A$, so the energy of the resulting {\it doublet} state does not depend on the phase. Since the superconducting current in the junction is given by $I = 2e\partial_\phi E /\hbar $, it is absent in the doublet state, this phenomenon being dubbed "quasiparticle poisoning". The doublet state is in fact a ground state at odd number of particles. Adding yet another quasiparticle brings the system to the {\it excited singlet} state. The phase-dependent part of the energy is $E_A$, so the supercurrent in this state is opposite to that in the ground state. This structure inspired theoretical proposals of two types of qubits:
i. the spin qubit~\cite{chtchelkatchevAndreevQuantumDots2003,padurariuTheoreticalProposalSuperconducting2010}, where the pair of qubit states are the two doublet states,
ii. the Andreev qubit~\cite{zazunovAndreevLevelQubit2003, zazunov2}, where the two qubit states are the ground and exited singlets.

This theory has been experimentally confirmed in a series of crucial experiments~\cite{zgirski2011, bretheauExcitingAndreevPairs2013, janvierCoherentManipulationAndreev2015, goffmanConductionChannelsInAsnanowire2017, pothierspinorbit2019, spinqubit1, spinqubit2} where the ABS energies have been measured with microwave spectroscopy, this being complemented with the critical current measurements. In recent experiments, sufficient accuracy has been achieved to go beyond the non-interacting approximation and identify the signatures of electron-electron interaction~\cite{signaturesfatemi2022,signaturespothier2022}.

However, the interaction addressed in the latter experiments was Coulomb interaction coming from the charge stored {\it inside} the junction, like that in superconducting quantum dots~\cite{Lee2014,vanDam2006}. While undoubtedly valid for the semiconducting nanowire junctions considered, this kind of interaction cannot be relevant for an intrinsic short junction that, in principle, has no inside. In this case, the interaction rather comes from {\it outside} the junction and is mediated by the external electromagnetic environment.

Such interactions have long been in the focus in studies of the Josephson effect~\cite{Josephson}. A Josephson junction can be regarded as a short-junction in the limit of small transmission coefficients, $T_n \ll 1$~\cite{NazarovBlanterQuantumTransport}. It has been understood~\cite{Legget1} and verified experimentally~\cite{Clarke} that interactions in a Josephson junction is induced by quantum fluctuations of the superconducting phase. Early studies concentrated on a dissipative environment and were essential for establishing the modern theory of dissipative quantum mechanics~\cite{Legget2,Weiss}. A seminal result was the interaction correction to the phase-dependent part of the ground-state energy~\cite{Schmid}, that reads $ 2 E_J \sin^2(\phi/2)$ without interaction,
\begin{equation}
\label{eq:Jocorrection}
\frac{\delta E_J}{E_J} = - 2 G_Q Z \ln \left[ \frac{\omega_{\rm H}}{\omega_{\rm L}} \right],
\end{equation}
where $G_Q = e^2/\pi \hbar$ and from now on we use units where $\hbar =1$.
This correction is proportional to the impedance of the external circuit $Z$ and diverges logarithmically upon decreasing the lower cut-off frequency $\omega_{\rm L}$ ($\omega_{\rm H}$ is the higher cut-off frequency).  The development of the logarithmic divergence leads to  the Schmid transition~\cite{Schmid}: the vanishing of the Josephson energy at a critical value of $Z$, $Z_c = 1/2G_Q$.
 While the prediction of Schmid is theoretically indisputable, the controversy concerning its experimental verification~\cite{Murani2020}  has been resolved only recently~\cite{Kuzmin2025,Pekola2026}.

A recent complementary theoretical development~\cite{oddparity, oddparity2} addresses Josephson quantum mechanics at odd parity, in particular, the interaction correction to doublet state energy. It has been shown that the bound energy of the doublet state, $\Omega \equiv \Delta - E_A = E_J \sin(\phi/2)$ without interaction, is corrected by
\begin{equation}
\label{eq:Jooddcorrection}
\frac{\delta \Omega}{\Omega} = - G_Q Z \ln \left[ \frac{\omega_{\rm H}}{\omega_{\rm L}} \right],
\end{equation}
one half compared to the correction to the even ground state energy (see Eq. \eqref{eq:Jocorrection}). Owing to this,  the Schmid transition for odd parity occurs at four times bigger critical impedance, so in an interval of impedances the Josephson current in odd state can be higher than in the even state. This is especially surprising owing to the fact that for a single-channel junction the current in odd state is completely poisoned without interaction, and is determined by the interaction correction at small $G_Q Z$.
In addition, at small $\phi$ the correction to the current exhibits a jump: the current is discontinuous at $\phi =0$. As explained in~\cite{oddparity}, this manifests the formation of {\it extra} ABSs in the narrow interval around  $\phi =0$. It remains to be investigated to which extent these results are relevant beyond the Josephson limit, that is, at higher transmission coefficients.

All this provides motivation for the goal of this article: to derive and comprehend general expressions for the first interaction correction to the ABS energies due to the electromagnetic environment. The environment is characterized by a frequency-dependent impedance $Z(\omega)$. In distinction from Josephson research, we need to consider frequencies of the order of $\Delta$ and take into account all virtual and real processes whereby localized and delocalized quasiparticles are generated by the fluctuating  phase.

We concentrate on the most interesting case of the single-channel short junction, but also explain how to extend the results to the multi-channel short junctions. We do not concern ourselves here with an energy-dependent scattering matrix and/or the semi-classical case of a quasi-continuous ABS spectrum: this requires different theoretical methods, see e.g.~\cite{Erdmanis}.

\begin{figure}[h]
    \includegraphics[width=1.0\columnwidth]{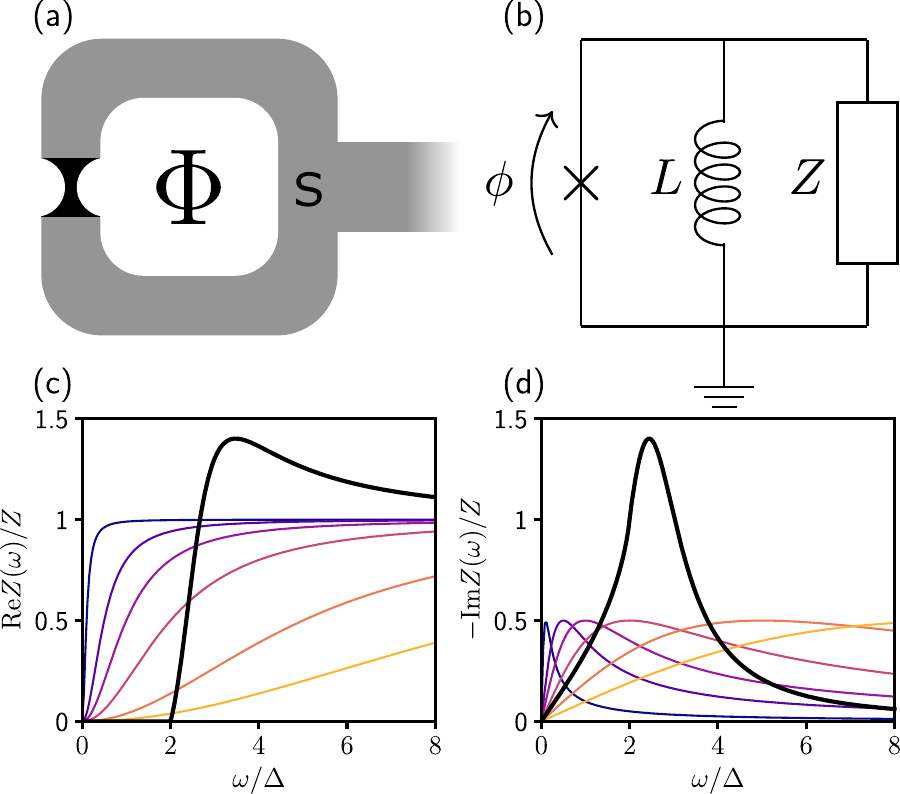}
	\caption{ The setup and impedance models. (a) The junction enclosed in a superconducting loop with flux $\Phi$, which is coupled to a circuit. (b) Circuit theory representation of the external impedance model. (c-d) Real and imaginary part of the frequency-dependent impedance. Colored lines: the external impedance model for $\omega_0/\Delta =[0.1, 0.5, 1.0, 2.0, 5.0, 10]$(from dark blue to yellow). Black line: The Mattis-Bardeen (MB) model.} \label{fig:environment}
\end{figure}

While we derive the expressions valid for arbitrary electromagnetic environment, for concrete illustrations and plots we utilize two models illustrated in (a) and (b) of Fig. \ref{fig:environment}. The first model is that of an {\it external impedance}. The frequency-dependent $Z(\omega)$ felt by the junction is a frequency-independent resistor $Z$ in parallel with an inductance $L$, $Z(\omega) = ((-i\omega L)^{-1} + Z^{-1})^{-1}$. Physically, it may be realized with a transmission line responsible for $Z$ that is connected to a SQUID loop of inductance $L$ utilized to fix the phase difference over the junction. The crossover between low-frequency inductive response and high-frequency resistive response occurs at $\omega \sim \omega_0 \equiv Z/L$. In the second model, the {\it MB model}, we attempt to take into account the microscopic peculiarities of the superconducting state. We assume that the environment of the junction in the {\it  normal state} is characterized by a single resistance $Z$. By neglecting the skin-effect and assuming thin-film leads we obtain the frequency-dependent impedance in the superconducting state $Z(\omega/\Delta)$ from the celebrated Mattis-Bardeen \cite{Mattis} formulas. In this model, the resistive response is absent at $\omega <2 \Delta$ and the inductive response at zero frequency corresponds to $L = Z/(\pi\Delta)$.

The first model is relevant if the dissipation does not occur in the neighboring superconducting leads but rather in more distant parts of the circuit, perhaps even beyond the chip. This corresponds to low normal resistance of the neighboring superconducting leads. The second model is relevant for high normal resistances of the leads, encompassing strongly disordered superconductors \cite{stronglydisordered} where sheet resistance of the leads may be of the order of $G_Q^{-1}$. In (c) and (d) of Fig. \ref{fig:environment} we plot the real and imaginary part respectively of $Z(\omega)$ for both models. The observation is that most results are qualitatively the same between the models provided $Z$ is the same and $\omega_0 \simeq 3-5 \Delta$.

We study three independent interaction corrections to: i. the phase-dependent part of the even ground state energy $\Delta E^{(e)}$, ii. the doublet addition energy $\delta^{\rm odd}$, iii. the singlet addition energy $\delta^{\rm even}$. It is constructive to separately discuss two of their linear combinations: the correction to the odd ground state energy $\Delta E^{(o)} \equiv \delta^{\rm odd} +  \Delta E^{(e)}$, and the interaction signature $ \delta^{\rm even} - 2\delta^{\rm odd}$. The latter is zero in absence of interactions, so its experimental quantification signifies the presence of interactions.

Let us briefly summarize the results. While a natural expectation is that the typical scale of the relative correction is of the order $G_Q Z$, we have found that in many cases the dimensionless scale $L G_Q \Delta$ provides a better estimate. We elucidate the logarithmic renormalizations of the energies by low- and high-frequency phase fluctuations, comparing the results with Eqs. \eqref{eq:Jocorrection} and \eqref{eq:Jooddcorrection}, and single out a non-superconducting contribution related to the renormalization of transmission eigenvalues~\cite{NazarovKindermann}. We find the peculiarities of the corrections at i. small phases, where the interaction causes a current jump for odd ground state and cusps in addition energies, ii. ABS energies close to the gap edge, $E_A \approx \Delta$, which also includes i., iii. ABS energies close to the fermi level, $E_A \ll \Delta$, iv. typical environmental frequencies $\omega_0 \gg \Delta$. We address these cases analytically and present numerical illustrations valid for a wide range of parameters. We identify parameter regions where the corrections may become comparable with the unperturbed energies even at $G_Q Z \ll 1$, and discuss the non-perturbative models adequate for these parameter regions. The interaction signature is given by a simple relation that only involves the inductances of the environment at zero frequency and at the frequency $2 E_A$.

The structure of the article is as follows. Since the expressions for the high-frequency junction admittance are crucial for our presentation, the second part of the introduction, Sec. \ref{sec:admittance}, briefly recites the results obtained in~\cite{kosFrequencydependentAdmittanceShort2013}. We explain the method of calculations in Sec. \ref{sec:method}. An overview of the results along with the most general expressions for the interaction corrections including the multi-channel case, are provided in Sec. \ref{sec:overview}. In Sec. \ref{sec:evenground} we provide the details for the even ground state energy correction. We address the odd ground state in Sec. \ref{sec:oddground}. The addition energies of the doublet and singlet are discussed in Sec. \ref{sec:add}, while the interaction signature is presented in Sec. \ref{sec:signature}. In Sec. \ref{sec:nonpert} we briefly discuss the limits of $\phi \to 0$ and $\phi \to \pi$, $T \to 1$: in these parameter regions, the correction competes with the non-interacting ABS energies so a non-perturbative approach is required.
We conclude in Sec. \ref{sec:conclusions}.

\section{Introduction: junction admittance}
\label{sec:admittance}
To present our results in a compact form it is convenient to use the junction admittance, the high-frequency response of a short junction on the a.c. voltage (or equivalently on the a.c. flux/phase). The detailed calculation of this has been presented in~\cite{kosFrequencydependentAdmittanceShort2013}, and here we give their results with slightly adjusted notations and some extra remarks. We only need the results at vanishing temperature.

The admittance $Y(\omega)$ is defined in terms of the current reaction to the applied voltage at frequency $\omega$,
\begin{equation}
I(\omega) = Y(\omega) V(\omega).
\end{equation}
It is a sum over contributions from all transport channels, so it suffices to give the expressions for one channel with the transmission coefficient $T$. In the limit of high frequency, $\omega \gg \Delta$, the superconducting response is negligible, and the admittance is just the normal conductance of the junction, $Y(\omega \to \infty) = G_Q T$. In the limit of low frequency, the response is purely inductive, $Y(\omega \to 0) = 1/(-i\omega L_J)$, where the junction inductance is expressed in terms of the phase dependent part of the ground state energy,
\begin{align}
1/L_J = -4 \pi G_Q \partial^2_{\phi} E_A.
\end{align}

The real part of the admittance in the even ground state is contributed by the processes where absorption of the energy quantum $\omega$ excites two quasiparticles with opposite spins.  This gives three contributions: i. both quasiparticles go to the leads, their energies $\epsilon_{1,2} > \Delta$, $\epsilon_1+\epsilon_2 = \omega$,
\begin{align}
\frac{{\rm Re} Y_1(\omega)}{G_Q T} = \frac{\theta(\omega - 2 \Delta)}{\omega} \int_{\Delta}^{\infty} d \epsilon_1 \rho(\epsilon_1)\int_{\Delta}^{\infty} d \epsilon_2 \rho(\epsilon_2) \nonumber \\
\delta(\omega - \epsilon_1 - \epsilon_2) \left ( 1 - \frac{\Delta^2(1+ (T-2) \sin^2(\phi/2))} {\epsilon_1 \epsilon_2} \right),
\end{align}
where $\rho(\epsilon) \equiv \epsilon \sqrt{\epsilon^2-\Delta^2}/(\epsilon^2 - E_A^2)$ is the effective density of states. ii. One quasipaticle with energy $\epsilon >\Delta$ goes to the leads while another is stuck in the ABS,
$E_A +\epsilon = \omega$,

\begin{align}
\frac{{\rm Re} Y_2(\omega)}{G_Q T} =  \frac{ \pi \Delta \sqrt{T} |\sin(\phi/2)|}{\omega} \theta(\omega - \Delta - E_A)\nonumber \\ \rho(\omega-E_A)  \left ( 1 - \frac{\Delta^2(1+ (T-2) \sin^2(\phi/2))} {E_A (\omega - E_A)} \right),
\end{align}
iii. both quasiparticles get stuck in the ABS, $\omega=2E_A$,
\begin{align}
\label{eq:y3}
&{\rm Re} Y_3(\omega)= 2E_A \bar{Y} \pi \delta(\omega - 2E_A);\\
&\bar{Y} \equiv \frac{\pi G_Q T^2(1-T)}{4} \left[\frac{\Delta \sin(\phi/2)}{E_A}\right]^4 \label{eq:bar_y}.
\end{align}
Thus, in the even ground state the admittance is given by
\begin{align}
Y^{(e)}(\omega) =  Y_1(\omega) + Y_2(\omega) + Y_3(\omega).
\end{align}
The imaginary part of $Y_{1-3}(\omega)$ is obtained from Kramers-Kronig relations. One has to be cautious applying the relations since the real part of admittance does not vanish at $\omega \to \infty$. A safe way implemented in  \cite{kosFrequencydependentAdmittanceShort2013} is to apply  Kramers-Kronig relations  to compute the quantity  $({\rm Im} Y^{(e)}(\omega) -1/\omega L_J)$ that vanishes at zero $\omega$. We prefer not to separate $1/\omega L_J$ and the rest of ${\rm Im} Y^{(e)}$. It is important for us to notice that at  $\omega \gg \Delta$ the imaginary part of impedance scales as $1/\omega$, ${\rm Im} Y(\omega) =1/(\omega L_{\infty})$. The Kramers-Kronig relations result in the following sum rule:
\begin{align}
\label{eq:sumRuleEven}
\frac{1}{L_{\infty}} -\frac{1}{L_J}  = \frac{2}{\pi} \int_0^{\infty} d\omega ({\rm Re} Y^{(e)}(\omega) - G_QT).
\end{align}
The concrete calculation gives
\begin{align}
\label{eq:Linfty}
\frac{1}{L_\infty} = -\pi G_Q \frac{\Delta^2- E_A^2}{E_A} = 2\pi G_Q T(1-T) \partial_T E_A.
\end{align}
We note that $L_\infty$ is always {\it negative}, while $L_J$ may be positive or negative depending on $\phi$.

\begin{figure}[h]
    \includegraphics[width=1.0\columnwidth]{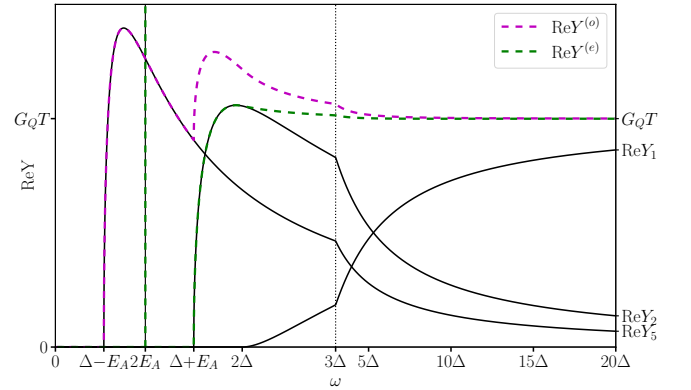}
	\caption{The state resolved junction admittance vs frequency for $T=0.9$ and $\phi = 3\pi/4$ which gives $E_A \approx 0.48 \Delta$. The purple and green dashed lines give the real part of the admittance in the odd and even ground state respectively. The black lines show the decomposition of the admittance into different dissipation channels ${\rm Re}Y_{1}, {\rm Re}Y_{2},$ and ${\rm Re}Y_{5}$ as indicated on the right vertical axis. The contributions from ${\rm Re}Y_{3}\propto\delta(\omega-2E_A)$ to ${\rm Re} Y^{(e)}$ is illustrated with the black and dashed green vertical lines at $\omega = 2E_A$. At $\omega = 3\Delta$ we change the horizontal scale as indicated by the dotted line.} \label{fig:admittance_illustration}
\end{figure}

In the odd ground state, there is an extra absorption process whereby a quasiparticle trapped in an ABS goes to the leads, the resulting quasiparticle energy being $\omega + E_A > \Delta$. The corresponding contribution to the admittance reads
\begin{align}
\label{eq:Y5}
\frac{{\rm Re} Y_5(\omega)}{G_Q T} = \theta(\omega+E_A -\Delta) \frac{\Delta \pi \sqrt{T} |\sin(\phi/2)|} {2 \omega}  \nonumber \\ \rho(\omega + E_A) \left( 1 + \frac{\Delta^2(1+ (T-2) \sin^2(\phi/2))} {E_A (\omega + E_A)}\right).
\end{align}
Besides, no absorption to the excited singlet state can take place, so the whole admittance in the odd ground state becomes (see~\cite{kosFrequencydependentAdmittanceShort2013} for a detailed explanation)
\begin{align}
Y^{(o)}(\omega) = Y_1(\omega) + \frac{Y_2(\omega)}{2} + Y_5(\omega).
\end{align}
It is worth noting that the zero-frequency inverse inductance vanishes in the odd ground state (quasiparticle poisoning). The calculation shows that the high-frequency inverse inductance also vanishes, so the analog of Eq. \eqref{eq:sumRuleEven} reads
\begin{align}
0 = \int_0^{\infty} ({\rm Re} Y^{(o)}(\omega) - G_QT).
\end{align}
While the Kramers-Kronig relations do not formally hold for the excited singlet state, we can define the absorption in this state,
\begin{align}
    {\rm Re} Y^{(ex)}(\omega) &= {\rm Re} Y_1(\omega)  - {\rm Re} Y_3(\omega)+ 2  {\rm Re} Y_5(\omega);\\
    &={\rm Re} (2Y^{(o)}(\omega)- Y^{(e)}(\omega)).
\end{align}
In Fig. \ref{fig:admittance_illustration} we illustrate the state-resolved admittance and show the decomposition into different dissipation channels ${\rm Re} Y_{1-3}$ and ${\rm Re} Y_{5}$.  Here $T=0.9$ and $\phi = 3 \pi /4$, which means $E_A \approx \Delta/2$, which was chosen to provide good visual separation between the threshold frequencies of ${\rm Re} Y_{1-3}$ and ${\rm Re} Y_5$. At large frequencies the contributions from the dissipation channels ${\rm Re} Y_{1}$, ${\rm Re} Y_{2}$, and ${\rm Re} Y_{5}$ converge as $1/\omega$ and $ {\rm Re} Y_2 \approx 2 {\rm Re} Y_5 \propto 1/\omega$ while ${\rm Re}Y_1 \approx G_QT -{\rm Re} Y_2$. Because of a cancellation between the $1/\omega$ asymptotes, the state resolved admittances ${\rm Re} Y^{(e)}$ and ${\rm Re} Y^{(o)}$ converge to the normal state admittance $G_Q T$ at a faster, integrable rate $\sim \ln(\omega/\Delta)/\omega^2 $.

There is a peculiarity in the above expressions to become important in further reasoning. The coefficient $\sqrt{T} |\sin(\phi/2)|$ in the expressions for $Y_2$ and $Y_5$ suggest the admittance is not analytical in $T$ or $\phi$ in the limit of $T\to 0$ and/or $\phi \to 0$. This is not true for  $Y^{(e)}(\omega)$: the analyticity in $T$ and $\phi$ in is restored for the sum $Y_1 + Y_2$.  This cancellation does not take place for $Y^{(o)}(\omega)$ which remains non-analytical.

\section{Method}
\label{sec:method}
To compute the corrections, we start with imaginary time action that describes a junction subject to the phase fluctuations $\chi(\tau)$ produced by a linear environment. The partition function is given by
\begin{align}
&{\cal Z}  = \int \prod_\tau d \chi(\tau) \exp (- S_{\rm J} - S_{\rm env}); \\
&S_{\rm env} = \frac{1}{8\pi G_Q}\sum_{\omega} \frac{|\omega|}{  Z(i |\omega|)}|\chi(\omega)|^2,
\end{align}
where the summation is over bosonic Matsubara frequencies, and $Z(\omega)$ is the impedance characterizing the environment.
The junction action for the short-junction model is known (see e.g.~\cite{NazarovBlanterQuantumTransport}). It is expressed in terms of Green's functions $\hat{G}_{1,2}$ of the left/right lead,
\begin{align}
&\hat{G}_{1,2}(\tau,\tau') = e^{ \pm i \tau_3 (\phi + \chi(\tau))/4 }   \hat{G}_0(\tau-\tau') e^{ \mp i \tau_3 (\phi + \chi(\tau'))/4 }; \nonumber \\
&\hat{G}_0(\epsilon) = \frac{ \Delta \tau_1 + \epsilon \tau_3}{\sqrt{\Delta^2 + \epsilon^2}},
\label{eq:GreenTau}
\end{align}
where $\phi$ is the constant phase difference and $\tau_{1,2,3}$ are Pauli matrices in Nambu space. A compact form of $S_{\rm J}$ treats $\hat{G}_{1, 2}$ as linear operators in both Nambu and imaginary time:
\begin{align}
\label{eq:junctionAction}
S_{\rm J} = -\frac{1}{2} \sum_n {\rm Tr} \left[ \ln\left( 1 - \frac{T_n}{4} (\hat{G}_1 - \hat{G}_2)^2\right)\right].
\end{align}
Here the trace is over the indices of $\hat{G}_{1, 2}$ and the summation is over transport channels, $T_n$ being the corresponding transmission coefficients.
\begin{figure}[b]
    \includegraphics[width=\columnwidth]{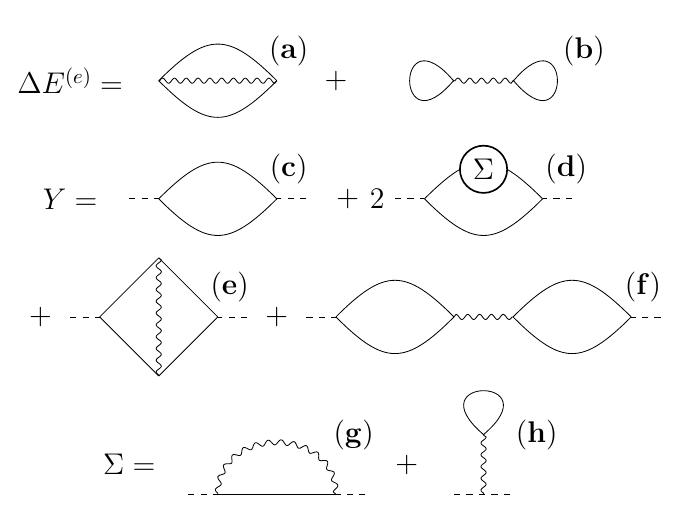}
	\caption{Diagrams related to the interaction corrections. The solid lines represent fermionic Green's functions originating from \eqref{eq:GreenTau} and \eqref{eq:GreenT}, while the wiggly lines represent the propagators of the external flux. (a-b) The corrections to the ground state energy, where (a) gives a convolution involving $Z(\omega)$, while (b) gives an inductive correction $\propto L$. (c) The bare junction admittance. The interaction correction to the admittance is given by the sum of $(d-f)$. $(d)$ The electron self-energy contribution, where '$\Sigma$' is the sum of the diagrams in $(g)$ and $(h)$. (e) A vertex correction. (f) A circuit-theory correction to the admittance.
	Comparing  $(a)$ and $(c)$, we see the relation between the admittance and the interaction correction to the ground state energy.
}
	\label{fig:Diagrams}
\end{figure}

The first perturbative correction in $Z$ give the correction to the energy $ - \ln{\cal Z}/\beta$ of the even ground state. It can be visualized as the diagrams in $(a)$ and $(b)$ in Fig. \ref{fig:Diagrams}. The diagram in $(a)$ contains a convolution of $Z(\omega)$ (wiggly line) and the second derivative of the junction action with respect to $\chi(\omega)$, that is, the admittance. The diagram in $(b)$ is proportional to $Z$ in the limit of $\omega \to 0$ and thus gives an {\it inductive} contribution to the energy.

To compute the corrections to the addition energies, and the odd ground state energy, we consider how these quantities are measured experimentally. For this, we need to excite the junction with a microwave source. The shift in the frequency position of the junction admittance pole at $\omega \approx 2 E_A$ gives the correction to the singlet addition energy. The shift of the absorption threshold at $\omega \approx E_A +\Delta$ gives the correction to the doublet addition energy.  In short, we need to address the interaction corrections to the junction {\it admittance}.

In principle, these corrections can be computed in the framework of imaginary time, however the required analytical continuation to real frequencies is technically cumbersome. Therefore, we resort to the Keldysh technique where the fluctuations of the phase are defined on two branches of the Keldysh contour, $\chi^{\pm}(t)$. The admittance and as well as all higher order response functions and correlators can be obtained from the generating functional
\begin{align}
{\cal Z}[I^+, I^-] = \int \prod_t d \chi^+(t) d \chi^{-}(t) e^{ i (S_{\rm J} +S_{\rm env}) } \nonumber \\
\times e^ {i\int dt (\chi^+(t) I^+(t) - \chi^{-}(t) I^-(t))},
\end{align}
which depends on the functions $I^{\pm}(t)$. Here
\begin{align}
	&S_{\rm env} =
	\frac{i}{4\pi G_Q}
	\int \frac{d\omega}{2\pi} \omega
	\begin{pmatrix}
	    \chi_{cl}(\omega)&\chi_q(\omega)
	\end{pmatrix}\nonumber \\
	&\times
	\begin{pmatrix}
	    0 & -\frac{1}{Z(\omega)^*}\\
		\frac{1}{Z(\omega) } & \Re\{\frac{2\coth(\frac{\beta \omega}{2})}{Z(\omega) }\}
	\end{pmatrix}
	\begin{pmatrix}
	    \chi_{cl}(\omega)\\
		\chi_q(\omega)
	\end{pmatrix}
\end{align}
where $\chi_{cl, q}(t) = (\chi^+(t) \pm \chi^-(t))/2$. At zero temperature $\coth(\frac{\beta \omega}{2}) =  {\rm sgn}(\omega)$. The junction action is given by a formula similar to Eq. \eqref{eq:junctionAction}, but with an addition of the Keldysh structure
\begin{align}
    S_{\rm J} = -\frac{i}{2} \sum_n {\rm Tr} \left[ \ln\left( 1 - \frac{T_n}{4} (\check{G}_1 - \check{G}_2)^2\right)\right].
\end{align}
The Keldysh Green's functions in the above expression are defined as
\begin{align}
&\check{G}_{1,2}(t,t') = e^{ \pm i \tau_3 (\phi + \check{\chi}(t))/4 }   \check{G}_0(t-t') e^{ \mp i \tau_3 (\phi + \check{\chi}(t'))/4 }.
\label{eq:GreenT}
\end{align}
Here the 'check' indicates the Keldysh matrix structure
\begin{align}
&\check{\chi}(t) =
\begin{pmatrix}
	\chi_{cl}(t) & \chi_{q}(t)\\
	\chi_{q}(t) & \chi_{cl}(t)\\
\end{pmatrix},\\
&  \check{G}_0(\epsilon) =
\begin{pmatrix}
	\hat{G}_R(\epsilon) & \tanh(\frac{\beta \epsilon}{2}) (\hat{G}_R(\epsilon) - \hat{G}_A(\epsilon))\\
	0 & \hat{G}_A(\epsilon)
\end{pmatrix},
\end{align}
and the advanced/retarded Green's functions read
\begin{equation}
	\hat{G}_{R/A}(\epsilon) = \frac{ \Delta \tau_1 -i \epsilon \tau_3}{\sqrt{\Delta^2 - (\epsilon\pm i0)^2}}.
\end{equation}
At zero temperature, $\tanh(\frac{\beta \epsilon}{2}) = {\rm sgn} (\epsilon)$.

The diagram for the bare admittance is given in Fig. \ref{fig:Diagrams} (c) and the corrections to first order in $Z$ are given in (d - f). The diagram in (f) represents a rather trivial contribution where the shift of the resonance can be described with elementary circuit theory, see \cite{atlongdistance} for details. The diagram in (d) is reduced to the corrections to electronic self-energy (g-h) and the factor of $2$ accounts for the symmetric diagram where the self energy appears on the lower fermion line. The diagram in (e) gives a vertex correction to the admittance. In all loops a summation over channel index is implied, so diagrams: (a), (c), (e), and (g) contain single sums over the channels, while: (b), (f), and (h) involve double sums.

We do not compute full correction to the admittance, but instead trace how the corrections shifts the positions of the singlet resonance and the threshold that defines the doublet addition energy.

\section{Overview of the general results}
\label{sec:overview}

In this section we present the results for the corrections to the ground state and addition energies in the most general form, postponing the analysis of limiting cases to later sections. The detailed derivations of all the level shifts are provided in the Supplementary Material~\cite{SupplementalMaterial}.

Let us start with a single channel model and the correction to the even ground state energy. It reads
\begin{align}
\Delta E^{(e)} = -\frac{1}{2}L I_{J}^2  +\frac{1}{2} \int_{0}^{\infty} \frac{d\omega}{\pi } {\rm Im } (Z(\omega)Y^{(e)}(\omega)).
\label{eq:even}
\end{align}
The first term is the trivial inductive energy corresponding to the diagram (b): the supercurrent $I_J = - 2e \partial_\phi E_A$ in the junction results in a phase drop over the external inductance $L$, this gives an energy decrease. The second term represents the effect of quantum fluctuations and, as we see, is expressed in terms of the junction admittance and the impedance of the environment.  In this convenient relation, we did not explicate the convergence of the integral at large or small frequencies. We do this with the following equivalent representations. The following form
\begin{align}
&\Delta E^{(e)} = -\frac{1}{2}L I_{J}^2  +\frac{\langle \hat{\Phi}^2\rangle }{2 L_{\infty}} -\int_0^\infty \frac{d\omega}{\pi \omega}{\rm Re}Z(\omega) \nonumber \\ & \times \int_0^\infty \frac{d\omega'}{\pi}{\rm Re}(Y^{(e)}(\omega')-G_QT)\frac{\omega'}{\omega + \omega'}
\label{eq:even-high}
\end{align}
gives an integral that converges at high frequencies even if $Z(\omega) \to {\rm const}$ for $\omega \to \infty$. The divergent part is contained in the second term where the total quantum fluctuation of flux is defined as
\begin{align}
\langle\hat{\Phi}^2\rangle  = \int_0^\infty \frac{ d\omega}{\pi \omega} {\rm Re} Z(\omega) \simeq \frac{Z}{\pi} \ln \left[ \frac{\omega_{\rm H}}{\omega_{\rm L}}\right].
\end{align}
The last relation stresses that for constant real $Z$ one needs to cut the integration at higher/lower frequencies, $\omega_{H/L}$. It is instructive to relate this term to the renormalization of the transmission coefficient discussed in~\cite{NazarovKindermann}. It has been shown that interaction correction to the transmission coeffcients in normal quantum transport can be presented as
\begin{align}
\delta T &= -  G_Q Z T(1-T)  \ln \left[ \frac{\omega_{\rm H}}{\omega_{\rm L}}\right] \nonumber \\
&= - G_Q T(1-T) \pi \langle\hat{\Phi}^2\rangle.
\end{align}
With this, and the definition of $L_{\infty}$ (Eq. \eqref{eq:Linfty}) it becomes evident that the corresponding correction $ \Delta_r E^{(e)} \equiv \langle \hat{\Phi}^2\rangle/2 L_{\infty}$ to the even ground state energy originates from the transmission renormalization of $E^{(e)} = - E_A$,
\begin{align}
\label{eq:transrenorm}
    \Delta_r E^{(e)} = - \delta T \  \partial_T E_A.
\end{align}
Since the transmission coefficients in a given structure are obtained from experimental observations (for instance, by fitting phase dependence of the supercurrent), and the observation gives already {\it renormalized} quantities, it is not evident $\Delta_r E^{(e)}$ can ever be observed. However, if there are means to change $Z(\omega)$, one can trace the change of the transmission renormalization as well.

The integral in Eq. \eqref{eq:even-high} may diverge at low frequencies. To clarify the nature of this divergence we use the sum rule from Eq. \eqref{eq:sumRuleEven} to give the equivalent representation
\begin{align}
\Delta E^{(e)} =-\frac{1}{2}L I_{J}^2  +\frac{\langle \hat{\Phi}^2\rangle }{2 L_J}  +\int_0^\infty \frac{d\omega}{\pi }{\rm Re}(Z(\omega)) \nonumber \\ \times \int_0^\infty \frac{d\omega'}{\pi}\frac{{\rm Re}(Y^{(e)}(\omega')-G_QT)}{\omega + \omega'}.
\label{eq:even-low}
\end{align}
The integral now converges at low frequencies and the low frequency divergence is contained in the second term. That term can be written as
\begin{align}
\frac{\langle \hat{\Phi}^2\rangle }{2 L_J} = - \frac{1}{2} \partial^2_{\Phi} E_A \langle \hat{\Phi}^2\rangle,
\end{align}
to recognize the contribution due to adiabatic fluctuations of the flux/phase.

Let us turn to the odd ground state. In a single-channel model, the superconducting current in this state is absent, so there is no inductive term, and the remaining correction is obtained by replacing the admittance $Y^{(e)} \to Y^{(o)}$,
\begin{align}
\label{eq:oddEnergy}
\Delta E^{(o)} = \frac{1}{2} \int_{0}^{\infty} \frac{d\omega}{\pi } {\rm Im}(Z(\omega)Y^{(o)}(\omega)).
\end{align}
We recall that the low/high frequency inverse inductance vanish in the odd ground state, so we do not need the equivalent forms like Eqs. \eqref{eq:even-high} and \eqref{eq:even-low}. This is also clear from the fact that without interaction the odd ground state energy depends neither on transmission nor flux: therefore, neither transmission renormalization nor adiabatic fluctuations manifest themselves in the correction to this energy.

The correction to the doublet addition energy, $\delta^{\rm odd}$, is given by the difference of the corrections to the odd and even ground state energies,
\begin{align}
&\delta^{\rm odd} = \frac{1}{2}L I_{J}^2 + \Lambda; \\
&\Lambda \equiv  \frac{1}{2} \int_{0}^{\infty} \frac{d\omega}{\pi } {\rm Im}(Z(\omega)(Y^{(o)}(\omega)-Y^{(e)}(\omega)).
\label{eq:sigma}
\end{align}
Similar to $\Delta E^{(e)}$, there are two equivalent representations of $\Lambda$ encapsulating the divergences at low/high frequencies
\begin{align}
&\Lambda = -\frac{\langle \hat{\Phi}^2\rangle }{2 L_J} +\int_0^\infty \frac{d\omega}{\pi }{\rm Re} Z(\omega) \nonumber \\ &\times \int_0^\infty \frac{d\omega'}{\pi}\frac{{\rm Re}(Y^{(o)}(\omega')-Y^{(e)}(\omega'))}{\omega + \omega'};
\label{eq:sigma-low} \\
&\Lambda=  -\frac{\langle \hat{\Phi}^2\rangle }{2 L_{\infty}} -\int_0^\infty \frac{d\omega}{\pi \omega}{\rm Re}Z(\omega)\} \nonumber \\ & \times \int_0^\infty \frac{d\omega'}{\pi}\frac{\omega'{\rm Re}(Y^{(o)}(\omega')-Y^{(e)}(\omega'))}{\omega+\omega'}.
\label{eq:sigma-high}
\end{align}

The correction to the singlet addition energy $\delta^{\rm even}$ reads
\begin{align}\label{eq:delta_even}
\delta^{\rm even} = 2 \Lambda +  2E_A \bar{Y}  {\rm Im} Z(2E_A),
\end{align}
where $\bar{Y}$ was defined in Eq. \eqref{eq:bar_y}.
There is no inductive contribution, since the square of the supercurrent is the same in the ground and excited singlet states. The first term reflects the correction to the single quasiparticle energy: it is doubled since two quasiparticles are added. The second term corresponds to diagram (d) in Fig \ref{fig:Diagrams}. It has a simple interpretation in terms of circuit theory: if we regard the resonance in the admittance at $\omega \approx 2 E_A$ as an oscillator, the second term gives the shift of oscillator frequency owing to the environment (see \cite{atlongdistance} for details).

Finally, we define the {\it interaction signature} as the difference between the singlet and twice the doublet addition energies. It only depends on the inductive and resonant energy terms:
\begin{align}
\label{eq:signature}
\delta^{\rm even} - 2 \delta^{\rm odd} = 2E_A \bar{Y}  {\rm Im}Z(2E_A) -L I_{J}^2.
\end{align}

Let us now extend the above expressions to the multi-channel case where there are a set of ABSs with energies $E_{A,n}$ and corresponding transmission coefficients $T_n$. While this is straightforward, the details for the inductive contributions are important. The correction to the even ground state energy is given by Eq. \ref{eq:even} upon substitution of the full current and full admittance contributed by all channels,
\begin{align}
&I_{J} \to I_{\rm f} \equiv \sum_n I_{J,n}; \\
&Y^{(e)} \to \sum_n Y^{(e)}_n.
\end{align}

The odd ground state is obtained by adding a quasiparticle to the ABS with the lowest energy, $n=1$. There is a finite current in this state from the channels with $n >1$, resulting in an inductive correction, so that
\begin{align}
\label{eq:oddMultiChannel}
&\Delta E^{(o)} = - \frac{L}{2} \left(I^{(o)}_{\rm f}\right)^2 + \int_{0}^{\infty} \frac{d\omega}{2\pi } \Im\{Z(\omega)Y_{\rm f}^{(o)}(\omega)\}; \\
&Y_{\rm f}^{(o)} \equiv Y^{(o)}_1 +\sum_{n>1} Y^{(e)}_n,\\
&I^{(o)}_{\rm f} \equiv \sum_{n>1} I_{J,n}. \nonumber
\end{align}

The correction to the doublet addition energy to the $N$-th ABS reads
\begin{align}
\delta^{\rm odd}_N = {L} I_{J,N}(I_{\rm f} - I_{J,N}/2) +\Lambda_N,
\end{align}
where $\Lambda_N$ is given by Eq. \eqref{eq:sigma} with $Y^{(e, o)}\rightarrow Y^{(e,o)}_N$.

The correction to the singlet addition energy with both quasiparticles excited to the $N$-th ABS becomes
\begin{align}
\delta^{\rm even}_{N,N} &= 2 {L} I_{J,N}(I_{\rm f} - I_{J,N}) + 2 \Lambda_N  \nonumber \\
&+2E_{A,N} \bar{Y}_{N}  {\rm Im}Z(2E_{A,N}) .
\end{align}

Two quasiparticles can also be  excited in two different channels $N \ne M$. For a short junction, these transitions can not be produced by a modulation of phase, so the resonant term does not contribute to the interaction correction to the energy of this excited state. Therefore,
\begin{align}
\label{eq:2multichannel}
\delta^{\rm even}_{N,M} &= L (I_{J,N} + I_{J,M}) (I_{\rm f} - (I_{J,N} + I_{J,M})/2) \nonumber \\
&+  \Lambda_N + \Lambda_M.
\end{align}

\section{Even ground state: Correction to supercurrent}
\label{sec:evenground}

In this section we consider correction to the even ground state energy. Since the phase-independent part of the correction is not observable, we mostly concentrate on its derivative with respect to the phase, that is, on the correction $\delta I$ to the supercurrent.

We start with the limit of small $\omega_0$, that is, large $L$. The dominant part of the correction is the inductive contribution $- L I^2_J/2$ (first term in  Eq. \eqref{eq:even-low}). The correction to the current reads
\begin{align}
\frac{\delta {I}}{I_J} = - \frac{L}{L_J},
\end{align}
which is a trivial flux division between junction and loop inductance. This can be subtracted, and the second term in Eq. \eqref{eq:even-low} becomes important. It diverges logarithmically at small $\omega_0$
\begin{align}
\label{eq:logdivergence}
\Delta E^{(e)} =- 2 G_Q Z \partial^2_{\phi} E_A \ln\left[\frac{\omega'}{\omega_0}\right].
\end{align}
In the case where the environment contains photons with energy $\gtrsim \Delta$, $\omega'$ is not a simple cutoff in the impedance, but is instead calculated self-consistently to account for the high frequency divergence (see Eq. \eqref{eq:omega_p}). We can rewrite Eq. \eqref{eq:logdivergence} in terms of the correction to the current,
\begin{align}
\delta I = 2 G_Q Z \partial^2_{\phi} I_J \ln\left[\frac{\omega'}{\omega_0}\right].
\end{align}
In the limit of low transparency, this reproduces Eq.\eqref{eq:Jocorrection} and eventually upon extension to $G_Q Z \simeq 1$ reproduces the same critical resistance for Schmid transition. We stress that we cannot make $\omega_0$ arbitrary small: the junction inductance should greatly exceed $L$, so it is limited by $Z/L_J$. This sets the maximum value of log-divergent correction:
\begin{align}
\delta I = 2 G_Q Z \partial^2_{\phi} I_J \ln\left[\frac{\omega' L_J}{Z}\right],
\end{align}
which is non-analytical in $Z$.

 To specify $\omega'$, we need to understand which frequency ranges contribute to renormalization at high frequencies $\omega \gg \Delta$. We recall the separation between the integrals convergent at lower and higher frequency (Eqs. \eqref{eq:even-high} and \eqref{eq:even-low}) and obtain
 \begin{align}
 \frac{\Delta E^{(e)}}
{Z/2\pi}  = \ln\left[\frac{\Delta}{\omega_0} \right] \frac{1}{L_J} + \ln\left[\frac{\omega_C}{\Delta} \right] \frac{1}{L_\infty},
\end{align}
which gives the combined scaling from both the higher and lower cutoff in impedance. Here $\omega_C \gg \Delta$ is a cut-off frequency of the impedance (for instance, due to shunting capacitance).
Comparing this with
Eq. \ref{eq:logdivergence}, we find
\begin{align}\label{eq:omega_p}
\omega' \simeq \Delta \left(\omega_C/\Delta\right)^{L_J/L_\infty}.
\end{align}

Let us turn to an opposite limit of large $\omega_0 \gg \Delta$. To leading order, the energy correction is dominated by the logarithmically divergent term
$\langle \hat{\Phi}^2\rangle/2 L_{\infty} = -\delta T \partial_T E_A$. As discussed, we can regard it as an (unobservable) renormalization of the transmission coefficient, so in further consideration we absorb it in the definition of $T$. With this, the leading correction comes from the frequency range  $\omega_0 \gg \omega \gg \Delta$. The environment impedance in this frequency range is purely imaginary, $Z(\omega) \simeq -i\omega L$. As for the real part of the admittance, it's asymptotic at $\omega \gg \Delta$ notably contains a logarithmic factor arising from $Y_1$, and reads
\begin{align}
\frac{{\rm Re} Y_1(\omega)}{G_Q T} \approx 1 -  \left( (2-T)\cos\phi +T\right) \left( \frac{\Delta}{\omega}\right)^2 \ln\left[\frac{\omega}{\Delta}\right],
\end{align}
with logarithmic accuracy. This gives the following asymptote of the energy correction,
\begin{align}
\label{eq:logsuareEnergy}
\Delta E^{(e)} \approx \frac{L G_Q \Delta^2}{4\pi}  T \left( (2-T)\cos\phi +T\right)  \ln^2\left[\frac{\omega_0}{\Delta}\right].
\end{align}
The corresponding relative current correction
\begin{align}
\label{eq:logsquare}
\frac{\delta I}{I_J} = - \frac{ L G_Q \Delta}{\pi} (2 - T) \frac{E_A}{\Delta}  \ln^2\left[\frac{\omega_0}{\Delta}\right]
\end{align}
is negative. At small $T$ ($E_A \approx \Delta$) the correction depends neither on phase nor on $T$.

\begin{figure}[h]
    \includegraphics[width=\columnwidth]{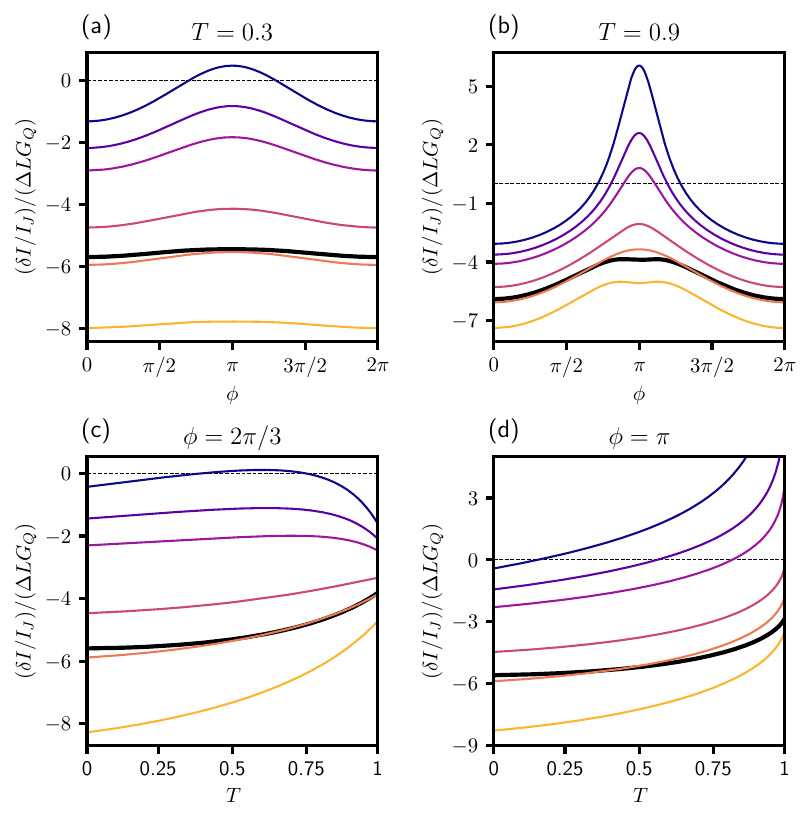}
	\caption{Relative correction to the even ground state superconducting current. In all plots, colored lines give the results for the external impedance model with $\omega_0/\Delta = [0.1, 0.5, 1, 3, 5, 10]$ (from dark blue to yellow). The black line gives the results for the MB model. The parameters are: (a) $T=0.3$ vs $\phi$, (b) $T=0.9$ vs $\phi$, (c) $\phi = 2\pi/3$ vs $T$, (d) $\phi = \pi$ vs $T$. In (d) the results for the external impedance model exhibit logarithmic divergence at $T\to 1$ (too weak to be visually resolved for large $\omega_0$), while the result for MB model approaches a finite limit $\approx - 2.86$. The black dashed line marks zero for reference.}
	\label{fig:dIe}
\end{figure}

So, if we disregard the renormalization of $T$ and log factors, the correction to the energy scales like $\Delta E^{(e)} \simeq L G_Q \Delta^2$, and the relative correction to the current $\delta I/I_J \simeq L G_Q \Delta$, for all $\omega_0$. We use this scaling to present our numerical illustrations (see Fig. \ref{fig:dIe}).

Before turning to the plots, let us address  the special case of $E_A \ll \Delta$. This is realized at close to ideal transmission ($R \ll 1;\, R \equiv 1-T$) and phases close to $\pi$ ($|\chi| \ll \pi;\, \chi \equiv \phi - \pi$). This defines a narrow parameter region ($|\chi| \simeq \sqrt{R}$) where $E_A = \Delta\sqrt{R + \chi^2/4}$. The superconducting current quickly switches between the opposite values $\pm e \Delta$ while the phase passes the region, so the junction inductance is negative and anomalously small, $ 1/L_J = -\pi G_Q R \Delta^4/E_A^3$, a factor of $\sqrt{R}$ smaller than the inverse inductance far from the region.

Let us turn to the interaction correction and first concentrate on the inductive part,
\begin{align}
\label{eq:indSmallEA}
 - \frac{L I_J^2}{2} &=  -  2\pi L G_Q \Delta^2 \frac{\chi^2}{4 \chi^2 +R}\\
&= - \frac{\pi L G_Q \Delta^2}{2}   + \Delta E^{(e)}_{\rm ind}; \\
\Delta E^{(e)}_{\rm ind} &\equiv \frac{\pi G_Q L\Delta^2}{2} \frac{\Delta^2 R}{E_A^2}.
\end{align}
Here we skipped a phase-independent term to define the energy $\Delta E^{(o)}_{\rm ind}$ which generates an anomalously large and narrow peak of the relative current correction in the region,
\begin{align}
\label{eq:slow}
\frac{\Delta I}{I_J} = \pi L G_Q \Delta \frac{R}{\left(\chi^2/4 + R\right)^{3/2}},
\end{align}
that is enhanced by a factor of $1/\sqrt{R}$ and can become of the order of one if the reduced junction resistance is comparable with the environment inductance $L$.

It turns out, however, that this spectacular peak persists only for "slow" environments where $\omega_0 \lesssim E_A$ and is canceled for faster environments. In the present limit there is a comparable contribution to the energy shift proportional to $Y_3$ (excitation of quasiparticle pairs with energy $2E_A$), which we combine with the inductive contribution to arrive at
\begin{align}
\Delta E^{(e)} = \Delta E^{(e)}_{\rm ind} (1-{\cal F}(2E_A)).
\label{eq:inductanceCancellation}
\end{align}
The function ${\cal F}(\omega)$ characterizing the environment generally expressed as
\begin{align}
{\cal F}(\omega) =\frac{1}{L} \int_0^\infty \frac{d\nu}{\pi \nu} \frac{2}{\nu + \omega}{\rm Re} Z(\nu),
\end{align}
which reduces to
\begin{align}
{\cal F}(\omega)  =  \frac{1 + 2 (\omega/\omega_0) \ln(\omega/\omega_0)/\pi}{1+ (\omega/\omega_0)^2}
\end{align}
for the external impedance model. The even part of ${\cal F}$ is immediately related to the ratio of the frequency-dependent and zero-frequency inductance, ${\cal F}(\omega) + {\cal F}(-\omega) = 2 L(\omega)/L$, where $L(\omega) \equiv -{\rm Im} Z(\omega) /\omega$.

\begin{figure}[h]
    \includegraphics[width=\columnwidth]{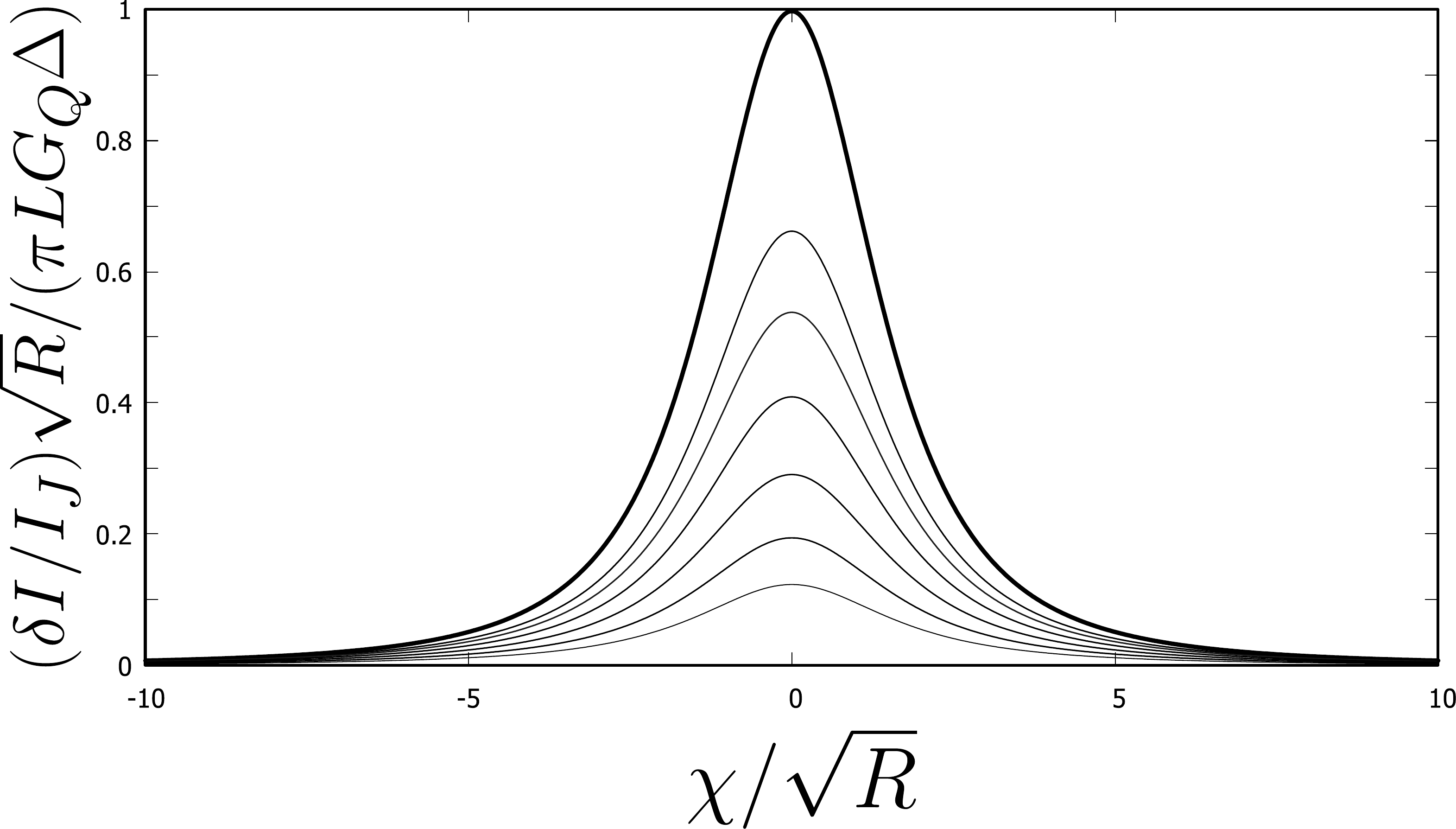}
	\caption{Interaction correction to the even ground state superconducting current at small $E_A$. The correction may be anomalously big ($\propto 1/\sqrt{R}$) in a narrow interval $\phi - \pi \equiv \chi \simeq \sqrt{R}$ at $\omega_0 \ll \sqrt{R} \Delta$, but is suppressed upon increasing $\omega_0$. For the upper thick curve, $\omega_0=0$. The subsequent curves correspond to $\omega_0/(\sqrt{R} \Delta)	= [0.5, 1.0, 2.0, 4.0, 8.0, 16.0]$ in descending order.} \label{fig:smallEA}
\end{figure}

The suppression of the peak feature with increasing $\omega_0 \simeq \sqrt{R} \Delta$ is illustrated in Fig. \ref{fig:smallEA}.
In the external impedance model, the suppressed peculiarity in the limit $\omega_0 \gg E_A$ is still a narrow peak
\begin{align}
\label{eq:fast}
\frac{\delta I}{I_J} = LG_Q \frac{\Delta^2}{\omega_0} \frac{2}{1 + \chi^2/4R}\ln\left[\frac{e^{1}\omega_0}{2\Delta\sqrt{R+\chi^2/4}}\right]
\end{align}
that diverges as $\ln(1/R)$ at $\chi=0$, for $R \to 0$. The divergence can be traced back to the existence of low frequency dissipation modes at $\omega \ll \Delta$, and therefore there is no such peak or divergence for the MB model.

Let us look at the plots in Fig. \ref{fig:dIe}. In all plots, we plot the ratio of the currents, $\delta I/I_J$. One should keep in mind that both $\delta I$ and $I_J$ are zero at $\phi =0, \pi$; and $T=0$. The ratio of the two however is finite upon approaching these values. The different color curves in the plots correspond to different $\omega_0$ in the external impedance model and the black line shows the results for the MB model. Qualitatively, the MB results are similar to that of external impedance model at $\omega_0 \simeq 5 \Delta$, with an important exception in Fig. \ref{fig:dIe} (d).

In Fig. \ref{fig:dIe} (a) we plot $\delta I/I_J$ at small $T=0.3$ versus $\phi$ at $\omega_0$ ranging from $0.1 \Delta$ to $5 \Delta$. We see that the results are of the same scale in the wide range of $\omega_0$, with a slow (logarithmic) decrease at larger $\omega_0$. For the smallest $\omega_0$, the energy correction mostly originates from the inductive term $-L I_J^2/2$, so $\delta I/I_J \approx - L/L_J$. Therefore, this correction is negative (positive) at $\phi = 0\, (\pi)$ where $L_J$ is positive (negative).  Upon increasing $\omega_0$, the shift becomes negative in the whole range of phases,
in accordance with Eq. \ref{eq:logsquare}. This also explains the small oscillating part $\propto - \cos\phi$. In Fig. \ref{fig:dIe} (b), we repeat the same plots, but at larger $T=0.9$, so $E_A$ is small near $\phi=\pi$. At small $\omega_0$, we observe the inductance peak which is gradually cancelled in accordance with Eq. \eqref{eq:inductanceCancellation} (see also Fig. \ref{fig:smallEA}).  At higher $\omega_0$ the peak is cancelled and the result in main follows Eq. \ref{eq:logsquare}. Note a small non-monotonous feature for small $\phi$ in the MB model and for $\omega_0 = 10 \Delta$.

In Fig. \ref{fig:dIe} (c-d) we fix $\phi$ and plot the relative correction versus $T$.
In (c) $\phi = 2\pi/3$. For small $\omega_0$ we see a non-monotous dependence on $T$, but for larger $\omega_0$ the relative shift increases monotonously with $T$ in accordance with the asymptote in Eq. \ref{eq:logsquare}. In (d) we fix $\phi = \pi$. The $T$-dependence of the correction manifests a logarithmic divergence as $T \to 1$ in agreement with Eq. \ref{eq:fast}. For external impedance model $\delta I/I_J \propto -\ln(1-T) \Delta/\omega_0$, while the MB result converges to $\approx - 2.38$ as $T\to 1$.

\section{Odd ground state: supercurrent}
\label{sec:oddground}
Let us turn to the interaction correction to the odd ground state. We reiterate that in the single-channel case, which we focus on, the odd ground state current is zero in absence of interactions. Therefore, the leading contribution comes from the "correction". In the multi-channel case, the odd ground state correction  is a superposition of $\Delta E^{(o)}$ in the highest transmission channel and $\Delta E^{(e)}$ in all other channels, augmented with the difference of inductive energies (see Eq. \eqref{eq:oddMultiChannel}).

The observable part of the energy is its phase-depended part, so we mostly concentrate on the supercurrent. In distinction from the correction to the even ground state, the odd ground state energy (Eq. \eqref{eq:oddEnergy})) does not contain the inductive contribution. It also does not contain the log divergent terms $\propto \langle \hat{\Phi}^2 \rangle$, since the low- and high-frequency inverse junction inductance is zero in the odd ground state. For this reason, the appropriate scales for the energy and supercurrent are: $I^{(o)} \simeq \Delta e T  (G_Q Z) $, $\Delta E^{(o)} \simeq \Delta T (G_Q Z)$.

Let us first address a special limit manifesting distinct interesting physics. We assume the ABS energy is close to the gap edge and define $\Omega \equiv  \Delta - E_A \ll \Delta$. This is realized at: i. low transmission $T \ll 1$,
ii. any transmission and $\phi \to 0$. Previous research (\cite{oddparity}, Supplemental Material) has addressed the low transmission limit, and here we extend this to arbitrary transmission. In either case $\Omega = (\Delta/2) T\sin^2\frac{\phi}{2}$ provided $\Omega \ll \Delta$.

We need the junction admittance at low frequencies $\omega \ll \Delta$. This is provided by $Y_5$ term (Eq. \eqref{eq:Y5}) responsible for the ionization of the quasiparticle trapped near the edge. The real part of the impedance therefore reads
\begin{align}
\frac{{\rm Re} Y^{(o)}(\omega) }{G_Q T} =
\theta(\omega -\Omega ) \pi \frac{ \Delta \sqrt{\Omega(\omega -\Omega)}}{\omega^2} \cos^2\frac{\phi}{2}.
\end{align}
To derive the corresponding imaginary part, we need to take into account that in the odd state the a.c. inverse inductance vanishes, $\lim_{\omega \to 0}{\rm Im}( Y(\omega) \omega) = 0$. With this,
\begin{align}
\label{eq:imY5}
&{\rm Im}Y^{(o)}(\omega) = G_Q T \pi \Delta \sqrt{\Omega} \cos^2\frac{\phi}{2} \nonumber \\
&\times \frac{\Theta(\Omega - \omega)\sqrt{\Omega -\omega}  - \sqrt{\Omega +\omega}}{\omega^2} + \frac{1}{\sqrt{\Omega} \omega}.
\end{align}
The last term shows that the inverse inductance absent at zero frequency, eventually restores to a constant value $L_{\Omega}^{-1}$ at relatively low frequencies $\omega \simeq \Omega$,
\begin{align}
\frac{1}{L_{\Omega}} \equiv G_Q T \pi \Delta \cos^2\frac{\phi}{2} = \frac{1}{2}\left( 1+ \frac{1}{\cos\phi}\right) \frac{1}{L_J}.\end{align}
This value is of the order of $1/L_J$ in the even ground state and exactly equals this as $\phi \to 0$.

The evaluation of interaction energy can be done analytically for the external impedance model with $\omega_0 \ll \Delta$, while $\omega$ and $\Omega$ can be of the same scale. It yields ($\omega_{\rm H}$ being an upper cut-off $\simeq \Delta$)
\begin{align}
\label{eq:oddOmegaCorrection}
\Delta E^{(o)} = G_Q Z \frac{T\Delta}{2} \cos^2\frac{\phi}{2} \left( \ln\left[\frac{\omega_{\rm H}}{\omega_0}\right] - X\left(\frac{\omega_0}{\Omega}\right)\right).
\end{align}
Here, $X(a)$ is a rather involved positive elementary function of $a>0$,
\begin{align}
&X(a) = \frac{2}{a} \left[ -{\rm arctanh}\left[\sqrt{\sqrt{1+a^2}-a}\right] \right. \nonumber \\
&\left. \times \sqrt{\sqrt{1+a^2}+a} \left(\sqrt{1+a^2} - a -1\right) + \right. \nonumber \\
&\left. {\rm arctan}\left[\sqrt{\sqrt{1+a^2}+a}\right] \right. \nonumber  \\
&\left. \times \sqrt{\sqrt{1+a^2}-a} \left(\sqrt{1+a^2} + a +1\right) -\frac{\pi}{2} \right],
\end{align}
with the following asymptotes:
\begin{align}
X(a) \to
\left\{ \begin{array}{ccc}
\ln\left[\frac{4 e^{1}}{a}\right] \ & {\rm as} \ & a \to 0; \\
\pi \sqrt{\frac{2}{a}} \ & \ {\rm as} \ & a \to \infty.\end{array} \right.
\end{align}

We see that the interaction correction is indeed $ \sim \Delta T  (G_Q Z) $ as expected. Let us look first  at the first term in  Eq. \eqref{eq:oddOmegaCorrection}. If we skip the energy-independent part, and associate $\Delta E^{(o)}$ with the renormalization $ - \delta\Omega$ of $\Omega$, it can be rewritten as
\begin{align}
\label{eq:wrongRenormalization}
\frac{\delta \Omega}{\Omega} =  G_Q Z \ln \left[ \frac{\omega_{\rm H}}{\omega_0} \right].
\end{align}
This resembles the low-frequency renormalizations given by Eqs. \ref{eq:Jocorrection}, \ref{eq:Jooddcorrection}, and \ref{eq:logdivergence}, $\omega_0$ being the lower cut-off. However, it comes with a different coefficient and eventually a wrong sign. So the analogy is superfluous. We come back to the renormalization in the next section.

We recall that for the odd ground state we do not expect a low-frequency renormalization since the a.c. inverse inductance vanishes. To confirm this, let us consider $\omega_0 \ll \Omega$. Taking the asymptote of $X(a)$ at small $a$, we obtain
\begin{align}
\label{eq:BigOmega0}
&\Delta E^{(o)}  = G_Q Z \frac{T\Delta}{2} \cos^2\frac{\phi}{2} \ln\left[\frac{\omega'_{\rm H}}{\Omega}\right]; \quad \frac{\omega'_{\rm H}}{\omega_{\rm H}} = 4 e^1,
\end{align}
which does not depend on $\omega_0$ at all, and the low-frequecy fluctuations are cut at $\Omega$. This is related to the inverse inductance persisting at frequencies $\gg \Omega$ and vanishing at zero frequency (see Eq. \eqref{eq:imY5}).

The current corresponding to Eq. \eqref{eq:BigOmega0} diverges as $\phi \to 0$,
\begin{align}
\label{eq:currentBigOmega0}
    I^{(o)} = -  2\Delta e T\frac{G_Q Z}{\phi},
\end{align}
and is also {\it negative} at positive phase. Thus, we need to explore the limit $\phi \to 0$ separately. Since $\Omega \propto \phi^2$ and $\omega_0 \gg \Omega$, we substitute the asymptote of $X(a)$ at small $a$ and skip the phase-independent term to arrive at
\begin{align}
\label{eq:SmallOmega0}
\Delta E^{(o)} = G_Q Z \frac{T\Delta}{2} \left(-\phi^2 \ln\left[\frac{ \omega_{\rm H}}{\omega}\right] - \pi \sqrt{ \frac{T\Delta}{\omega_0}} \frac{|\phi|}{2} \right).
\end{align}
The second term dominates for small $\phi$ and gives rise to the {\it current jump} predicted in \cite{oddparity}. The current approaches a finite value ${\rm sgn}(\phi) I_{\rm hj}$ as $\phi \to  0$, where
\begin{align}
\label{eq:enhancedCurrentJump}
    \frac{I_{\rm hj}}{ \Delta e T} = -\frac{\pi G_Q Z }{2 }  \sqrt{\frac{T \Delta}{\omega_0}} = -\frac{\sqrt{\pi G_Q Z} }{2} \sqrt{\frac{L}{L_J}}.
\end{align}
For $\omega_0 \simeq \Delta$, $I_{\rm hj} \simeq -\Delta e T^{3/2} (G_Q Z) $, while for smaller $\omega_0$ (large inductance) the current is enhanced $\sim \sqrt{\Delta/\omega_0}$. Since $L$ cannot exceed $L_J$, the enhancement is limited by $I_{\rm hj} \simeq \sqrt{G_Q Z} \Delta e T$. The current enhancement extends to a narrow region of phases $\phi \simeq \phi_c \equiv \sqrt{8 \omega_0/\Delta T}$. There is a universal non-trivial dependence of the current on phase in this region
\begin{figure}
    \includegraphics[width=\columnwidth]{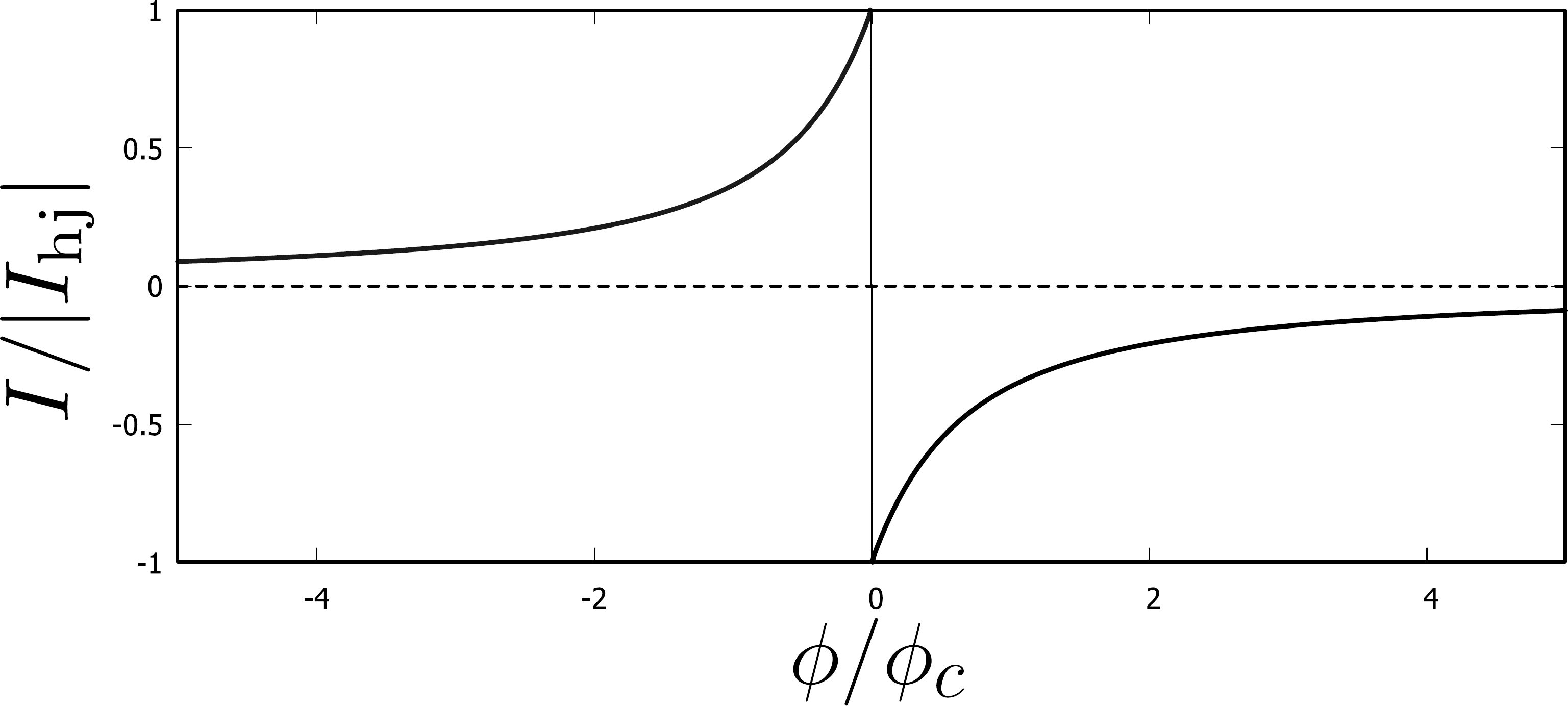}
    \caption{Odd ground state current jump at $\phi=0$. The solid line shows the universal dependence (Eq. \eqref{eq:universalCurrentJump}) of the current enhanced at $\omega_0 \ll \Delta$ and concentrated in a narrow interval $\phi_c \ll 1$.}
    \label{fig:UniversalCurrent}
\end{figure}

\begin{align}
\label{eq:universalCurrentJump}
&\frac{I}{|I_{\rm hj}|} = U\left(\frac{\phi}{\phi_c}\right); \nonumber \\
& U(a) = \frac{\sqrt{2} X'(a^{-2})}{\pi a^3}.
\end{align}
The graph of $U$ is shown in Fig. \ref{fig:UniversalCurrent}. For small $a$, $U(a) \rightarrow -{\rm sgn}(a)$ which gives the current jump. The large $a$ asymptote is
\begin{equation}
	U(a) \to -  \frac{\sqrt{2}}{\pi a}\quad {\rm as} \quad a\to \infty,
\end{equation}
which corresponds to Eq. \eqref{eq:currentBigOmega0}.

\begin{figure}[h]
    \includegraphics[width=\columnwidth]{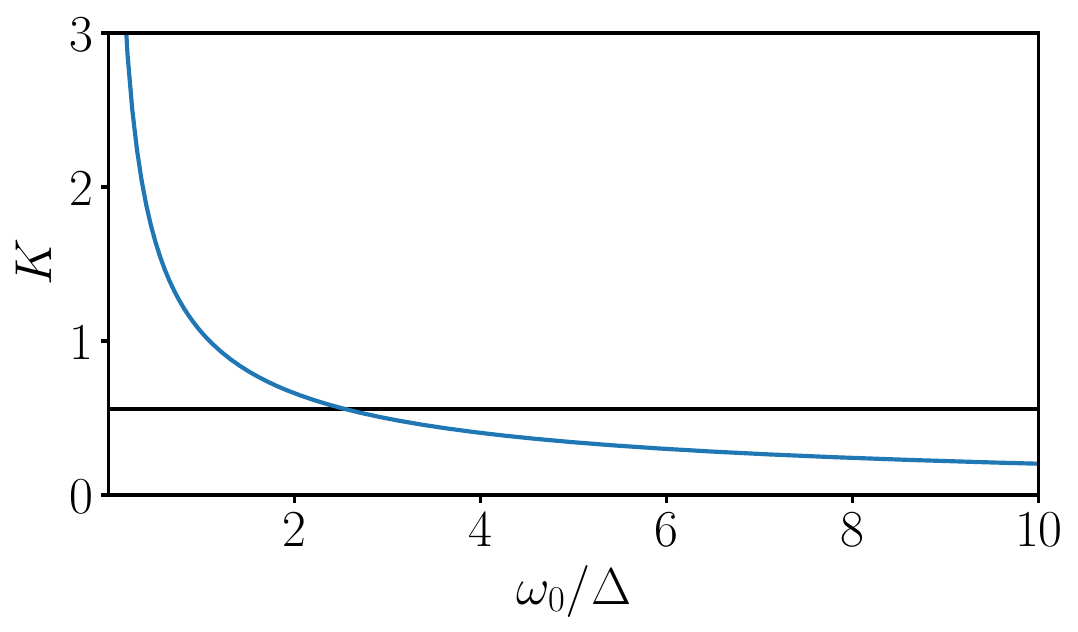}
	\caption{The dimensionless coefficient $K$ defining the odd ground state current jump (Eq. \ref{eq:currentJumpGeneral}). In blue, $K$ vs frequency $\omega_0/\Delta$ of the external impedance model. For reference the result for the MB model is given by the black line fixed at $K\approx 0.56$. } \label{fig:K}
\end{figure}

Let us next turn to the case of an arbitrary impedance. The current jump can be expressed
\begin{align}
\label{eq:currentJumpGeneral}
    I_{\rm hj}= - \Delta eT^{3/2}  G_Q Z   K,
\end{align}
where the dependence on the environment is fully contained in the positive dimensionless coefficient $K$, given by
\begin{align}
\label{eq:forK}
K=\int_0^\infty \frac{d \omega \Delta  {\rm Re} Z(\omega)}{\pi \omega^2 Z } ( {\cal I}\left(\frac{\omega}{\Delta} + 1\right) - {\cal I}\left(\frac{\omega}{\Delta} - 1\right) ),
\end{align}
and the dimensionless function ${\cal I}(x)$ is defined as
\begin{align}
{\cal I}(x) =
\left\{
\begin{array}{lll}
    - \sqrt{1-x^2} \ {\rm arctan}\left[\sqrt{\frac{1-x}{1+x}}\right] &{\rm for} \ & |x|<1; \\
     \sqrt{x^2-1} \ {\rm arctanh}  \left[\sqrt{\frac{x-1}{x+1}}\right]&{\rm for} \ & x>1.
\end{array}
\right.
\end{align}
We note that $Z$, which we have only defined in relation to the MB and external impedance models, cancels between Eqs. \eqref{eq:currentJumpGeneral} and \eqref{eq:forK}. Therefore, for a general impedance model it can be an arbitrary impedance that characterizes the scale.

In Fig \ref{fig:K} we plot $K$ versus $\omega_0$ for the external impedance model. The asymptotes of $K$ are:
\begin{align}
\label{eq:Kasymptots}
K = \left\{ \begin{array}{lll}
\frac{\Delta}{2 \omega_0} \ln\left[\frac{2e^{1}\omega_0}{\Delta}\right] &{\rm for} \ & \omega_0 \gg \Delta; \\
 \frac{\pi}{2} \sqrt{\frac{\Delta}{\omega_0}}&{\rm for} \ & \omega_0 \ll \Delta,
 \end{array} \right.
\end{align}
where the latter limit corresponds to Eq. \eqref{eq:enhancedCurrentJump}. For the MB model, $K \approx 0.56$.

The leading order contribution to $\Delta E^{(o)}$  for $\omega_0 \gg \Delta$ is the same as for the even ground state (see Eq. \ref{eq:logsuareEnergy}), and therefore we obtain the leading contribution to the current from Eq. \eqref{eq:logsquare},
\begin{align}
\label{eq:logsquare_odd}
\frac{I^{(o)}}{\Delta e T }= - Z G_Q \frac{ 2 - T}{ 2\pi}   \frac{\Delta}{\omega_0} \ln^2\left[\frac{\omega_0}{\Delta}\right] \sin \phi.
\end{align}
It vanishes at $\phi=0$ so it does not contribute to the current jump. This is consistent with the asymptote in Eq. \eqref{eq:Kasymptots}, which originates from a lower order $\simeq (\Delta/\omega_0) \ln(\omega_0/\Delta)$ contribution to the energy shift. The ratio of the current jump to the typical current is thus $\simeq \ln(\omega_0/\Delta)^{-1}$.

\begin{figure}[h]
    \includegraphics[width=\columnwidth]{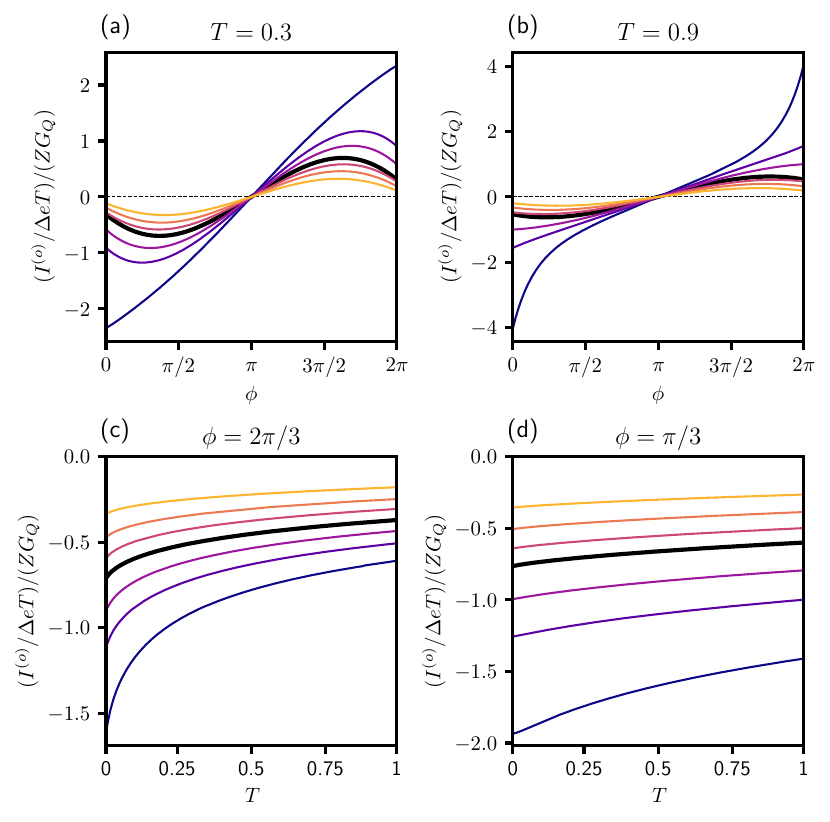}
	\caption{Odd ground state superconducting current normalized to the typical scale $ \Delta e T (G_Q Z) $. The colored lines show the current for the external impedance model with  $\omega_0/\Delta = [0.1, 0.5, 1, 3, 5, 10]$ (from dark blue to yellow). The black curve gives the current for the MB model. The parameters corresponding to the panes are: (a) $T=0.3$ vs $\phi$, (b) $T=0.9$ vs $\phi$, (c) $\phi=2\pi/3$ vs $T$, (d) $\phi=\pi/3$ vs $T$. }
	\label{fig:dIo}
\end{figure}

The odd ground state supercurrent is plotted vs phase/transmission in the panes of Fig. \ref{fig:dIo}. In (a-b) we fix $T$ and plot the current vs phase various $\omega_0$. In both panes we see the current jump at $\phi=0,\, 2\pi$ which is significant for all the curves, but enhanced at small $\omega_0$ in accordance with Eqs. \eqref{eq:enhancedCurrentJump} and \eqref{eq:currentJumpGeneral}. For large $\omega_0 \gg \Delta$ the current is $\propto 1/\omega_0$ up to logarithmic prefactors, and approaches a sine curve, consistent with Eq. \ref{eq:logsquare_odd}. Comparing (a) and (b), we see that the magnitude of the current at $\phi=0$ increases with $T$. Note that the current is normalized by $T$, so the ratio of the normalized current jumps are $\sqrt{3} \approx 1.7$ according to Eq. \eqref{eq:currentJumpGeneral}. In (b) we see that at small $\omega_0$ the current is concentrated at small $\phi$, matching Eq. \ref{eq:universalCurrentJump} and Fig. \ref{fig:UniversalCurrent}.

In panes (c-d) in Fig. \ref{fig:dIo} we explore the $T$ dependence at fixed phases $\phi=2\pi/3, \, \pi/3$. For large $\omega_0$ the curves becomes flat with a weak positive slope, consistent with the asymptote Eq. \eqref{eq:logsquare_odd}. For smaller $\omega_0$ the curves bend more strongly close to $T=0$, which is especially noticeable in (c). This indicates an enhanced contribution $\propto T^{3/2}$. Note however, the prefactor of the $T^{3/2}$ contribution appears to have opposite sign compared to the current at $\phi = 0^+$.

\section{Corrections to addition energies}
\label{sec:add}
In this section, we consider the corrections to addition energies, mostly concentrating on the correction to the doublet addition energy $\delta^{\rm odd}$.

Let us first consider the logarithmic renormalization of this quantity by quantum fluctuations. We assume the ABS energy close to the gap edge, defining bound energy $\Omega \equiv \Delta-E_A$ and its correction $\delta \Omega = - \delta^{\rm odd}$. We combine the renormalizations of odd (Eq. \eqref{eq:BigOmega0}) and even (Eq. \eqref{eq:logdivergence}) ground state energies to obtain the dependence of $\delta \Omega$ on the upper cutoff $\omega_H$:
\begin{align}
\label{eq:fixedRenormalization}
\frac{\delta{\Omega}}{\Omega} = G_Q Z \ln \left[ \frac{\omega_{\rm H}}{\Omega} \right] - 2 G_Q Z \ln \left[ \frac{\omega_{\rm H}}{\omega_0} \right].
\end{align}
If we look at the dependence on high-energy cut-off, we fix the problem with Eq. \eqref{eq:wrongRenormalization} and recover Eq. \eqref{eq:Jooddcorrection} that was utilized in \cite{oddparity} to predict that the Schmid transition in the odd state occurs at a larger critical impedance. Note the different low energy cutoffs in two terms in Eq. \eqref{eq:fixedRenormalization}. This implies that if $\Omega$ is initially of the order of $\Delta$, the renormalization at $G_Q Z \simeq 1$ proceeds in two stages. First, the renormalization reduces $\Omega$ to values $\ll \Delta$ at the "rate" $-2G_QZ$. Second, the further reduction takes place at the half rate, $-G_QZ$.

Further on, we follow the line we choose while discussing even ground state and absorb the renormalization of $T$ in the definition of $T$. With this, it is important to assess the value of the correction at $\phi=0$. Indeed, near this value phase the energy of the ABS approaches the gap edge with the orthogonal wave functions at positive/negative phase. A non-zero correction would mean something drastic: either a destruction of the bound state if positive, or the mixing of orthogonal states if negative. It is important to state that in general $\delta^{\rm odd}=0$ at $\phi =0$. Indeed, the inductive contribution cancels since $I_J=0$. The remaining contribution is proportional to $Y^{(o)}-Y^{(e)} =  Y_5 + Y_3 -Y_2/2$, and all these admittances $\to 0$ as $\phi \to 0$. The dominant contribution at small $\phi$ is therefore $\propto |\phi|$: it is the current jump discussed in the previous section. Therefore,
\begin{align}
\label{eq:smallPhi}
\delta^{\rm odd}  = - \frac{|\phi|}{2} \Delta \ G_Q Z T^{3/2} K;
\end{align}
where $K$ is defined in Eq. \eqref{eq:forK}.

In another interesting limit of $E_A \ll \Delta$, the behavior of $\delta^{\rm odd}$ is dominated by the corrections to the even ground state energy discussed in Sec. \ref{sec:evenground} (See Eqs. \eqref{eq:indSmallEA} and \eqref{eq:inductanceCancellation}):
\begin{align}
\label{eq:deltaOddInductanceCancellation}
\delta^{\rm odd} = \frac{\pi L G_Q \Delta^2}{2} \left[ 1 - \frac{ 1 -{\cal F}(2E_A)}{1 + \chi^2/4R} \right].
\end{align}
For "slow" environments, this gives a narrow ($\chi \simeq \sqrt{R}$) dip of height $\pi L G_Q \Delta^2/2$ and flattens to a constant value for "faster" environments.

In the limit of $\omega_0 \gg \Delta$, the correction is dominated by the $\simeq 1/\omega^2$ tails of the admittances and reads
\begin{align}
\label{eq:deltaOddLargeOmega}
&\delta^{\rm odd} = \frac{L G_Q \Delta^2T^{1/2}|\sin\frac{\phi}{2}|}{2}  \nonumber \\ & \times \left[\frac{2(1-T)\Delta}{E_A}+ \frac{(T - 2)E_A}{\Delta} \right] \ln\left[\frac{\omega_0}{\Delta}\right],
\end{align}
with log accuracy. This expression exhibits a current jump at $\phi=0$ and no singularity at $E_A=0$. This asymptote changes sign at $\tan(\phi/2) = 1/\sqrt{1-T}$ (that is, $\phi \in (\pi/2,3\pi/2)$).

\begin{figure}[h]
    \includegraphics[width=\columnwidth]{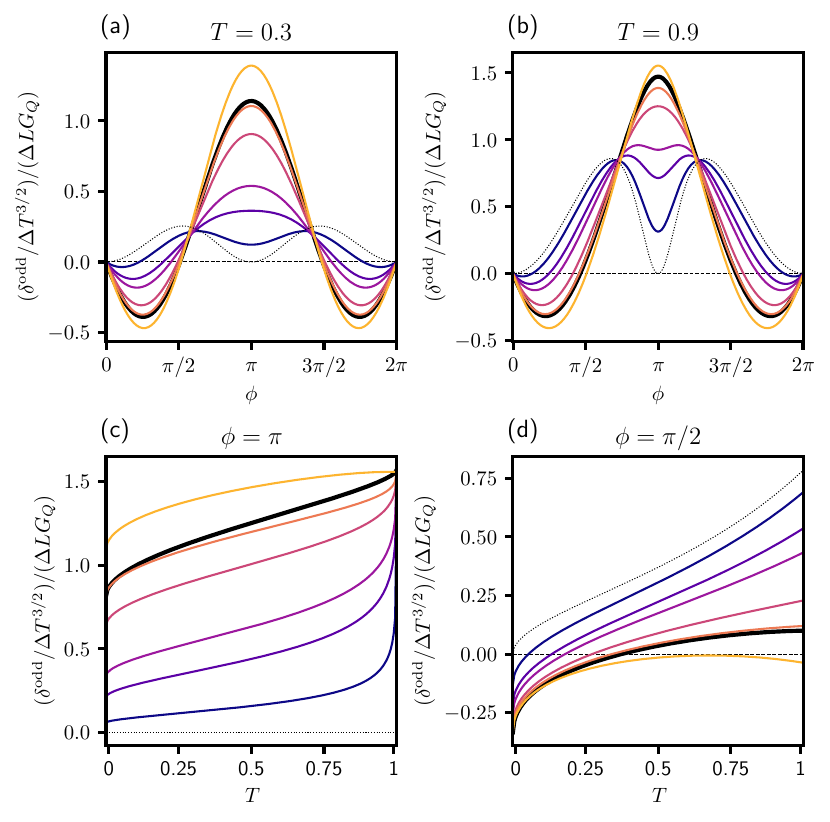}
	\caption{Correction to the doublet addition energy. The colored lines correspond to the external impedance model with $\omega_0/\Delta = [0.1, 0.5, 1, 3, 5, 10]$ (from dark blue to yellow), and the solid black line gives the shift for the MB model. The parameters corresponding to the panes are as follows: (a) $T=0.3$ vs $\phi$, (b) $T=0.9$ vs $\phi$, (c) $\phi=\pi$ vs $T$, (d) $\phi = \pi/2$ vs $T$. The dotted black line shows the inductive contribution to the shift, which is the only part that survives in the limit $\omega_0 \rightarrow 0$. In pane (c) all the solid lines approach $\pi/2$ as $T\to 1$. The dashed black line shows zero for reference.}
\label{fig:delta_odd}
\end{figure}

\begin{figure}[h]
    \includegraphics[width=\columnwidth]{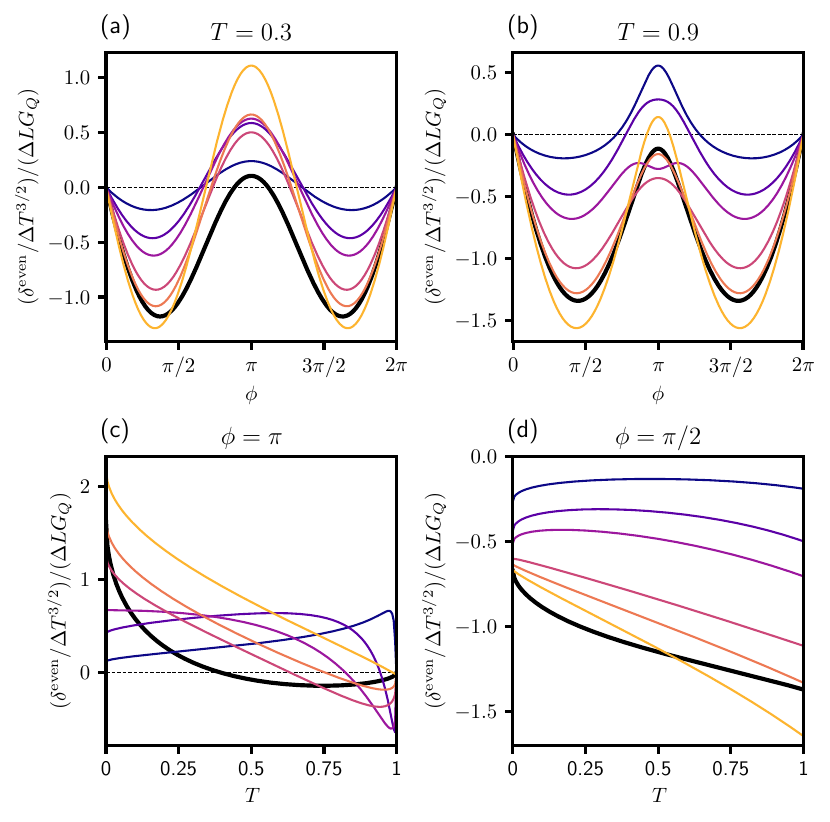}
	\caption{Correction to the singlet addition energy. The colored lines correspond to the external impedance model with $\omega_0/\Delta = [0.1, 0.5, 1, 3, 5, 10]$ (from dark blue to yellow), and the solid black line gives the shift for the MB model. The parameters corresponding to the panes are as follows: (a) $T=0.3$ vs $\phi$, (b) $T=0.9$ vs $\phi$, (c) $\phi=\pi$ vs $T$, (d) $\phi = \pi/2$ vs $T$. Note that in (c) the curves develop sharp negative peaks close to $T=1$ (not resolved for the dark blue curve) before they settle at $0$ for $T=1$. The dashed black line shows zero for reference.}
	\label{fig:delta_even}
\end{figure}

We illustrate the odd ground state shift for various parameters with the plots in Fig. \ref{fig:delta_odd}. Because $\delta^{\rm odd} \propto T^{3/2}$ at small $T$ we normalize $\delta^{\rm odd}$  to the typical scale $T^{3/2} L G_Q \Delta^2$. At this scale, the contribution related to the current jump are $\propto \sqrt{ \omega_0}$ at small $\omega_0$, while inductive contribution (shown by the dashed lines) remains constant.

In Fig. \ref{fig:delta_odd} (a) we fix ($T=0.3$) and the inductive contribution ($\propto T^2$) is small and even for the smallest $\omega_0$ it competes with contribution related to the current jump $\propto T^{3/2} \sqrt{\omega_0}$. This can be seen from the significant difference between the dark blue and dashed curve. At larger $\omega_0$ the shift increases in magnitude, consistent with Eq. \eqref{eq:deltaOddLargeOmega}. According to Eq. $\eqref{eq:deltaOddLargeOmega}$ the shift in panes (a-b) in Fig. \ref{fig:delta_odd} should change sign at $\pm 2 {\rm arctan}(1/\sqrt{1-T})$ in the limit of $\omega_0 \rightarrow \infty$. However, in the figure there are clear deviations from this prediction since the asymptote is logarithmic and the next to the leading constant gives a significant contribution, even for the largest $\omega_0 = 10 \Delta $. In Fig. \ref{fig:delta_odd} (b) we fix ($T=0.9$), and the inductive contribution dominates at small $\omega_0$, which is seen be the convergence of the colored lines to the dashed line for small $\omega_0$. Upon increasing $\omega_0$, consistent with the asymptote in Eq. \eqref{eq:deltaOddLargeOmega}, the curves develop a peak at $\phi=\pi$ and a zero crossing that approach the interval $(\pi/2, 3\pi/2)$, similar to in (a).

In Fig. \ref{fig:delta_odd} (c) we fix $\phi=\pi$ and see a rather featureless dependence on $T$ except for $T\to 1$ were we approach the regime of small $E_A$. All the curves eventually settle at $\pi/2$, as seen from Eq. \eqref{eq:deltaOddInductanceCancellation}, for $E_A \rightarrow 0$. The curves increase monotonously with $\omega_0$ in agreement with Eq. \eqref{eq:deltaOddLargeOmega} which is positive for $\phi=\pi/2$ (except at $T=1$ where the asymptote is $0$). In Fig. \ref{fig:delta_odd} (d) we see monotonic increase with $T$ except in the curve with the largest $\omega_0$. This curve is also always negative in contrast to other results that change sign at a certain $T$ that depends on $\omega_0$. For the value $\phi = \pi/2$ the asymptote \eqref{eq:deltaOddLargeOmega} is negative except at $T=0$, where it crosses zero. This explains why the curves with larger $\omega_0$ are pushed down and eventually becomes negative in the whole interval.

We do not talk about the correction to the singlet addition energy $\delta^{\rm even}$ in detail and instead thoroughly discuss the interaction signature $ \delta^{\rm even} - 2\delta^{\rm odd}$ in the next section. Let still briefly comment on the difference between the plots for $\delta^{\rm odd}$ in Fig. \ref{fig:delta_odd} with those for $\delta^{\rm even}$ with the same parameters in Fig. \ref{fig:delta_even}. The most striking difference is that the curves no longer converge to the inductive energy for small $\omega_0$, which does not contribute to $\delta^{\rm even}$.

Another striking feature is in Fig. \ref{fig:delta_even} (c), where the curves develop very sharp peaks as $T \rightarrow 1$. This parameter regime corresponds to small $E_A \ll \Delta$, and we can show that the curves approach
\begin{equation}
	\frac{\delta^{\rm even}}{\Delta^2 L G_Q} \rightarrow \frac{2x\ln(x)}{x^2 + 1}; \quad x \equiv  \sqrt{1-T} \frac{2\Delta }{\omega_0}.
\end{equation}
This indeed changes sign at a potentially small $1-T=\omega_0^2/(4\Delta^2)$,  before $\delta^{\rm even}$ develops a peak of order $\simeq \Delta^2 L G_Q $, and eventually settle at $\delta^{\rm even}=0$ for $T=1$.  For $\omega_0 = 0.1 \Delta$, $\delta^{\rm even}$ changes sign at $1-T =0.0025$, which is not resolved in Fig. \ref{fig:delta_even} (c). This regime should be compared to the peaks close to $\phi=\pi$ in Fig. \ref{fig:signature} (b), which we come back in the next section.

Otherwise, the curves in all the panes of Fig. \ref{fig:delta_even} seem to develop similar asymptotes at large $\omega_0$, and indeed the large $\omega_0$ asymptote for $\delta^{\rm even}$ is given by two times the asymptote for $\delta^{\rm odd}$ in Eq. \eqref{eq:deltaOddLargeOmega}.

\section{Interaction signature}
\label{sec:signature}

A straightforward way to reveal the interaction experimentally is to compare the addition energy of one and two quasiparticles: if the quasiparticles do not interact, the latter is two times bigger. Thus, the difference $\delta^{\rm even} - 2 \delta^{\rm odd}$ signifies the interaction. It is given by a simple formula \eqref{eq:signature} that combines an inductive contribution with a circuit-theory correction due to the resonance of the Andreev level qubit. It is instructive to bring the {\it interaction signature} to the following form:
\begin{align}
\label{eq:signaturesimple}
    \delta^{\rm even} - 2 \delta^{\rm odd} = -L(2E_A) I^2_\perp - L I_{J}^2,
\end{align}
where $L(\omega) \equiv -\Im Z(\omega) /\omega$ is the frequency dependent inductance and
\begin{align}
&I^2_\perp = \pi G_Q T^2(1-T) \frac{(\Delta \sin\frac{\phi}{2})^4}{E_A^2};\\
&\frac{I_\perp^2}{I_J^2} = (1-T) \tan^2\frac{\phi}{2},
\end{align}
gives the intensity of current oscillations in a superposition state of the Andreev level qubit. In our impedance models $L(\omega)$ is always positive, therefore the signature is always negative, corresponding to {\it attraction} between the quasiparticles trapped in an ABS. Yet this simple conclusion does not hold in general. For instance, if the environment contains a narrow-line oscillator, $L(2 E_A)$ diverges and changes sign when $2 E_A$ approaches the oscillator frequency, so the sign of the signature may be arbitrary.

The first term in Eq. \eqref{eq:signaturesimple} vanishes for "slow" environments, and the second term does not depend on the environment frequency $\omega_0$. We note that the signature has a familiar scale $L G_Q \Delta^2$ and vanishes as $\phi \to 0$. For small $T$, the signature is $\propto T^2$.

\begin{figure}[h]
    \includegraphics[width=1.0\columnwidth]{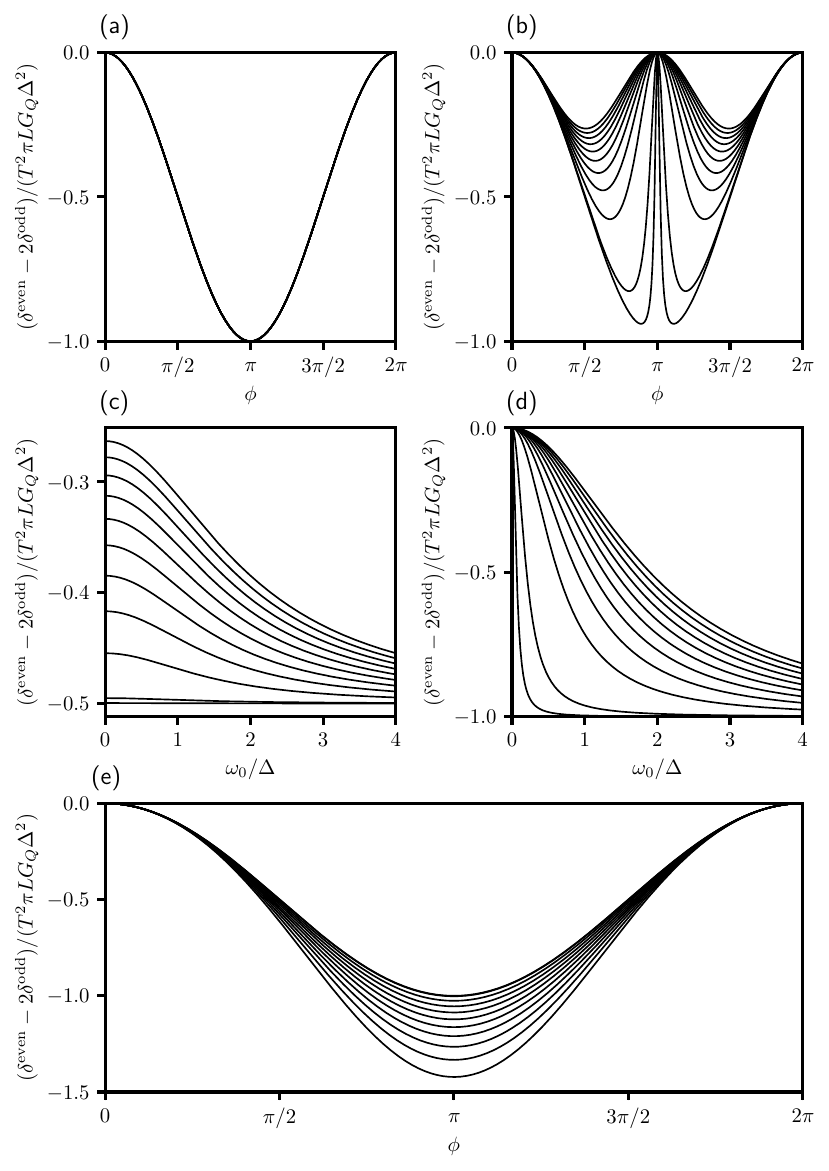}
	\caption{The interaction signature. In all plots, different curves correspond to T = [0.1,0.2,0.3,0.4,0.5, 0.6, 0.7, 0.8,0.9,0.99,0.999]. All the curves coincide in (a), in (b-d) the order is from upper to lower curve, and in (e) the order is from lower to upper curve. The signature is plotted vs phase in: (a) for the "fast" environment ($L(\omega)=L$), in (b) for the "slow" environment (only the inductive term), and in (e) for the MB model. In (c) and (d) the signature is plotted vs $\omega_0/\Delta$ for $\phi=\pi/2$ and $\phi=\pi$ respectively.}
	\label{fig:signature}
\end{figure}

We illustrate the interaction signature with the plots in Fig. \ref{fig:signature}. To bring the curves for different $T$ to a similar scale we normalize by  $T^2$. This works especially well in Fig. \ref{fig:signature} (a), where we plot the signature versus phase for the "fast" environment ($\omega_0 \to \infty$, $L(\omega)=L$). The signature becomes $\propto T^2$,
\begin{align} \label{eq:signature_fast}
    \delta^{\rm even} - 2 \delta^{\rm odd} = -\pi G_Q T^2 L \Delta^2 \sin^2\frac{\phi}{2},
\end{align}
so all the curves for different $T$ coincide. In Fig. \ref{fig:signature} (b) we plot the signature for a "slow" environment ($\omega_0 \to 0$), where only the inductive term matters. The narrow peak around $\phi=\pi$ for $T \to 1$ is a signature of the quickly changing $I_J$.  In Figs. \ref{fig:signature} (c) and (d) we illustrate the transition between "slow" and "fast" environments, plotting the interaction signature versus $\omega_0$: it always decreases with increasing $\omega_0$, except for $T \approx 1$ and $\phi \ne \pi$ where it remains constant. In (a-d) of Fig. \ref{fig:signature} the normalized signature is never smaller than $-1$. This can be understood as an artifact of $L(2E_A)<L$ in the external impedance model. In Fig. \ref{fig:signature} (e) we plot the signature for the MB model. Because $L(2E_A)\geq L$ in the MB model we see that the normalized signature can become smaller than $-1$. Moreover, this results in a "flipped" ordering compared to (a-d), so the lowest line corresponds to the smallest $T$ in (e). The sharp peak from the inductive contribution $-L I_J^2/2$ around $\phi=\pi$ for $T \to 1$ (seen in Fig. \ref{fig:signature} (b)) is cancelled by another peak-like contribution from the "qubit term" $-L(2E_A) I_\perp^2/2$. This results in smooth lines around $\phi=\pi$ as seen in Fig. \ref{fig:signature} (e). This cancellation is quite general because $E_A \rightarrow 0$ whenever $\phi\to \pi$ and $T \to 1$. Therefore, for an impedance that is smoothly changing around zero (i.e. not a "slow" environment) $L(2E_A) \rightarrow L$, and we recover Eq. \eqref{eq:signature_fast}.

\section{When the corrections are significant}
\label{sec:nonpert}
Let us point out two parameter regimes where the corrections become significant even at small $L, Z$, so perturbation theory fails and the solution of a dedicated model is required to capture the essential physics. We do not provide the solutions, which is beyond the scope of this work, but rather formulate the problems for more detailed research.

\begin{figure}[h]
    \includegraphics[width=\columnwidth]{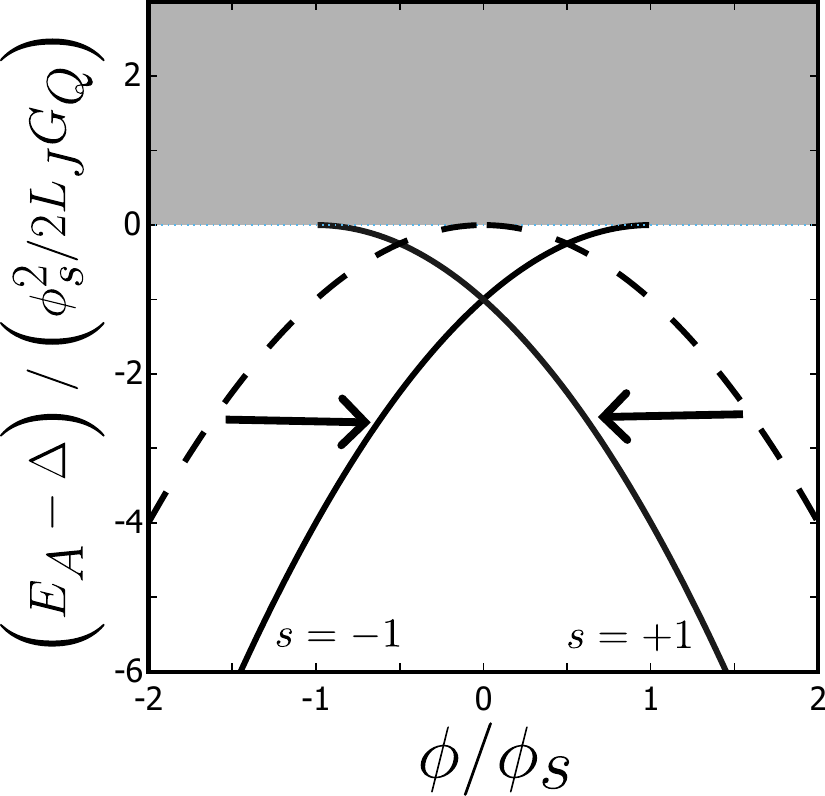}
	\caption{ABS energies near $\phi =0$ according to \cite{oddparity}. Dashed lines: two branches of different superpositions $s$ without interaction. Solid lines: the branches are shifted by interaction by $s \phi_s$. }
	\label{fig:shifts}
\end{figure}

The first parameter region is that of small $\phi$, where ABS energies approach the gap edge. There, the corrections to the odd ground state energy and addition energies are proportional to $|\phi|$ and thus exceed the non-interacting values $\propto \phi^2$ at sufficiently small $\phi$. It has been suggested in~\cite{oddparity,oddparity2} that this complex many-body problem has a simple solution. In fact, the bound state wave functions at positive and negative $\phi$ are different corresponding to different ($s =\pm 1$) superpositions of quasiparticles on the two sides of the junction. Without interaction, at $|\phi| \ll \pi$ they form two parabolic branches (see Fig. \ref{fig:shifts}) that touch each other and the gap edge precisely at $\phi=0$. According to \cite{oddparity}, the interaction shifts these branches in phase in opposite directions, $s \phi_s$ being the shifts (Fig. \ref{fig:shifts}). This reproduces a current jump in the odd state which we also see in our calculations.

The shift corresponding to the correction \eqref{eq:smallPhi} reads:
\begin{align}
\label{eq:phis}
\phi_s = 2 T^{1/2} K G_Q Z.
\end{align}
Its typical value is $\phi_s \simeq G_Q Z$. For "slow" environments this can be increased to $\phi_s \simeq \sqrt{G_Q Z}$ (see Eq. \eqref{eq:enhancedCurrentJump}).

A qualitative difference is the presence of two ABSs in the interval $|\phi| < \phi_s$. The present research indicates that the same picture holds for two-particle excitations, and the interaction between the quasiparticles can be neglected	at small $\phi$ (Sec. \ref{sec:signature}). This gives 6 possible two-quasiparticle bound states in this interval: two singlets with quasiparticles at the same branches, a singlet and a triplet with quasiparticles at opposite branches.

This certainly motivates a concentrated research on interaction effects in this parameter region. We stress that there are many physical factors affecting the rather fragile bound states close to the gap edge: i. weak energy dependence of scattering matrix~\cite{Yokoyama2015}, ii. spin-orbit splitting~\cite{Yokoyama2015}, iii. slightly different superconducting energy gaps on the two sides of the junction~\cite{Yip}, iv. disorder-driven modification of the spectral properties of the bulk superconductors near the gap edge \cite{Larkin1972}. So this has to be a multifactor investigation.

Another parameter regime is $T \to 1$ and $\phi \to \pi$, where $E_A \ll \Delta$. This region has already been noticed in Ref.~\cite{zazunov2} where the authors identified the relevant model as a spin-boson \cite{Legget2} Hamiltonian. In our notations, it reads
\begin{align}
\hat{H}_{\rm SB} = \Delta\left(\frac{\chi + \hat{\phi}}{2} \sigma_z +\sqrt{R} \sigma_x \right) + \hat{H}_{\rm env},
\end{align}
where $\sigma_{z,x}$ are Pauli matrices in the space of the ground/excited singlets and $\hat{\phi}$ represents fluctuations of phase induced by the environment described by the bosonic $\hat{H}_{\rm env}$. The conclusion of the authors was that the effect of the environment can be incorporated into a small renormalization of $R$.

However, there is an alternative view on the same problem, by default involving no quantum mechanics~\cite{Houzet}. Since the junction is in series with the inductance $L$, the actual $\tilde{\chi}$ dropping at the junction is not the bias value $\chi$ but has to be determined from the minimization of an energy involving both the junction and the inductance,
\begin{align}
E = - \Delta \sqrt{R + \tilde{\chi}^2/4} + \frac{(\chi -\tilde{\chi})^2}{8 \pi L G_Q}.
\end{align}
If $L \ll L_J$, then $\chi \approx \tilde{\chi}$. However, this condition may be violated since the junction inductance is anomalously small at small $\chi$ and $R$. If $L G_Q \Delta \simeq \sqrt{R}$, then $\tilde{\chi}$ differs significantly from $\chi$, and the energy may even have multiple minima indicating classical bistability. Our analysis confirms this estimate for "slow" environments (see Eq. \eqref{eq:slow}), but gives a more relaxed estimate $L G_Q \Delta \simeq \omega_0/\Delta$ for faster environments $\omega_0 \gg \sqrt{R} \Delta$ (see Eq. \eqref{eq:fast}). The parameter range in question thus deserves a more detailed investigation.

\section{Conclusions}
\label{sec:conclusions}
As a part of the conclusions, let us discuss the possibilities of experimental observation of our results. As mentioned, the addition energies can be extracted from spectroscopic measurements and the supercurrents in the even or odd states can either be measured directly or be extracted from the microwave response as a flux-dependent inductance. For a generic environment, the typical scale of the external impedance is $ZG_Q \simeq 10^{-2}$. This sets the accuracy required for the measurements. To enhance the correction scale, one can also increase the inductance $L$. In recent years, much experimental effort was given to realization of high inductance environment (see e.g.~\cite{Manucharyan,Astafiev}) that can be realized by either chains of Josephson junctions or using strongly disordered superconductors. With this, the correction can be boosted to values $\simeq 1$. Interesting parameter regimes exhibiting novel physics are identified in Sec. \ref{sec:nonpert}. It is interesting that the interaction signature (Sec. \ref{sec:signature}) is given by a simple formula with all parameters directly extractable from the experimental observation: that can serve for the verification of the approach.

Overall, we present here a detailed theoretical investigation of the environment-induced interaction correction to the ground states and addition energies in a short superconducting junction. We derive general formulas  for these corrections, linking them to the frequency-dependent admittance of the junction and the environment impedance (Eqs. \eqref{eq:even}-\eqref{eq:2multichannel}). In particular, we provide numerical results for two relevant environmental models: that of an external impedance and the Mattis-Bardeen model, assuming in both cases small dimensionless impedances $G_Q Z \ll 1$.

While an expectation is that the typical scale of the relative correction is of the order $G_Q Z$, we have found that another dimensionless scale $L G_Q \Delta$ provides a better estimate for: the corrections to the even ground state energy, the addition energies, and the interaction signature. We discuss the logarithmic renormalizations of the energies by low- and high-frequency phase fluctuations (Eqs. \eqref{eq:logdivergence} and \eqref{eq:BigOmega0}), and compare those with previous results (Eqs. \eqref{eq:Jocorrection} and \eqref{eq:Jooddcorrection}).  A part of the correction is related to the renormalization of transmission eigenvalues, which we single out in Eq. \eqref{eq:transrenorm}. In detail, we investigate the peculiarities of the corrections at small phases, where the interaction causes a current jump in the odd ground state and a cusp in addition energies (Eqs. \eqref{eq:currentJumpGeneral} and \eqref{eq:smallPhi}). We study ABS energies close to the gap edge (Eq. \eqref{eq:oddOmegaCorrection}) and verify the results of \cite{oddparity,oddparity2}, and extend their conjecture about the non-perturbative interaction modification of the ABSs to arbitrary transmission coefficients (Eq. \eqref{eq:phis}). We analyze the potential enhancement of the corrections for $E_A \ll \Delta$ (Eq. \eqref{eq:slow}), and find that this enhancement cancels for "fast" environments (Eq. \eqref{eq:inductanceCancellation}), which motivates further non-perturbative research of this parameter regime. Last but not least, we provide a simple relation for the interaction signature (Eq. \eqref{eq:signaturesimple}).

\begin{acknowledgments}
We acknowledge useful motivating discussions with: H.~Pothier, A.~L.~Yeyati, L.~Glazman, J.~Meyer, A.~Balatsky, K.~Mertiri, and M.~Houzet.
\end{acknowledgments}

\section*{Data availability}
Code and data to reproduce the figures in this manuscript are available at Zenodo~\cite{samuelsenCodeFiguresData2026}.

\bibliography{main}

\begin{thebibliography}{49}%
\makeatletter
\providecommand \@ifxundefined [1]{%
 \@ifx{#1\undefined}
}%
\providecommand \@ifnum [1]{%
 \ifnum #1\expandafter \@firstoftwo
 \else \expandafter \@secondoftwo
 \fi
}%
\providecommand \@ifx [1]{%
 \ifx #1\expandafter \@firstoftwo
 \else \expandafter \@secondoftwo
 \fi
}%
\providecommand \natexlab [1]{#1}%
\providecommand \enquote  [1]{``#1''}%
\providecommand \bibnamefont  [1]{#1}%
\providecommand \bibfnamefont [1]{#1}%
\providecommand \citenamefont [1]{#1}%
\providecommand \href@noop [0]{\@secondoftwo}%
\providecommand \href [0]{\begingroup \@sanitize@url \@href}%
\providecommand \@href[1]{\@@startlink{#1}\@@href}%
\providecommand \@@href[1]{\endgroup#1\@@endlink}%
\providecommand \@sanitize@url [0]{\catcode `\\12\catcode `\$12\catcode
  `\&12\catcode `\#12\catcode `\^12\catcode `\_12\catcode `\%12\relax}%
\providecommand \@@startlink[1]{}%
\providecommand \@@endlink[0]{}%
\providecommand \url  [0]{\begingroup\@sanitize@url \@url }%
\providecommand \@url [1]{\endgroup\@href {#1}{\urlprefix }}%
\providecommand \urlprefix  [0]{URL }%
\providecommand \Eprint [0]{\href }%
\providecommand \doibase [0]{https://doi.org/}%
\providecommand \selectlanguage [0]{\@gobble}%
\providecommand \bibinfo  [0]{\@secondoftwo}%
\providecommand \bibfield  [0]{\@secondoftwo}%
\providecommand \translation [1]{[#1]}%
\providecommand \BibitemOpen [0]{}%
\providecommand \bibitemStop [0]{}%
\providecommand \bibitemNoStop [0]{.\EOS\space}%
\providecommand \EOS [0]{\spacefactor3000\relax}%
\providecommand \BibitemShut  [1]{\csname bibitem#1\endcsname}%
\let\auto@bib@innerbib\@empty
\bibitem [{\citenamefont {Jeanneret}\ and\ \citenamefont
  {Benz}(2009)}]{jeanneretApplicationJosephsonEffect2009}%
  \BibitemOpen
  \bibfield  {author} {\bibinfo {author} {\bibfnamefont {B.}~\bibnamefont
  {Jeanneret}}\ and\ \bibinfo {author} {\bibfnamefont {S.~P.}\ \bibnamefont
  {Benz}},\ }\bibfield  {title} {\bibinfo {title} {Application of the
  {{Josephson}} effect in electrical metrology},\ }\href
  {https://doi.org/10.1140/epjst/e2009-01050-6} {\bibfield  {journal} {\bibinfo
   {journal} {The European Physical Journal Special Topics}\ }\textbf {\bibinfo
  {volume} {172}},\ \bibinfo {pages} {181} (\bibinfo {year}
  {2009})}\BibitemShut {NoStop}%
\bibitem [{\citenamefont {Kjaergaard}\ \emph {et~al.}(2020)\citenamefont
  {Kjaergaard}, \citenamefont {Schwartz}, \citenamefont {Braumueller},
  \citenamefont {Krantz}, \citenamefont {Wang}, \citenamefont {Gustavsson},\
  and\ \citenamefont {Oliver}}]{kjaergaardSuperconductingQubitsCurrent2020}%
  \BibitemOpen
  \bibfield  {author} {\bibinfo {author} {\bibfnamefont {M.}~\bibnamefont
  {Kjaergaard}}, \bibinfo {author} {\bibfnamefont {M.~E.}\ \bibnamefont
  {Schwartz}}, \bibinfo {author} {\bibfnamefont {J.}~\bibnamefont
  {Braumueller}}, \bibinfo {author} {\bibfnamefont {P.}~\bibnamefont {Krantz}},
  \bibinfo {author} {\bibfnamefont {J.~I.-J.}\ \bibnamefont {Wang}}, \bibinfo
  {author} {\bibfnamefont {S.}~\bibnamefont {Gustavsson}},\ and\ \bibinfo
  {author} {\bibfnamefont {W.~D.}\ \bibnamefont {Oliver}},\ }\bibfield  {title}
  {\bibinfo {title} {Superconducting {{Qubits}}: {{Current State}} of
  {{Play}}},\ }\href {https://doi.org/10.1146/annurev-conmatphys-031119-050605}
  {\bibfield  {journal} {\bibinfo  {journal} {Annual Review of Condensed Matter
  Physics}\ }\textbf {\bibinfo {volume} {11}},\ \bibinfo {pages} {369}
  (\bibinfo {year} {2020})}\BibitemShut {NoStop}%
\bibitem [{\citenamefont
  {Beenakker}(2013)}]{beenakkerSearchMajoranaFermions2013}%
  \BibitemOpen
  \bibfield  {author} {\bibinfo {author} {\bibfnamefont {C.~W.~J.}\
  \bibnamefont {Beenakker}},\ }\bibfield  {title} {\bibinfo {title} {Search for
  {{Majorana Fermions}} in {{Superconductors}}},\ }\href
  {https://doi.org/10.1146/annurev-conmatphys-030212-184337} {\bibfield
  {journal} {\bibinfo  {journal} {Annual Review of Condensed Matter Physics}\
  }\textbf {\bibinfo {volume} {4}},\ \bibinfo {pages} {113} (\bibinfo {year}
  {2013})}\BibitemShut {NoStop}%
\bibitem [{\citenamefont {Bardeen}\ \emph {et~al.}(1957)\citenamefont
  {Bardeen}, \citenamefont {Cooper},\ and\ \citenamefont
  {Schrieffer}}]{bardeenTheorySuperconductivity1957}%
  \BibitemOpen
  \bibfield  {author} {\bibinfo {author} {\bibfnamefont {J.}~\bibnamefont
  {Bardeen}}, \bibinfo {author} {\bibfnamefont {L.~N.}\ \bibnamefont
  {Cooper}},\ and\ \bibinfo {author} {\bibfnamefont {J.~R.}\ \bibnamefont
  {Schrieffer}},\ }\bibfield  {title} {\bibinfo {title} {Theory of
  {{Superconductivity}}},\ }\href {https://doi.org/10.1103/PhysRev.108.1175}
  {\bibfield  {journal} {\bibinfo  {journal} {Physical Review}\ }\textbf
  {\bibinfo {volume} {108}},\ \bibinfo {pages} {1175} (\bibinfo {year}
  {1957})}\BibitemShut {NoStop}%
\bibitem [{\citenamefont
  {Anderson}(1959)}]{andersonTheoryDirtySuperconductors1959}%
  \BibitemOpen
  \bibfield  {author} {\bibinfo {author} {\bibfnamefont {P.~W.}\ \bibnamefont
  {Anderson}},\ }\bibfield  {title} {\bibinfo {title} {Theory of dirty
  superconductors},\ }\href {https://doi.org/10.1016/0022-3697(59)90036-8}
  {\bibfield  {journal} {\bibinfo  {journal} {Journal of Physics and Chemistry
  of Solids}\ }\textbf {\bibinfo {volume} {11}},\ \bibinfo {pages} {26}
  (\bibinfo {year} {1959})}\BibitemShut {NoStop}%
\bibitem [{\citenamefont {Blais}\ \emph {et~al.}(2021)\citenamefont {Blais},
  \citenamefont {Grimsmo}, \citenamefont {Girvin},\ and\ \citenamefont
  {Wallraff}}]{blaisCircuitQuantumElectrodynamics2021}%
  \BibitemOpen
  \bibfield  {author} {\bibinfo {author} {\bibfnamefont {A.}~\bibnamefont
  {Blais}}, \bibinfo {author} {\bibfnamefont {A.~L.}\ \bibnamefont {Grimsmo}},
  \bibinfo {author} {\bibfnamefont {S.~M.}\ \bibnamefont {Girvin}},\ and\
  \bibinfo {author} {\bibfnamefont {A.}~\bibnamefont {Wallraff}},\ }\bibfield
  {title} {\bibinfo {title} {Circuit quantum electrodynamics},\ }\href
  {https://doi.org/10.1103/RevModPhys.93.025005} {\bibfield  {journal}
  {\bibinfo  {journal} {Reviews of Modern Physics}\ }\textbf {\bibinfo {volume}
  {93}},\ \bibinfo {pages} {025005} (\bibinfo {year} {2021})}\BibitemShut
  {NoStop}%
\bibitem [{\citenamefont
  {Andreev}(1964)}]{andreevThermalConductivityIntermediate1964}%
  \BibitemOpen
  \bibfield  {author} {\bibinfo {author} {\bibfnamefont {A.}~\bibnamefont
  {Andreev}},\ }\bibfield  {title} {\bibinfo {title} {The {{Thermal
  Conductivity}} of the {{Intermediate State}} in {{Superconductors}}},\
  }\href@noop {} {\bibfield  {journal} {\bibinfo  {journal} {Soviet Physics
  JETP-USSR}\ }\textbf {\bibinfo {volume} {19}},\ \bibinfo {pages} {1228}
  (\bibinfo {year} {1964})}\BibitemShut {NoStop}%
\bibitem [{\citenamefont
  {Beenakker}(1991)}]{beenakkerUniversalLimitCriticalcurrent1991}%
  \BibitemOpen
  \bibfield  {author} {\bibinfo {author} {\bibfnamefont {C.~W.~J.}\
  \bibnamefont {Beenakker}},\ }\bibfield  {title} {\bibinfo {title} {Universal
  limit of critical-current fluctuations in mesoscopic {{Josephson}}
  junctions},\ }\href {https://doi.org/10.1103/PhysRevLett.67.3836} {\bibfield
  {journal} {\bibinfo  {journal} {Physical Review Letters}\ }\textbf {\bibinfo
  {volume} {67}},\ \bibinfo {pages} {3836} (\bibinfo {year}
  {1991})}\BibitemShut {NoStop}%
\bibitem [{\citenamefont {Nazarov}\ and\ \citenamefont
  {Blanter}(2009)}]{NazarovBlanterQuantumTransport}%
  \BibitemOpen
  \bibfield  {author} {\bibinfo {author} {\bibfnamefont {Y.~V.}\ \bibnamefont
  {Nazarov}}\ and\ \bibinfo {author} {\bibfnamefont {Y.~M.}\ \bibnamefont
  {Blanter}},\ }\href {https://cds.cern.ch/record/1186214} {\emph {\bibinfo
  {title} {{Quantum transport}}}}\ (\bibinfo  {publisher} {Cambridge Univ.
  Press},\ \bibinfo {address} {Cambridge},\ \bibinfo {year} {2009})\BibitemShut
  {NoStop}%
\bibitem [{\citenamefont {Chtchelkatchev}\ and\ \citenamefont
  {Nazarov}(2003)}]{chtchelkatchevAndreevQuantumDots2003}%
  \BibitemOpen
  \bibfield  {author} {\bibinfo {author} {\bibfnamefont {N.~M.}\ \bibnamefont
  {Chtchelkatchev}}\ and\ \bibinfo {author} {\bibfnamefont {{\relax Yu}.~V.}\
  \bibnamefont {Nazarov}},\ }\bibfield  {title} {\bibinfo {title} {Andreev
  {{Quantum Dots}} for {{Spin Manipulation}}},\ }\href
  {https://doi.org/10.1103/PhysRevLett.90.226806} {\bibfield  {journal}
  {\bibinfo  {journal} {Physical Review Letters}\ }\textbf {\bibinfo {volume}
  {90}},\ \bibinfo {pages} {226806} (\bibinfo {year} {2003})}\BibitemShut
  {NoStop}%
\bibitem [{\citenamefont {Padurariu}\ and\ \citenamefont
  {Nazarov}(2010)}]{padurariuTheoreticalProposalSuperconducting2010}%
  \BibitemOpen
  \bibfield  {author} {\bibinfo {author} {\bibfnamefont {C.}~\bibnamefont
  {Padurariu}}\ and\ \bibinfo {author} {\bibfnamefont {{\relax Yu}.~V.}\
  \bibnamefont {Nazarov}},\ }\bibfield  {title} {\bibinfo {title} {Theoretical
  proposal for superconducting spin qubits},\ }\href
  {https://doi.org/10.1103/PhysRevB.81.144519} {\bibfield  {journal} {\bibinfo
  {journal} {Physical Review B}\ }\textbf {\bibinfo {volume} {81}},\ \bibinfo
  {pages} {144519} (\bibinfo {year} {2010})}\BibitemShut {NoStop}%
\bibitem [{\citenamefont {Zazunov}\ \emph {et~al.}(2003)\citenamefont
  {Zazunov}, \citenamefont {Shumeiko}, \citenamefont {Bratus'}, \citenamefont
  {Lantz},\ and\ \citenamefont {Wendin}}]{zazunovAndreevLevelQubit2003}%
  \BibitemOpen
  \bibfield  {author} {\bibinfo {author} {\bibfnamefont {A.}~\bibnamefont
  {Zazunov}}, \bibinfo {author} {\bibfnamefont {V.~S.}\ \bibnamefont
  {Shumeiko}}, \bibinfo {author} {\bibfnamefont {E.~N.}\ \bibnamefont
  {Bratus'}}, \bibinfo {author} {\bibfnamefont {J.}~\bibnamefont {Lantz}},\
  and\ \bibinfo {author} {\bibfnamefont {G.}~\bibnamefont {Wendin}},\
  }\bibfield  {title} {\bibinfo {title} {Andreev {{Level Qubit}}},\ }\href
  {https://doi.org/10.1103/PhysRevLett.90.087003} {\bibfield  {journal}
  {\bibinfo  {journal} {Physical Review Letters}\ }\textbf {\bibinfo {volume}
  {90}},\ \bibinfo {pages} {087003} (\bibinfo {year} {2003})}\BibitemShut
  {NoStop}%
\bibitem [{\citenamefont {Zazunov}\ \emph {et~al.}(2005)\citenamefont
  {Zazunov}, \citenamefont {Shumeiko}, \citenamefont {Wendin},\ and\
  \citenamefont {Bratus'}}]{zazunov2}%
  \BibitemOpen
  \bibfield  {author} {\bibinfo {author} {\bibfnamefont {A.}~\bibnamefont
  {Zazunov}}, \bibinfo {author} {\bibfnamefont {V.~S.}\ \bibnamefont
  {Shumeiko}}, \bibinfo {author} {\bibfnamefont {G.}~\bibnamefont {Wendin}},\
  and\ \bibinfo {author} {\bibfnamefont {E.~N.}\ \bibnamefont {Bratus'}},\
  }\bibfield  {title} {\bibinfo {title} {Dynamics and phonon-induced
  decoherence of andreev level qubit},\ }\href
  {https://doi.org/10.1103/PhysRevB.71.214505} {\bibfield  {journal} {\bibinfo
  {journal} {Phys. Rev. B}\ }\textbf {\bibinfo {volume} {71}},\ \bibinfo
  {pages} {214505} (\bibinfo {year} {2005})}\BibitemShut {NoStop}%
\bibitem [{\citenamefont {Zgirski}\ \emph {et~al.}(2011)\citenamefont
  {Zgirski}, \citenamefont {Bretheau}, \citenamefont {Le~Masne}, \citenamefont
  {Pothier}, \citenamefont {Esteve},\ and\ \citenamefont
  {Urbina}}]{zgirski2011}%
  \BibitemOpen
  \bibfield  {author} {\bibinfo {author} {\bibfnamefont {M.}~\bibnamefont
  {Zgirski}}, \bibinfo {author} {\bibfnamefont {L.}~\bibnamefont {Bretheau}},
  \bibinfo {author} {\bibfnamefont {Q.}~\bibnamefont {Le~Masne}}, \bibinfo
  {author} {\bibfnamefont {H.}~\bibnamefont {Pothier}}, \bibinfo {author}
  {\bibfnamefont {D.}~\bibnamefont {Esteve}},\ and\ \bibinfo {author}
  {\bibfnamefont {C.}~\bibnamefont {Urbina}},\ }\bibfield  {title} {\bibinfo
  {title} {Evidence for long-lived quasiparticles trapped in superconducting
  point contacts},\ }\href {https://doi.org/10.1103/PhysRevLett.106.257003}
  {\bibfield  {journal} {\bibinfo  {journal} {Phys. Rev. Lett.}\ }\textbf
  {\bibinfo {volume} {106}},\ \bibinfo {pages} {257003} (\bibinfo {year}
  {2011})}\BibitemShut {NoStop}%
\bibitem [{\citenamefont {Bretheau}\ \emph {et~al.}(2013)\citenamefont
  {Bretheau}, \citenamefont {Girit}, \citenamefont {Pothier}, \citenamefont
  {Esteve},\ and\ \citenamefont {Urbina}}]{bretheauExcitingAndreevPairs2013}%
  \BibitemOpen
  \bibfield  {author} {\bibinfo {author} {\bibfnamefont {L.}~\bibnamefont
  {Bretheau}}, \bibinfo {author} {\bibfnamefont {{\c C}.~{\"O}.}\ \bibnamefont
  {Girit}}, \bibinfo {author} {\bibfnamefont {H.}~\bibnamefont {Pothier}},
  \bibinfo {author} {\bibfnamefont {D.}~\bibnamefont {Esteve}},\ and\ \bibinfo
  {author} {\bibfnamefont {C.}~\bibnamefont {Urbina}},\ }\bibfield  {title}
  {\bibinfo {title} {Exciting {{Andreev}} pairs in a superconducting atomic
  contact},\ }\href {https://doi.org/10.1038/nature12315} {\bibfield  {journal}
  {\bibinfo  {journal} {Nature}\ }\textbf {\bibinfo {volume} {499}},\ \bibinfo
  {pages} {312} (\bibinfo {year} {2013})}\BibitemShut {NoStop}%
\bibitem [{\citenamefont {Janvier}\ \emph {et~al.}(2015)\citenamefont
  {Janvier}, \citenamefont {Tosi}, \citenamefont {Bretheau}, \citenamefont
  {Girit}, \citenamefont {Stern}, \citenamefont {Bertet}, \citenamefont
  {Joyez}, \citenamefont {Vion}, \citenamefont {Esteve}, \citenamefont
  {Goffman}, \citenamefont {Pothier},\ and\ \citenamefont
  {Urbina}}]{janvierCoherentManipulationAndreev2015}%
  \BibitemOpen
  \bibfield  {author} {\bibinfo {author} {\bibfnamefont {C.}~\bibnamefont
  {Janvier}}, \bibinfo {author} {\bibfnamefont {L.}~\bibnamefont {Tosi}},
  \bibinfo {author} {\bibfnamefont {L.}~\bibnamefont {Bretheau}}, \bibinfo
  {author} {\bibfnamefont {{\c C}.~{\"O}.}\ \bibnamefont {Girit}}, \bibinfo
  {author} {\bibfnamefont {M.}~\bibnamefont {Stern}}, \bibinfo {author}
  {\bibfnamefont {P.}~\bibnamefont {Bertet}}, \bibinfo {author} {\bibfnamefont
  {P.}~\bibnamefont {Joyez}}, \bibinfo {author} {\bibfnamefont
  {D.}~\bibnamefont {Vion}}, \bibinfo {author} {\bibfnamefont {D.}~\bibnamefont
  {Esteve}}, \bibinfo {author} {\bibfnamefont {M.~F.}\ \bibnamefont {Goffman}},
  \bibinfo {author} {\bibfnamefont {H.}~\bibnamefont {Pothier}},\ and\ \bibinfo
  {author} {\bibfnamefont {C.}~\bibnamefont {Urbina}},\ }\bibfield  {title}
  {\bibinfo {title} {Coherent manipulation of {{Andreev}} states in
  superconducting atomic contacts},\ }\href
  {https://doi.org/10.1126/science.aab2179} {\bibfield  {journal} {\bibinfo
  {journal} {Science}\ }\textbf {\bibinfo {volume} {349}},\ \bibinfo {pages}
  {1199} (\bibinfo {year} {2015})}\BibitemShut {NoStop}%
\bibitem [{\citenamefont {Goffman}\ \emph {et~al.}(2017)\citenamefont
  {Goffman}, \citenamefont {Urbina}, \citenamefont {Pothier}, \citenamefont
  {Nyg{\aa}rd}, \citenamefont {Marcus},\ and\ \citenamefont
  {Krogstrup}}]{goffmanConductionChannelsInAsnanowire2017}%
  \BibitemOpen
  \bibfield  {author} {\bibinfo {author} {\bibfnamefont {M.~F.}\ \bibnamefont
  {Goffman}}, \bibinfo {author} {\bibfnamefont {C.}~\bibnamefont {Urbina}},
  \bibinfo {author} {\bibfnamefont {H.}~\bibnamefont {Pothier}}, \bibinfo
  {author} {\bibfnamefont {J.}~\bibnamefont {Nyg{\aa}rd}}, \bibinfo {author}
  {\bibfnamefont {C.~M.}\ \bibnamefont {Marcus}},\ and\ \bibinfo {author}
  {\bibfnamefont {P.}~\bibnamefont {Krogstrup}},\ }\bibfield  {title} {\bibinfo
  {title} {Conduction channels of an {{InAs-Al}} nanowire {{Josephson}} weak
  link},\ }\href {https://doi.org/10.1088/1367-2630/aa7641} {\bibfield
  {journal} {\bibinfo  {journal} {New Journal of Physics}\ }\textbf {\bibinfo
  {volume} {19}},\ \bibinfo {pages} {092002} (\bibinfo {year}
  {2017})}\BibitemShut {NoStop}%
\bibitem [{\citenamefont {Tosi}\ \emph {et~al.}(2019)\citenamefont {Tosi},
  \citenamefont {Metzger}, \citenamefont {Goffman}, \citenamefont {Urbina},
  \citenamefont {Pothier}, \citenamefont {Park}, \citenamefont {Yeyati},
  \citenamefont {Nyg\aa{}rd},\ and\ \citenamefont
  {Krogstrup}}]{pothierspinorbit2019}%
  \BibitemOpen
  \bibfield  {author} {\bibinfo {author} {\bibfnamefont {L.}~\bibnamefont
  {Tosi}}, \bibinfo {author} {\bibfnamefont {C.}~\bibnamefont {Metzger}},
  \bibinfo {author} {\bibfnamefont {M.~F.}\ \bibnamefont {Goffman}}, \bibinfo
  {author} {\bibfnamefont {C.}~\bibnamefont {Urbina}}, \bibinfo {author}
  {\bibfnamefont {H.}~\bibnamefont {Pothier}}, \bibinfo {author} {\bibfnamefont
  {S.}~\bibnamefont {Park}}, \bibinfo {author} {\bibfnamefont {A.~L.}\
  \bibnamefont {Yeyati}}, \bibinfo {author} {\bibfnamefont {J.}~\bibnamefont
  {Nyg\aa{}rd}},\ and\ \bibinfo {author} {\bibfnamefont {P.}~\bibnamefont
  {Krogstrup}},\ }\bibfield  {title} {\bibinfo {title} {Spin-orbit splitting of
  andreev states revealed by microwave spectroscopy},\ }\href
  {https://doi.org/10.1103/PhysRevX.9.011010} {\bibfield  {journal} {\bibinfo
  {journal} {Phys. Rev. X}\ }\textbf {\bibinfo {volume} {9}},\ \bibinfo {pages}
  {011010} (\bibinfo {year} {2019})}\BibitemShut {NoStop}%
\bibitem [{\citenamefont {Hays}\ \emph {et~al.}(2021)\citenamefont {Hays},
  \citenamefont {Fatemi}, \citenamefont {Bouman}, \citenamefont {Cerrillo},
  \citenamefont {Diamond}, \citenamefont {Serniak}, \citenamefont {Connolly},
  \citenamefont {Krogstrup}, \citenamefont {Nygård}, \citenamefont {Yeyati},
  \citenamefont {Geresdi},\ and\ \citenamefont {Devoret}}]{spinqubit1}%
  \BibitemOpen
  \bibfield  {author} {\bibinfo {author} {\bibfnamefont {M.}~\bibnamefont
  {Hays}}, \bibinfo {author} {\bibfnamefont {V.}~\bibnamefont {Fatemi}},
  \bibinfo {author} {\bibfnamefont {D.}~\bibnamefont {Bouman}}, \bibinfo
  {author} {\bibfnamefont {J.}~\bibnamefont {Cerrillo}}, \bibinfo {author}
  {\bibfnamefont {S.}~\bibnamefont {Diamond}}, \bibinfo {author} {\bibfnamefont
  {K.}~\bibnamefont {Serniak}}, \bibinfo {author} {\bibfnamefont
  {T.}~\bibnamefont {Connolly}}, \bibinfo {author} {\bibfnamefont
  {P.}~\bibnamefont {Krogstrup}}, \bibinfo {author} {\bibfnamefont
  {J.}~\bibnamefont {Nygård}}, \bibinfo {author} {\bibfnamefont {A.~L.}\
  \bibnamefont {Yeyati}}, \bibinfo {author} {\bibfnamefont {A.}~\bibnamefont
  {Geresdi}},\ and\ \bibinfo {author} {\bibfnamefont {M.~H.}\ \bibnamefont
  {Devoret}},\ }\bibfield  {title} {\bibinfo {title} {Coherent manipulation of
  an andreev spin qubit},\ }\href {https://doi.org/10.1126/science.abf0345}
  {\bibfield  {journal} {\bibinfo  {journal} {Science}\ }\textbf {\bibinfo
  {volume} {373}},\ \bibinfo {pages} {430} (\bibinfo {year}
  {2021})}\BibitemShut {NoStop}%
\bibitem [{\citenamefont {Pita-Vidal}\ \emph {et~al.}(2023)\citenamefont
  {Pita-Vidal}, \citenamefont {Bargerbos}, \citenamefont {{\v{Z}}itko},
  \citenamefont {Splitthoff}, \citenamefont {Gr{\"u}nhaupt}, \citenamefont
  {Wesdorp}, \citenamefont {Liu}, \citenamefont {Kouwenhoven}, \citenamefont
  {Aguado}, \citenamefont {van Heck}, \citenamefont {Kou},\ and\ \citenamefont
  {Andersen}}]{spinqubit2}%
  \BibitemOpen
  \bibfield  {author} {\bibinfo {author} {\bibfnamefont {M.}~\bibnamefont
  {Pita-Vidal}}, \bibinfo {author} {\bibfnamefont {A.}~\bibnamefont
  {Bargerbos}}, \bibinfo {author} {\bibfnamefont {R.}~\bibnamefont
  {{\v{Z}}itko}}, \bibinfo {author} {\bibfnamefont {L.~J.}\ \bibnamefont
  {Splitthoff}}, \bibinfo {author} {\bibfnamefont {L.}~\bibnamefont
  {Gr{\"u}nhaupt}}, \bibinfo {author} {\bibfnamefont {J.~J.}\ \bibnamefont
  {Wesdorp}}, \bibinfo {author} {\bibfnamefont {Y.}~\bibnamefont {Liu}},
  \bibinfo {author} {\bibfnamefont {L.~P.}\ \bibnamefont {Kouwenhoven}},
  \bibinfo {author} {\bibfnamefont {R.}~\bibnamefont {Aguado}}, \bibinfo
  {author} {\bibfnamefont {B.}~\bibnamefont {van Heck}}, \bibinfo {author}
  {\bibfnamefont {A.}~\bibnamefont {Kou}},\ and\ \bibinfo {author}
  {\bibfnamefont {C.~K.}\ \bibnamefont {Andersen}},\ }\bibfield  {title}
  {\bibinfo {title} {Direct manipulation of a superconducting spin qubit
  strongly coupled to a transmon qubit},\ }\href
  {https://doi.org/10.1038/s41567-023-02071-x} {\bibfield  {journal} {\bibinfo
  {journal} {Nature Physics}\ }\textbf {\bibinfo {volume} {19}},\ \bibinfo
  {pages} {1110} (\bibinfo {year} {2023})}\BibitemShut {NoStop}%
\bibitem [{\citenamefont {Fatemi}\ \emph {et~al.}(2022)\citenamefont {Fatemi},
  \citenamefont {Kurilovich}, \citenamefont {Hays}, \citenamefont {Bouman},
  \citenamefont {Connolly}, \citenamefont {Diamond}, \citenamefont {Frattini},
  \citenamefont {Kurilovich}, \citenamefont {Krogstrup}, \citenamefont
  {Nyg\aa{}rd}, \citenamefont {Geresdi}, \citenamefont {Glazman},\ and\
  \citenamefont {Devoret}}]{signaturesfatemi2022}%
  \BibitemOpen
  \bibfield  {author} {\bibinfo {author} {\bibfnamefont {V.}~\bibnamefont
  {Fatemi}}, \bibinfo {author} {\bibfnamefont {P.~D.}\ \bibnamefont
  {Kurilovich}}, \bibinfo {author} {\bibfnamefont {M.}~\bibnamefont {Hays}},
  \bibinfo {author} {\bibfnamefont {D.}~\bibnamefont {Bouman}}, \bibinfo
  {author} {\bibfnamefont {T.}~\bibnamefont {Connolly}}, \bibinfo {author}
  {\bibfnamefont {S.}~\bibnamefont {Diamond}}, \bibinfo {author} {\bibfnamefont
  {N.~E.}\ \bibnamefont {Frattini}}, \bibinfo {author} {\bibfnamefont {V.~D.}\
  \bibnamefont {Kurilovich}}, \bibinfo {author} {\bibfnamefont
  {P.}~\bibnamefont {Krogstrup}}, \bibinfo {author} {\bibfnamefont
  {J.}~\bibnamefont {Nyg\aa{}rd}}, \bibinfo {author} {\bibfnamefont
  {A.}~\bibnamefont {Geresdi}}, \bibinfo {author} {\bibfnamefont {L.~I.}\
  \bibnamefont {Glazman}},\ and\ \bibinfo {author} {\bibfnamefont {M.~H.}\
  \bibnamefont {Devoret}},\ }\bibfield  {title} {\bibinfo {title} {Microwave
  susceptibility observation of interacting many-body andreev states},\ }\href
  {https://doi.org/10.1103/PhysRevLett.129.227701} {\bibfield  {journal}
  {\bibinfo  {journal} {Phys. Rev. Lett.}\ }\textbf {\bibinfo {volume} {129}},\
  \bibinfo {pages} {227701} (\bibinfo {year} {2022})}\BibitemShut {NoStop}%
\bibitem [{\citenamefont {Matute-Ca\~nadas}\ \emph {et~al.}(2022)\citenamefont
  {Matute-Ca\~nadas}, \citenamefont {Metzger}, \citenamefont {Park},
  \citenamefont {Tosi}, \citenamefont {Krogstrup}, \citenamefont {Nyg\aa{}rd},
  \citenamefont {Goffman}, \citenamefont {Urbina}, \citenamefont {Pothier},\
  and\ \citenamefont {Yeyati}}]{signaturespothier2022}%
  \BibitemOpen
  \bibfield  {author} {\bibinfo {author} {\bibfnamefont {F.~J.}\ \bibnamefont
  {Matute-Ca\~nadas}}, \bibinfo {author} {\bibfnamefont {C.}~\bibnamefont
  {Metzger}}, \bibinfo {author} {\bibfnamefont {S.}~\bibnamefont {Park}},
  \bibinfo {author} {\bibfnamefont {L.}~\bibnamefont {Tosi}}, \bibinfo {author}
  {\bibfnamefont {P.}~\bibnamefont {Krogstrup}}, \bibinfo {author}
  {\bibfnamefont {J.}~\bibnamefont {Nyg\aa{}rd}}, \bibinfo {author}
  {\bibfnamefont {M.~F.}\ \bibnamefont {Goffman}}, \bibinfo {author}
  {\bibfnamefont {C.}~\bibnamefont {Urbina}}, \bibinfo {author} {\bibfnamefont
  {H.}~\bibnamefont {Pothier}},\ and\ \bibinfo {author} {\bibfnamefont {A.~L.}\
  \bibnamefont {Yeyati}},\ }\bibfield  {title} {\bibinfo {title} {Signatures of
  interactions in the andreev spectrum of nanowire josephson junctions},\
  }\href {https://doi.org/10.1103/PhysRevLett.128.197702} {\bibfield  {journal}
  {\bibinfo  {journal} {Phys. Rev. Lett.}\ }\textbf {\bibinfo {volume} {128}},\
  \bibinfo {pages} {197702} (\bibinfo {year} {2022})}\BibitemShut {NoStop}%
\bibitem [{\citenamefont {Lee}\ \emph {et~al.}(2014)\citenamefont {Lee},
  \citenamefont {Jiang}, \citenamefont {Houzet}, \citenamefont {Aguado},
  \citenamefont {Lieber},\ and\ \citenamefont {De~Franceschi}}]{Lee2014}%
  \BibitemOpen
  \bibfield  {author} {\bibinfo {author} {\bibfnamefont {E.~J.~H.}\
  \bibnamefont {Lee}}, \bibinfo {author} {\bibfnamefont {X.}~\bibnamefont
  {Jiang}}, \bibinfo {author} {\bibfnamefont {M.}~\bibnamefont {Houzet}},
  \bibinfo {author} {\bibfnamefont {R.}~\bibnamefont {Aguado}}, \bibinfo
  {author} {\bibfnamefont {C.~M.}\ \bibnamefont {Lieber}},\ and\ \bibinfo
  {author} {\bibfnamefont {S.}~\bibnamefont {De~Franceschi}},\ }\bibfield
  {title} {\bibinfo {title} {Spin-resolved andreev levels and parity crossings
  in hybrid superconductor--semiconductor nanostructures},\ }\href
  {https://doi.org/10.1038/nnano.2013.267} {\bibfield  {journal} {\bibinfo
  {journal} {Nature Nanotechnology}\ }\textbf {\bibinfo {volume} {9}},\
  \bibinfo {pages} {79} (\bibinfo {year} {2014})}\BibitemShut {NoStop}%
\bibitem [{\citenamefont {van Dam}\ \emph {et~al.}(2006)\citenamefont {van
  Dam}, \citenamefont {Nazarov}, \citenamefont {Bakkers}, \citenamefont
  {De~Franceschi},\ and\ \citenamefont {Kouwenhoven}}]{vanDam2006}%
  \BibitemOpen
  \bibfield  {author} {\bibinfo {author} {\bibfnamefont {J.~A.}\ \bibnamefont
  {van Dam}}, \bibinfo {author} {\bibfnamefont {Y.~V.}\ \bibnamefont
  {Nazarov}}, \bibinfo {author} {\bibfnamefont {E.~P. A.~M.}\ \bibnamefont
  {Bakkers}}, \bibinfo {author} {\bibfnamefont {S.}~\bibnamefont
  {De~Franceschi}},\ and\ \bibinfo {author} {\bibfnamefont {L.~P.}\
  \bibnamefont {Kouwenhoven}},\ }\bibfield  {title} {\bibinfo {title}
  {Supercurrent reversal in quantum dots},\ }\href
  {https://doi.org/10.1038/nature05018} {\bibfield  {journal} {\bibinfo
  {journal} {Nature}\ }\textbf {\bibinfo {volume} {442}},\ \bibinfo {pages}
  {667} (\bibinfo {year} {2006})}\BibitemShut {NoStop}%
\bibitem [{\citenamefont {Josephson}(1962)}]{Josephson}%
  \BibitemOpen
  \bibfield  {author} {\bibinfo {author} {\bibfnamefont {B.}~\bibnamefont
  {Josephson}},\ }\bibfield  {title} {\bibinfo {title} {Possible new effects in
  superconductive tunnelling},\ }\href
  {https://doi.org/https://doi.org/10.1016/0031-9163(62)91369-0} {\bibfield
  {journal} {\bibinfo  {journal} {Physics Letters}\ }\textbf {\bibinfo {volume}
  {1}},\ \bibinfo {pages} {251} (\bibinfo {year} {1962})}\BibitemShut {NoStop}%
\bibitem [{\citenamefont {Leggett}(1980)}]{Legget1}%
  \BibitemOpen
  \bibfield  {author} {\bibinfo {author} {\bibfnamefont {A.~J.}\ \bibnamefont
  {Leggett}},\ }\bibfield  {title} {\bibinfo {title} {Macroscopic quantum
  systems and the quantum theory of measurement},\ }\href
  {https://doi.org/10.1143/PTP.69.80} {\bibfield  {journal} {\bibinfo
  {journal} {Progress of Theoretical Physics Supplement}\ }\textbf {\bibinfo
  {volume} {69}},\ \bibinfo {pages} {80} (\bibinfo {year} {1980})}\BibitemShut
  {NoStop}%
\bibitem [{\citenamefont {Clarke}\ \emph {et~al.}(1988)\citenamefont {Clarke},
  \citenamefont {Cleland}, \citenamefont {Devoret}, \citenamefont {Esteve},\
  and\ \citenamefont {Martinis}}]{Clarke}%
  \BibitemOpen
  \bibfield  {author} {\bibinfo {author} {\bibfnamefont {J.}~\bibnamefont
  {Clarke}}, \bibinfo {author} {\bibfnamefont {A.~N.}\ \bibnamefont {Cleland}},
  \bibinfo {author} {\bibfnamefont {M.~H.}\ \bibnamefont {Devoret}}, \bibinfo
  {author} {\bibfnamefont {D.}~\bibnamefont {Esteve}},\ and\ \bibinfo {author}
  {\bibfnamefont {J.~M.}\ \bibnamefont {Martinis}},\ }\bibfield  {title}
  {\bibinfo {title} {Quantum mechanics of a macroscopic variable: The phase
  difference of a josephson junction},\ }\href
  {https://doi.org/10.1126/science.239.4843.992} {\bibfield  {journal}
  {\bibinfo  {journal} {Science}\ }\textbf {\bibinfo {volume} {239}},\ \bibinfo
  {pages} {992} (\bibinfo {year} {1988})}\BibitemShut {NoStop}%
\bibitem [{\citenamefont {Leggett}\ \emph {et~al.}(1987)\citenamefont
  {Leggett}, \citenamefont {Chakravarty}, \citenamefont {Dorsey}, \citenamefont
  {Fisher}, \citenamefont {Garg},\ and\ \citenamefont {Zwerger}}]{Legget2}%
  \BibitemOpen
  \bibfield  {author} {\bibinfo {author} {\bibfnamefont {A.~J.}\ \bibnamefont
  {Leggett}}, \bibinfo {author} {\bibfnamefont {S.}~\bibnamefont
  {Chakravarty}}, \bibinfo {author} {\bibfnamefont {A.~T.}\ \bibnamefont
  {Dorsey}}, \bibinfo {author} {\bibfnamefont {M.~P.~A.}\ \bibnamefont
  {Fisher}}, \bibinfo {author} {\bibfnamefont {A.}~\bibnamefont {Garg}},\ and\
  \bibinfo {author} {\bibfnamefont {W.}~\bibnamefont {Zwerger}},\ }\bibfield
  {title} {\bibinfo {title} {Dynamics of the dissipative two-state system},\
  }\href {https://doi.org/10.1103/RevModPhys.59.1} {\bibfield  {journal}
  {\bibinfo  {journal} {Rev. Mod. Phys.}\ }\textbf {\bibinfo {volume} {59}},\
  \bibinfo {pages} {1} (\bibinfo {year} {1987})}\BibitemShut {NoStop}%
\bibitem [{\citenamefont {Weiss}(2021)}]{Weiss}%
  \BibitemOpen
  \bibfield  {author} {\bibinfo {author} {\bibfnamefont {U.}~\bibnamefont
  {Weiss}},\ }\href {https://doi.org/10.1142/12402} {\emph {\bibinfo {title}
  {Quantum Dissipative Systems}}}\ (\bibinfo  {publisher} {World Scientific},\
  \bibinfo {year} {2021})\BibitemShut {NoStop}%
\bibitem [{\citenamefont {Schmid}(1983)}]{Schmid}%
  \BibitemOpen
  \bibfield  {author} {\bibinfo {author} {\bibfnamefont {A.}~\bibnamefont
  {Schmid}},\ }\bibfield  {title} {\bibinfo {title} {Diffusion and localization
  in a dissipative quantum system},\ }\href
  {https://doi.org/10.1103/PhysRevLett.51.1506} {\bibfield  {journal} {\bibinfo
   {journal} {Phys. Rev. Lett.}\ }\textbf {\bibinfo {volume} {51}},\ \bibinfo
  {pages} {1506} (\bibinfo {year} {1983})}\BibitemShut {NoStop}%
\bibitem [{\citenamefont {Murani}\ \emph {et~al.}(2020)\citenamefont {Murani},
  \citenamefont {Bourlet}, \citenamefont {le~Sueur}, \citenamefont {Portier},
  \citenamefont {Altimiras}, \citenamefont {Esteve}, \citenamefont {Grabert},
  \citenamefont {Stockburger}, \citenamefont {Ankerhold},\ and\ \citenamefont
  {Joyez}}]{Murani2020}%
  \BibitemOpen
  \bibfield  {author} {\bibinfo {author} {\bibfnamefont {A.}~\bibnamefont
  {Murani}}, \bibinfo {author} {\bibfnamefont {N.}~\bibnamefont {Bourlet}},
  \bibinfo {author} {\bibfnamefont {H.}~\bibnamefont {le~Sueur}}, \bibinfo
  {author} {\bibfnamefont {F.}~\bibnamefont {Portier}}, \bibinfo {author}
  {\bibfnamefont {C.}~\bibnamefont {Altimiras}}, \bibinfo {author}
  {\bibfnamefont {D.}~\bibnamefont {Esteve}}, \bibinfo {author} {\bibfnamefont
  {H.}~\bibnamefont {Grabert}}, \bibinfo {author} {\bibfnamefont
  {J.}~\bibnamefont {Stockburger}}, \bibinfo {author} {\bibfnamefont
  {J.}~\bibnamefont {Ankerhold}},\ and\ \bibinfo {author} {\bibfnamefont
  {P.}~\bibnamefont {Joyez}},\ }\bibfield  {title} {\bibinfo {title} {Absence
  of a dissipative quantum phase transition in josephson junctions},\ }\href
  {https://doi.org/10.1103/PhysRevX.10.021003} {\bibfield  {journal} {\bibinfo
  {journal} {Phys. Rev. X}\ }\textbf {\bibinfo {volume} {10}},\ \bibinfo
  {pages} {021003} (\bibinfo {year} {2020})}\BibitemShut {NoStop}%
\bibitem [{\citenamefont {Kuzmin}\ \emph {et~al.}(2025)\citenamefont {Kuzmin},
  \citenamefont {Mehta}, \citenamefont {Grabon}, \citenamefont {Mencia},
  \citenamefont {Burshtein}, \citenamefont {Goldstein},\ and\ \citenamefont
  {Manucharyan}}]{Kuzmin2025}%
  \BibitemOpen
  \bibfield  {author} {\bibinfo {author} {\bibfnamefont {R.}~\bibnamefont
  {Kuzmin}}, \bibinfo {author} {\bibfnamefont {N.}~\bibnamefont {Mehta}},
  \bibinfo {author} {\bibfnamefont {N.}~\bibnamefont {Grabon}}, \bibinfo
  {author} {\bibfnamefont {R.~A.}\ \bibnamefont {Mencia}}, \bibinfo {author}
  {\bibfnamefont {A.}~\bibnamefont {Burshtein}}, \bibinfo {author}
  {\bibfnamefont {M.}~\bibnamefont {Goldstein}},\ and\ \bibinfo {author}
  {\bibfnamefont {V.~E.}\ \bibnamefont {Manucharyan}},\ }\bibfield  {title}
  {\bibinfo {title} {Observation of the schmid--bulgadaev dissipative quantum
  phase transition},\ }\href {https://doi.org/10.1038/s41567-024-02695-7}
  {\bibfield  {journal} {\bibinfo  {journal} {Nature Physics}\ }\textbf
  {\bibinfo {volume} {21}},\ \bibinfo {pages} {132} (\bibinfo {year}
  {2025})}\BibitemShut {NoStop}%
\bibitem [{\citenamefont {Subero}\ \emph {et~al.}(2026)\citenamefont {Subero},
  \citenamefont {Chang}, \citenamefont {Monteiro}, \citenamefont {Chen},\ and\
  \citenamefont {Pekola}}]{Pekola2026}%
  \BibitemOpen
  \bibfield  {author} {\bibinfo {author} {\bibfnamefont {D.}~\bibnamefont
  {Subero}}, \bibinfo {author} {\bibfnamefont {Y.-C.}\ \bibnamefont {Chang}},
  \bibinfo {author} {\bibfnamefont {M.}~\bibnamefont {Monteiro}}, \bibinfo
  {author} {\bibfnamefont {Z.-Y.}\ \bibnamefont {Chen}},\ and\ \bibinfo
  {author} {\bibfnamefont {J.~P.}\ \bibnamefont {Pekola}},\ }\bibfield  {title}
  {\bibinfo {title} {Revisiting dissipation-driven phase transition in a
  josephson junction},\ }\href {https://doi.org/10.1103/4f3c-jv43} {\bibfield
  {journal} {\bibinfo  {journal} {Phys. Rev. Lett.}\ }\textbf {\bibinfo
  {volume} {136}},\ \bibinfo {pages} {116301} (\bibinfo {year}
  {2026})}\BibitemShut {NoStop}%
\bibitem [{\citenamefont {Houzet}\ \emph {et~al.}(2024)\citenamefont {Houzet},
  \citenamefont {Meyer},\ and\ \citenamefont {Nazarov}}]{oddparity}%
  \BibitemOpen
  \bibfield  {author} {\bibinfo {author} {\bibfnamefont {M.}~\bibnamefont
  {Houzet}}, \bibinfo {author} {\bibfnamefont {J.~S.}\ \bibnamefont {Meyer}},\
  and\ \bibinfo {author} {\bibfnamefont {Y.~V.}\ \bibnamefont {Nazarov}},\
  }\bibfield  {title} {\bibinfo {title} {Josephson quantum mechanics at odd
  parity},\ }\href {https://doi.org/10.1103/PhysRevB.110.L020504} {\bibfield
  {journal} {\bibinfo  {journal} {Phys. Rev. B}\ }\textbf {\bibinfo {volume}
  {110}},\ \bibinfo {pages} {L020504} (\bibinfo {year} {2024})}\BibitemShut
  {NoStop}%
\bibitem [{\citenamefont {Houzet}\ \emph {et~al.}(2026)\citenamefont {Houzet},
  \citenamefont {Meyer},\ and\ \citenamefont {Nazarov}}]{oddparity2}%
  \BibitemOpen
  \bibfield  {author} {\bibinfo {author} {\bibfnamefont {M.}~\bibnamefont
  {Houzet}}, \bibinfo {author} {\bibfnamefont {J.~S.}\ \bibnamefont {Meyer}},\
  and\ \bibinfo {author} {\bibfnamefont {Y.~V.}\ \bibnamefont {Nazarov}},\
  }\bibfield  {title} {\bibinfo {title} {Andreev bound states in a
  superconducting qubit at odd parity},\ }\href
  {https://doi.org/10.1007/s10909-026-03398-4} {\bibfield  {journal} {\bibinfo
  {journal} {Journal of Low Temperature Physics}\ }\textbf {\bibinfo {volume}
  {222}},\ \bibinfo {pages} {71} (\bibinfo {year} {2026})}\BibitemShut
  {NoStop}%
\bibitem [{\citenamefont {Erdmanis}\ \emph {et~al.}(2022)\citenamefont
  {Erdmanis}, \citenamefont {Luk\'acs},\ and\ \citenamefont
  {Nazarov}}]{Erdmanis}%
  \BibitemOpen
  \bibfield  {author} {\bibinfo {author} {\bibfnamefont {J.}~\bibnamefont
  {Erdmanis}}, \bibinfo {author} {\bibfnamefont {A.}~\bibnamefont {Luk\'acs}},\
  and\ \bibinfo {author} {\bibfnamefont {Y.~V.}\ \bibnamefont {Nazarov}},\
  }\bibfield  {title} {\bibinfo {title} {Drastic effect of weak interaction
  near special points in semiclassical multiterminal superconducting
  nanostructures},\ }\href {https://doi.org/10.1103/PhysRevB.106.125422}
  {\bibfield  {journal} {\bibinfo  {journal} {Phys. Rev. B}\ }\textbf {\bibinfo
  {volume} {106}},\ \bibinfo {pages} {125422} (\bibinfo {year}
  {2022})}\BibitemShut {NoStop}%
\bibitem [{\citenamefont {Mattis}\ and\ \citenamefont
  {Bardeen}(1958)}]{Mattis}%
  \BibitemOpen
  \bibfield  {author} {\bibinfo {author} {\bibfnamefont {D.~C.}\ \bibnamefont
  {Mattis}}\ and\ \bibinfo {author} {\bibfnamefont {J.}~\bibnamefont
  {Bardeen}},\ }\bibfield  {title} {\bibinfo {title} {Theory of the anomalous
  skin effect in normal and superconducting metals},\ }\href
  {https://doi.org/10.1103/PhysRev.111.412} {\bibfield  {journal} {\bibinfo
  {journal} {Phys. Rev.}\ }\textbf {\bibinfo {volume} {111}},\ \bibinfo {pages}
  {412} (\bibinfo {year} {1958})}\BibitemShut {NoStop}%
\bibitem [{\citenamefont {Sacépé}\ \emph {et~al.}(2020)\citenamefont
  {Sacépé}, \citenamefont {Feigel’man},\ and\ \citenamefont
  {Klapwijk}}]{stronglydisordered}%
  \BibitemOpen
  \bibfield  {author} {\bibinfo {author} {\bibfnamefont {B.}~\bibnamefont
  {Sacépé}}, \bibinfo {author} {\bibfnamefont {M.}~\bibnamefont
  {Feigel’man}},\ and\ \bibinfo {author} {\bibfnamefont {T.}~\bibnamefont
  {Klapwijk}},\ }\bibfield  {title} {\bibinfo {title} {Quantum breakdown of
  superconductivity in low-dimensional materials},\ }\href
  {https://doi.org/10.1038/s41567-020-0905-x} {\bibfield  {journal} {\bibinfo
  {journal} {Nature Physics}\ }\textbf {\bibinfo {volume} {16}},\ \bibinfo
  {pages} {734} (\bibinfo {year} {2020})}\BibitemShut {NoStop}%
\bibitem [{\citenamefont {Kindermann}\ and\ \citenamefont
  {Nazarov}(2003)}]{NazarovKindermann}%
  \BibitemOpen
  \bibfield  {author} {\bibinfo {author} {\bibfnamefont {M.}~\bibnamefont
  {Kindermann}}\ and\ \bibinfo {author} {\bibfnamefont {Y.~V.}\ \bibnamefont
  {Nazarov}},\ }\bibfield  {title} {\bibinfo {title} {Interaction effects on
  counting statistics and the transmission distribution},\ }\href
  {https://doi.org/10.1103/PhysRevLett.91.136802} {\bibfield  {journal}
  {\bibinfo  {journal} {Phys. Rev. Lett.}\ }\textbf {\bibinfo {volume} {91}},\
  \bibinfo {pages} {136802} (\bibinfo {year} {2003})}\BibitemShut {NoStop}%
\bibitem [{\citenamefont {Kos}\ \emph {et~al.}(2013)\citenamefont {Kos},
  \citenamefont {Nigg},\ and\ \citenamefont
  {Glazman}}]{kosFrequencydependentAdmittanceShort2013}%
  \BibitemOpen
  \bibfield  {author} {\bibinfo {author} {\bibfnamefont {F.}~\bibnamefont
  {Kos}}, \bibinfo {author} {\bibfnamefont {S.~E.}\ \bibnamefont {Nigg}},\ and\
  \bibinfo {author} {\bibfnamefont {L.~I.}\ \bibnamefont {Glazman}},\
  }\bibfield  {title} {\bibinfo {title} {Frequency-dependent admittance of a
  short superconducting weak link},\ }\href
  {https://doi.org/10.1103/PhysRevB.87.174521} {\bibfield  {journal} {\bibinfo
  {journal} {Physical Review B}\ }\textbf {\bibinfo {volume} {87}},\ \bibinfo
  {pages} {174521} (\bibinfo {year} {2013})}\BibitemShut {NoStop}%
\bibitem [{\citenamefont {Samuelsen}\ and\ \citenamefont
  {Nazarov}(2025)}]{atlongdistance}%
  \BibitemOpen
  \bibfield  {author} {\bibinfo {author} {\bibfnamefont {E.~S.}\ \bibnamefont
  {Samuelsen}}\ and\ \bibinfo {author} {\bibfnamefont {Y.~V.}\ \bibnamefont
  {Nazarov}},\ }\bibfield  {title} {\bibinfo {title} {Formation of andreev
  molecules at long distance using an embedding circuit},\ }\href
  {https://doi.org/10.1103/hbtb-8hds} {\bibfield  {journal} {\bibinfo
  {journal} {Phys. Rev. B}\ }\textbf {\bibinfo {volume} {112}},\ \bibinfo
  {pages} {214523} (\bibinfo {year} {2025})}\BibitemShut {NoStop}%
\bibitem [{Sup()}]{SupplementalMaterial}%
  \BibitemOpen
  \href@noop {} {}\bibinfo {note} {See Supplemental Material at [URL will be
  inserted by publisher] for the detailed derivation.}\BibitemShut {Stop}%
\bibitem [{\citenamefont {Yokoyama}\ and\ \citenamefont
  {Nazarov}(2015)}]{Yokoyama2015}%
  \BibitemOpen
  \bibfield  {author} {\bibinfo {author} {\bibfnamefont {T.}~\bibnamefont
  {Yokoyama}}\ and\ \bibinfo {author} {\bibfnamefont {Y.~V.}\ \bibnamefont
  {Nazarov}},\ }\bibfield  {title} {\bibinfo {title} {Singularities in the
  andreev spectrum of a multiterminal josephson junction},\ }\href
  {https://doi.org/10.1103/PhysRevB.92.155437} {\bibfield  {journal} {\bibinfo
  {journal} {Phys. Rev. B}\ }\textbf {\bibinfo {volume} {92}},\ \bibinfo
  {pages} {155437} (\bibinfo {year} {2015})}\BibitemShut {NoStop}%
\bibitem [{\citenamefont {Yip}(2003)}]{Yip}%
  \BibitemOpen
  \bibfield  {author} {\bibinfo {author} {\bibfnamefont {S.~K.}\ \bibnamefont
  {Yip}},\ }\bibfield  {title} {\bibinfo {title} {Supercurrent and noise in
  point contacts between two different superconductors},\ }\href
  {https://doi.org/10.1103/PhysRevB.68.024511} {\bibfield  {journal} {\bibinfo
  {journal} {Phys. Rev. B}\ }\textbf {\bibinfo {volume} {68}},\ \bibinfo
  {pages} {024511} (\bibinfo {year} {2003})}\BibitemShut {NoStop}%
\bibitem [{\citenamefont {Larkin}\ and\ \citenamefont
  {Ovchinnikov}(1972)}]{Larkin1972}%
  \BibitemOpen
  \bibfield  {author} {\bibinfo {author} {\bibfnamefont {A.~I.}\ \bibnamefont
  {Larkin}}\ and\ \bibinfo {author} {\bibfnamefont {Y.~N.}\ \bibnamefont
  {Ovchinnikov}},\ }\bibfield  {title} {\bibinfo {title} {Density of states in
  inhomogeneous superconductors},\ }\href@noop {} {\bibfield  {journal}
  {\bibinfo  {journal} {Sov. Phys. JETP}\ }\textbf {\bibinfo {volume} {34}},\
  \bibinfo {pages} {1144} (\bibinfo {year} {1972})},\ \bibinfo {note} {zh.
  Eksp. Teor. Fiz. 61, 2147--2153 (1972)}\BibitemShut {NoStop}%
\bibitem [{\citenamefont {Houzet}(2012)}]{Houzet}%
  \BibitemOpen
  \bibfield  {author} {\bibinfo {author} {\bibfnamefont {M.}~\bibnamefont
  {Houzet}},\ }\href@noop {} {}\bibinfo {howpublished} {workshop talk}
  (\bibinfo {year} {2012}),\ \bibinfo {note} {email exchange}\BibitemShut
  {NoStop}%
\bibitem [{\citenamefont {Manucharyan}\ \emph {et~al.}(2012)\citenamefont
  {Manucharyan}, \citenamefont {Masluk}, \citenamefont {Kamal}, \citenamefont
  {Koch}, \citenamefont {Glazman},\ and\ \citenamefont
  {Devoret}}]{Manucharyan}%
  \BibitemOpen
  \bibfield  {author} {\bibinfo {author} {\bibfnamefont {V.~E.}\ \bibnamefont
  {Manucharyan}}, \bibinfo {author} {\bibfnamefont {N.~A.}\ \bibnamefont
  {Masluk}}, \bibinfo {author} {\bibfnamefont {A.}~\bibnamefont {Kamal}},
  \bibinfo {author} {\bibfnamefont {J.}~\bibnamefont {Koch}}, \bibinfo {author}
  {\bibfnamefont {L.~I.}\ \bibnamefont {Glazman}},\ and\ \bibinfo {author}
  {\bibfnamefont {M.~H.}\ \bibnamefont {Devoret}},\ }\bibfield  {title}
  {\bibinfo {title} {Evidence for coherent quantum phase slips across a
  josephson junction array},\ }\href
  {https://doi.org/10.1103/PhysRevB.85.024521} {\bibfield  {journal} {\bibinfo
  {journal} {Phys. Rev. B}\ }\textbf {\bibinfo {volume} {85}},\ \bibinfo
  {pages} {024521} (\bibinfo {year} {2012})}\BibitemShut {NoStop}%
\bibitem [{\citenamefont {Antonov}\ \emph {et~al.}(2026)\citenamefont
  {Antonov}, \citenamefont {Shaikhaidarov}, \citenamefont {Kim}, \citenamefont
  {Golubev}, \citenamefont {Linzen}, \citenamefont {Il'ichev}, \citenamefont
  {Antonov},\ and\ \citenamefont {Astafiev}}]{Astafiev}%
  \BibitemOpen
  \bibfield  {author} {\bibinfo {author} {\bibfnamefont {I.}~\bibnamefont
  {Antonov}}, \bibinfo {author} {\bibfnamefont {R.~S.}\ \bibnamefont
  {Shaikhaidarov}}, \bibinfo {author} {\bibfnamefont {K.~H.}\ \bibnamefont
  {Kim}}, \bibinfo {author} {\bibfnamefont {D.}~\bibnamefont {Golubev}},
  \bibinfo {author} {\bibfnamefont {S.}~\bibnamefont {Linzen}}, \bibinfo
  {author} {\bibfnamefont {E.~V.}\ \bibnamefont {Il'ichev}}, \bibinfo {author}
  {\bibfnamefont {V.~N.}\ \bibnamefont {Antonov}},\ and\ \bibinfo {author}
  {\bibfnamefont {O.~V.}\ \bibnamefont {Astafiev}},\ }\bibfield  {title}
  {\bibinfo {title} {The microwave phase locking in bloch transistor},\ }\href
  {https://doi.org/10.1038/s41467-025-67735-z} {\bibfield  {journal} {\bibinfo
  {journal} {Nature Communications}\ }\textbf {\bibinfo {volume} {17}},\
  \bibinfo {pages} {1264} (\bibinfo {year} {2026})}\BibitemShut {NoStop}%
\bibitem [{\citenamefont {Samuelsen}(2026)}]{samuelsenCodeFiguresData2026}%
  \BibitemOpen
  \bibfield  {author} {\bibinfo {author} {\bibfnamefont {E.~S.}\ \bibnamefont
  {Samuelsen}},\ }\href {https://doi.org/10.5281/ZENODO.22238533} {\bibinfo
  {title} {Code, figures, and data for: "{{Interaction}} effects on {{Andreev}}
  states from an electromagnetic environment"}} (\bibinfo {year}
  {2026})\BibitemShut {NoStop}%
\end{thebibliography}%


\begin{thebibliography}{3}%
\makeatletter
\providecommand \@ifxundefined [1]{%
 \@ifx{#1\undefined}
}%
\providecommand \@ifnum [1]{%
 \ifnum #1\expandafter \@firstoftwo
 \else \expandafter \@secondoftwo
 \fi
}%
\providecommand \@ifx [1]{%
 \ifx #1\expandafter \@firstoftwo
 \else \expandafter \@secondoftwo
 \fi
}%
\providecommand \natexlab [1]{#1}%
\providecommand \enquote  [1]{``#1''}%
\providecommand \bibnamefont  [1]{#1}%
\providecommand \bibfnamefont [1]{#1}%
\providecommand \citenamefont [1]{#1}%
\providecommand \href@noop [0]{\@secondoftwo}%
\providecommand \href [0]{\begingroup \@sanitize@url \@href}%
\providecommand \@href[1]{\@@startlink{#1}\@@href}%
\providecommand \@@href[1]{\endgroup#1\@@endlink}%
\providecommand \@sanitize@url [0]{\catcode `\\12\catcode `\$12\catcode
  `\&12\catcode `\#12\catcode `\^12\catcode `\_12\catcode `\%12\relax}%
\providecommand \@@startlink[1]{}%
\providecommand \@@endlink[0]{}%
\providecommand \url  [0]{\begingroup\@sanitize@url \@url }%
\providecommand \@url [1]{\endgroup\@href {#1}{\urlprefix }}%
\providecommand \urlprefix  [0]{URL }%
\providecommand \Eprint [0]{\href }%
\providecommand \doibase [0]{https://doi.org/}%
\providecommand \selectlanguage [0]{\@gobble}%
\providecommand \bibinfo  [0]{\@secondoftwo}%
\providecommand \bibfield  [0]{\@secondoftwo}%
\providecommand \translation [1]{[#1]}%
\providecommand \BibitemOpen [0]{}%
\providecommand \bibitemStop [0]{}%
\providecommand \bibitemNoStop [0]{.\EOS\space}%
\providecommand \EOS [0]{\spacefactor3000\relax}%
\providecommand \BibitemShut  [1]{\csname bibitem#1\endcsname}%
\let\auto@bib@innerbib\@empty
\bibitem [{\citenamefont {Kos}\ \emph {et~al.}(2013)\citenamefont {Kos},
  \citenamefont {Nigg},\ and\ \citenamefont
  {Glazman}}]{kosFrequencydependentAdmittanceShort2013b}%
  \BibitemOpen
  \bibfield  {author} {\bibinfo {author} {\bibfnamefont {F.}~\bibnamefont
  {Kos}}, \bibinfo {author} {\bibfnamefont {S.~E.}\ \bibnamefont {Nigg}},\ and\
  \bibinfo {author} {\bibfnamefont {L.~I.}\ \bibnamefont {Glazman}},\
  }\bibfield  {title} {\bibinfo {title} {Frequency-dependent admittance of a
  short superconducting weak link},\ }\href
  {https://doi.org/10.1103/PhysRevB.87.174521} {\bibfield  {journal} {\bibinfo
  {journal} {Physical Review B}\ }\textbf {\bibinfo {volume} {87}},\ \bibinfo
  {pages} {174521} (\bibinfo {year} {2013})}\BibitemShut {NoStop}%
\bibitem [{\citenamefont {Snyman}\ and\ \citenamefont
  {Nazarov}(2008)}]{snymanKeldyshActionMultiterminal2008}%
  \BibitemOpen
  \bibfield  {author} {\bibinfo {author} {\bibfnamefont {I.}~\bibnamefont
  {Snyman}}\ and\ \bibinfo {author} {\bibfnamefont {Y.~V.}\ \bibnamefont
  {Nazarov}},\ }\bibfield  {title} {\bibinfo {title} {Keldysh action of a
  multiterminal time-dependent scatterer},\ }\href
  {https://doi.org/10.1103/PhysRevB.77.165118} {\bibfield  {journal} {\bibinfo
  {journal} {Physical Review B}\ }\textbf {\bibinfo {volume} {77}},\ \bibinfo
  {pages} {165118} (\bibinfo {year} {2008})}\BibitemShut {NoStop}%
\bibitem [{\citenamefont {Kindermann}\ and\ \citenamefont
  {Nazarov}(2003)}]{kindermannInteractionEffectsCounting2003}%
  \BibitemOpen
  \bibfield  {author} {\bibinfo {author} {\bibfnamefont {M.}~\bibnamefont
  {Kindermann}}\ and\ \bibinfo {author} {\bibfnamefont {{\relax Yu}.~V.}\
  \bibnamefont {Nazarov}},\ }\bibfield  {title} {\bibinfo {title} {Interaction
  {{Effects}} on {{Counting Statistics}} and the {{Transmission
  Distribution}}},\ }\href {https://doi.org/10.1103/PhysRevLett.91.136802}
  {\bibfield  {journal} {\bibinfo  {journal} {Physical Review Letters}\
  }\textbf {\bibinfo {volume} {91}},\ \bibinfo {pages} {136802} (\bibinfo
  {year} {2003})}\BibitemShut {NoStop}%
\end{thebibliography}%

\end{document}


\title{Supplementary material for: 'Interaction effects on Andreev states from an electromagnetic environment'}

\author{E. S. Samuelsen}

\affiliation{Kavli Institute of Nanoscience, Delft University of Technology, 2628 CJ Delft, The Netherlands}

\author{Y. V. Nazarov}

\affiliation{Kavli Institute of Nanoscience, Delft University of Technology, 2628 CJ Delft, The Netherlands}

\maketitle

\section{Overview of the supplementary material}
In this supplementary material we present the derivation of the energy shifts of bound states in Josephson junctions coupled to an external environment impedance $Z(\omega)$. Here we give short outline of the structure of this note.
\begin{itemize}
    \item In section \ref{sec:matsubara} we present a derivation of the ground state energy shift $\Delta E^{(e)}$ based on the Matsubara system action. For simplicity, we treat the action abstractly and obtain results in terms of the junction admittance which we obtain from \cite{kosFrequencydependentAdmittanceShort2013b}.
    \item In section \ref{sec:Keldysh} we obtain both the doubled addition shift $\delta^{\rm odd}$ and the singlet addition shift $\delta^{\rm even}$ by calculating the pole shift of the fermion greens function and admittance respectively in an interacting environment. While this can in principle be done with the Matsubara action, we prefer to work with the Keldysh action to avoid a cumbersome analytical continuation. This is the most technical section since we use the explicit expression for the junction action \cite{snymanKeldyshActionMultiterminal2008}. The advantage of this is that we obtain direct results in terms of junction parameters, without relying on other external results than the action.
    \item In section \ref{sec:Hamiltonian} we present a parallel derivation of all the energy shifts based on second order preturbation theory using the quasiparticle Hamiltonian. We use an abstract microscopic model in terms of apriori unknown matrix elements which we relate to the junction admittance and circuit impedance. This derivation is simpler than what is presented in section \ref{sec:Keldysh}, but relies more heavily on results derived elsewhere. In particular, it requires the state resolved junction admittance from \cite{kosFrequencydependentAdmittanceShort2013b}, and the $T$ renormalization from UV photons obtained in \cite{kindermannInteractionEffectsCounting2003}.
\end{itemize}

\section{Ground state energy shifts from the Matsubara action}\label{sec:matsubara}
\begin{figure}[h]
    \centering
    \includegraphics[width=0.7\textwidth]{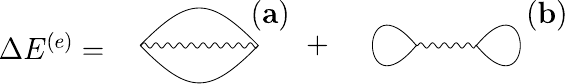}
    \caption{The Feynman diagrams that contribute to the ground state shift at first order in $Z$.}
    \label{fig:ground_state}
\end{figure}

In this section we use the Matsubara action to obtain the free energy shift $\Delta F$ due to the coupling of the junction to an external environment. At zero temperature this reduces to the ground state shift. We first introduce the Gaussian action
\begin{equation}
	S_{\rm env}[\Phi] =\frac{1}{2}\sum_n \frac{|\Phi_{n}|^2}{D(i\omega_n)},
\end{equation}
where $\Phi(\tau) \equiv \beta^{-1}\sum_n e^{i\omega_n \tau } \Phi_{n}$ is the imaginary time fluctuations in the system flux and $\Phi_{-n}=\Phi_{n}^{*}$. By analytical continuation to the upper half plane we relate $D$ to the circuit impedance
\begin{equation}
	Z(\omega) = i\omega D(\omega+i0).
\end{equation}
The junction action $S_{J}[\Phi]$ also depends on the fluctuating flux and we can express the free energy shift due to fluctuations by averaging the coupling $S_{J}[\Phi] - S_J[0]$ over fluctuations
\begin{equation}\label{eq:free_energy_shift}
	\Delta F = -\frac{1}{\beta}\ln\Bigl( \frac{\int \mathcal{D}[\Phi]e^{- S_{\rm env}[\Phi] -(S_J[\Phi] -S_J[0]) }}{\int \mathcal{D}[\Phi]e^{ - S_{\rm env}[\Phi] }} \Bigr).
\end{equation}
We stress that $\Phi$ only represents fluctuations on top of a fixed background flux, so $\Delta F$ depends on the junction phase $\phi$ and can be differentiated to obtain the fluctuation induced shift in current. The junction action generates time ordered current correlators according to
\begin{equation}
	 S_J[\Phi] = -\ln \langle \mathcal{T} e^{-\int_0^\beta d\tau \hat{H}_J(\Phi(\tau))} \rangle,
\end{equation}
where $\hat{H}_J[\Phi]$ is the junction Hamiltonian coupled to an imaginary time flux source.
By differentiating both sides we obtain
\begin{align}
	S_J[\Phi] &= S_J[0]+ \langle \hat{I}\rangle \Phi_0 + \frac{1}{2}\sum_n \Pi(i\omega_n)\Phi_n^2 + \mathcal{O}(\Phi^3),
\end{align}
where we used that $\hat{I}= \partial \hat{H}_J /\delta \Phi$. After analytical continuation to the upper half plane we relate $\Pi$ to the junction admittance
\begin{equation}
	Y(\omega) = \frac{i}{\omega}\Pi(\omega + i0).
\end{equation}
Next we insert the expansion for the junction action in Eq. \eqref{eq:free_energy_shift} and perform the Gaussian integral to obtain
\begin{align}
	\Delta F &= - \frac{1}{2}D(0) \langle \hat{I}\rangle^2 +  \frac{1}{2 \beta} \sum_{n} D(i\omega_n)\Pi(i\omega_n)\\
	&= - \frac{1}{2}L \langle \hat{I}\rangle^2   + \frac{1}{2} \int_0^\infty \frac{d\omega}{\pi } N_\omega {\rm Im} (Z(\omega) Y(\omega)),
\end{align}
where $N_{\omega}\equiv \coth(\beta \omega/2)$ and at zero temperature
\begin{equation}\label{eq:delta_E_g}
	\Delta E^{(e)} = - \frac{1}{2}L I_J^2  + \frac{1}{2} \int_0^\infty \frac{d\omega}{\pi } {\rm Im} (Z(\omega) Y^{(e)}(\omega)),
\end{equation}
where and $Y^{(e)}$ is the junction admittance in the even ground state, and $I_J$ the Josephson current. In Fig \ref{fig:ground_state} we represent the different contributions with Feynman diagrams. The straight lines give the fermion propagators while the wiggly lines represent the photon propagator $\sim Z$. Therefore, the circles with one fixed vertex gives a current so (b) in Fig. \ref{fig:ground_state} corresponds to $-LI_J^2/2$. The bubble diagrams with two vertices corresponds to an admittance so the second term in Eq. \eqref{eq:delta_E_g} can be illustrated by diagram (a).

For one transmission channel $I_J=-2e\partial_{\phi} E_A$, where
\begin{equation}
	E_A \equiv \Delta \sqrt{1-T \sin^2(\phi/2)}.
\end{equation}
Here $\Delta$ is the superconducting gap, $T$ the transmission coefficient, and $\phi$ the junction phase drop. The explicit expression for $Y^{(e)}$ is provided in~\cite{kosFrequencydependentAdmittanceShort2013b} and reads
\begin{equation}
	{\rm Re} Y^{(e)}(\omega) = {\rm Re} (Y_1(\omega) + Y_2(\omega) + Y_3(\omega) ),
\end{equation}
where
\begin{align}
{\rm Re} Y_1(\omega)&= G_Q T\frac{\theta(\omega - 2 \Delta)}{\omega} \int_{\Delta}^{\infty} d \epsilon_1 \rho(\epsilon_1)\int_{\Delta}^{\infty} d \epsilon_2 \rho(\epsilon_2)
\delta(\omega - \epsilon_1 - \epsilon_2) \left ( 1 - \frac{\Delta^2(1+ (T-2) \sin^2(\phi/2))} {\epsilon_1 \epsilon_2} \right),\label{eq:y1}\\
{\rm Re} Y_2(\omega)&=  G_Q T\frac{ \pi \Delta \sqrt{T} |\sin(\phi/2)|}{\omega} \theta(\omega - \Delta - E_A) \rho(\omega-E_A)  \left ( 1 - \frac{\Delta^2(1+ (T-2) \sin^2(\phi/2))} {E_A (\omega - E_A)} \right),\label{eq:y2}\\
{\rm Re} Y_3(\omega) &= G_Q T\frac{\Delta^4\pi^2}{2}  \frac{ T(1-T) \sin^4(\phi/2)}{E_A^3}
\delta(\omega - 2E_A),\label{eq:y3}
\end{align}
and $\rho(\epsilon) \equiv \epsilon \sqrt{\epsilon^2-\Delta^2}/(\epsilon^2 - E_A^2)$. Here $G_Q = e^2 / \pi$ is the conductance quantum and $e$ the electron charge. We use $\hbar = 1$.

\section{Addition energy shifts from the Keldysh action} \label{sec:Keldysh}
In this section we use the Keldysh action to obtain the fermion pole shift $\delta^{odd}$ of the fermion pole at $\epsilon = E_A$, and the pole shift $\delta^{even}$ of the admittance pole at $\omega = 2E_A$. In section \ref{sec:Keldysh_action} we introduce the different contributions to the action. Then in section \ref{sec:doublet_shift} and \ref{sec:singlet_shift} we use it to obtain $\delta^{odd}$ and $\delta^{even}$ respectively.

\subsection{The Keldysh action } \label{sec:Keldysh_action}
The full Keldysh action can be separated in an environment action $S_{env}$ that and a junction action $S_{J}$
\begin{equation}\label{eq:full_action}
	S = S_J + S_{\rm env}.
\end{equation}
The environment action controls fluctuations in the junction phase due to circuit impedance, while the junction action depends on the phase through the lead greens functions $\check{G}_i$.

\subsubsection{The environment action}
The Keldysh environment action is
\begin{equation}\label{eq:environment}
	S_{\rm env} =
	\frac{1}{2}
	\int \frac{d\omega}{2\pi}
	\begin{pmatrix}
	    \chi_{cl}(\omega)&\chi_q(\omega)
	\end{pmatrix}
	\begin{pmatrix}
	    D^K(\omega) & D^R(\omega)\\
		D^A(\omega) & 0
	\end{pmatrix}^{-1}
	\begin{pmatrix}
	    \chi_{cl}(\omega)\\
		\chi_q(\omega)
	\end{pmatrix},
\end{equation}
where we work in the basis of the classical and quantum fields
\begin{equation}
	\chi_{cl, q}(t) \equiv \frac{1}{2}(\chi_{+}(t) \pm \chi_-(t)).
\end{equation}
Here $\chi_{\pm}(t)$ is the fluctuating phase evaluated at the $\pm$ Keldysh branches. For convenience, we define $\chi_\alpha(t)$ with an extra factor of $2$ compared to the main text, so the flux is $\Phi_\alpha (t) =  \chi_\alpha(t)/e$.
To correctly reproduce the fluctuations of the circuit impedance we have
\begin{equation}
	D^R(\omega) = -i \pi G_Q \frac{Z(\omega) }{2\omega },
\end{equation}
while $D^A(\omega)=D^R(\omega)^{*}$ and we assume thermal equilibrium where the Keldysh component is
\begin{equation}
	D^K(\omega) = N_\omega (D^R(\omega) - D^A(\omega)).
\end{equation}

\subsubsection{The junction action}
The junction action $S_J[\chi]$ generates time ordered current correlators on the Keldysh contour and can be expressed
\begin{equation}\label{eq:keldysh_definition}
    S_J[\chi] = -i\ln\langle \mathcal{T} e^{-i \oint dt \hat{H}_J(\chi(t)) }\rangle
\end{equation}
where $\hat{H}_J$ is the junction Hamiltonian and the time integral goes around the Keldysh contour. The junction action can be expressed~\cite{snymanKeldyshActionMultiterminal2008}
\begin{equation}\label{eq:keldysh_lead_version}
	S_J = -\frac{i}{2} \tr \ln(1-\frac{T}{4}(\check{G}_{\phi+2\check{\chi}} - \check{G}_{-\phi-2\check{\chi}})^2),
\end{equation}
where $\phi$ is the fixed phase of junctions and for simplicity of notation we assume one transmission channel with transmission coefficient $T$. Here $\check{G}_i$ are greens function characterizing the leads. It has matrix structure in: frequency, Nambu, and Keldysh. The trace runs over all matrix labels. Let us use $\tau_i$ as a Pauli basis for the Nambu structure and $\sigma_i$ as a basis for the Keldysh structure. Now the lead greens functions can be expressed

\begin{equation}
    \check{G}_{\phi+2\check{\chi}}(t;t') \equiv e^{-i \tau_3\check{\chi}(t)/2 }   \check{G}_{\phi_1}(t-t')e^{i \tau_3\check{\chi}(t')/2 } = e^{-i \tau_3(\phi+2\check{\chi}(t))/4 }   \check{G}_{0}(t-t')e^{i \tau_3(\phi +2\check{\chi}(t'))/4 }
\end{equation}
and similarly for $\check{G}_{-\phi-\check{\chi}}$, where
\begin{align}
    &\check{\chi}(t) =
\begin{pmatrix}
	\chi_{cl}(t) & \chi_{q}(t)\\
	\chi_{q}(t) & \chi_{cl}(t)
\end{pmatrix} = \sigma_0\chi_{cl}(t) + \sigma_1 \chi_{q}(t) \equiv \sum_{\alpha} \sigma_\alpha \chi_\alpha(t) \\
&\check{G}_0(\epsilon) =
\begin{pmatrix}
	\hat{G}_R(\epsilon) & F_\epsilon (\hat{G}_R(\epsilon) - \hat{G}_A(\epsilon))\\
	0 & \hat{G}_A(\epsilon)
\end{pmatrix} =
\check{M}^{-1}(\epsilon)
\begin{pmatrix}
	\hat{G}_R(\epsilon) & 0\\
	0 & \hat{G}_A(\epsilon)
\end{pmatrix}
\check{M}(\epsilon); \quad \check{M}(\epsilon) \equiv
	\begin{pmatrix}
		1 & F_\epsilon\\
		0 & 1
	\end{pmatrix}\\
&\hat{G}_{R/A}(\epsilon) = \frac{ \Delta \tau_1 -i (\epsilon\pm i0) \tau_3}{\sqrt{\Delta^2 - (\epsilon\pm i0)^2}}.
\end{align}
Here we use the convention where $\check{G}_0$ and $\check{\chi}$ without a label are matrices in frequency and time and with a label they are projected, so $\bra{\epsilon} \check{G}_0 \ket{\epsilon'} \equiv \check{G}_0(\epsilon) \delta(\epsilon - \epsilon')$, $\bra{t} \check{G}_0 \ket{t'} \equiv \check{G}_0(t-t')$, and similarly for $\check{\chi}$ which is diagonal in time. We use the convention
\begin{align}
    &1 = \int dt \ket{t}\bra{t} =  \int \frac{d\epsilon}{2\pi}\, \ket{\epsilon }\bra{\epsilon},\quad \quad \quad \braket{t}{\epsilon} = e^{-i\epsilon t}.
\end{align}
A usefull property of the Greens function is that they are normalized $(\check{G}_{1,2})^2 = 1$. It is convenient to introduce the matrix $\check{\epsilon}$ which is diagonal in frequency
\begin{equation}
	\bra{\epsilon'}\check{\epsilon} \ket{\epsilon''} \equiv \check{M}^{-1}(\epsilon') (\epsilon' + i0 \sigma_3)\check{M}(\epsilon') \delta(\epsilon'-\epsilon''),
\end{equation}
in order to write out the Keldysh structure in a compact way
\begin{equation}
\check{G}_0 = \frac{ \Delta \tau_1 -i \check{\epsilon}\tau_3}{\sqrt{\Delta^2 - \check{\epsilon}^2}},
\end{equation}
From Eq. \eqref{eq:keldysh_definition} it is clear that we obtain the junction current by expanding to first order in $\chi$ and the admittance from the second order expansion. To obtain the lowest order fluctuation correction to the admittance we therefore have to expand to forth order in $\chi$. In principle, we could proceed like this and obtain $\delta^{even}$ from the pole shift of the admittance close to $\omega = 2E_A$, and $\delta^{odd}$ would appear as a shift in the threshold frequency $\omega = \Delta + E_A $ of $Y_2(\omega)$. This program turns out to be unnecessarily cumbersome.

We know there is a fermion pole at $\epsilon = E_A$, and it would be more convenient to directly calculate the shift in this to second order in $\chi$, instead of expanding the whole action to forth order in order to calculate shift in the admittance. This pole is in fact a pole of the matrix
\begin{equation}\label{eq:Q_matrix}
	\check{Q} \equiv [1-\frac{T}{4}(\check{G}_{\phi} - \check{G}_{-\phi})^2]^{-1} = \frac{\check{\epsilon}^2-\Delta^2}{\check{\epsilon}^2 - E_A^2 }.
\end{equation}
However, on the form of Eq. \eqref{eq:keldysh_lead_version} the fluctuating phase $\check{\chi}$ enters $\check{Q}$ in a complicated way, and it is not clear how to organize the pole expansion. The resolution to this is that we should introduce a scattering matrix with a matrix label corresponding to 'lead structure'. This gives us an alternative representation of the junction action where we separate the fluctuations $\check{\chi}(t)$ from a 'bare' greens function $G \propto \check{Q}$ characterizing the junction in absence of fluctuations. By using this we obtain $\delta^{odd}$ by identifying the self energy in a Dyson expansion (see \ref{sec:doublet_shift}). The alternative representation of the action also simplifies the calculation of the admittance in presence of fluctuations since we can put the phase $e^{-i \tau_3\check{\chi}(t)/4 }$ on the frequency independent scattering matrix and avoid generating terms on the form $\sim [\check{\chi}, [\check{\chi}, [\check{\chi}, [\check{\chi}, \check{G}_{\phi} ]]]]$.

We relegate the somewhat technical derivation of the alternative representation of the junction action to appendix \ref{sec:alternative_action} and present the results in the following.

\subsubsection{Alternative representation of the Junction action}\label{sec:summary}
In appendix \ref{sec:alternative_action} we show that the junction action can be written
\begin{align}\label{eq:keldysh_final}
    S_J = -\frac{i}{2}\tr \ln(1 - G v).
\end{align}
In addition to energy/time, the trace is over $4$ different two-dimensional matrix labels for which we use 4 different Pauli bases: $\eta_i$ for lead, $\tau_i$ for Nambu, $\sigma_i$ for Keldysh and $\zeta_i$ for the projection onto the $\pm1$ eigenvalues of $\check{G}_0$. The matrix structure in $\zeta_i$ purely Abelian and we only use $\zeta_3 \equiv \zeta$. It is only relevant for the sign $\zeta=\pm1$ it contributes in the trace. The interaction vertex is defined
\begin{align}
	v & = -\sqrt{T}(e^{i \tau_3 \eta_3 \check{\chi}} -1)e^{i \tau_3 \eta_3 \varphi/2}\eta_1 \propto \check{\chi}\label{eq:v}.
\end{align}
The greens function $G$ can be written as a direct product of, a matrix $\mathcal{P}$ with Nambu structure, a matrix $\mathcal{S}$ with lead structure, and $\check{Q}$ which contains the bound state pole structure
\begin{align}
    G &\equiv  \mathcal{P} \mathcal{S}  \check{Q}, \label{eq:G}\\
    \mathcal{P} & \equiv \frac{1}{2}(1 -i\tau_3\zeta\check{E} + \tau_1\zeta\check{D}),\\
    \mathcal{S} &\equiv \sqrt{1-T}\eta_3  + \sqrt{T} c\eta_1+i\sqrt{T}s \eta_2 \zeta \check{E}, \label{eq:S} \\
    \check{E}&\equiv \frac{ \check{\epsilon}}{\sqrt{\Delta^2 - \check{\epsilon}^2}}, \quad \quad \quad \check{ D} \equiv \frac{  \Delta}{\sqrt{\Delta^2 - \check{\epsilon}^2}},\quad \quad \quad s \equiv \sin(\phi/2), \quad \quad \quad c \equiv \cos(\phi/2)\label{eq:ED}.
\end{align}
Some useful properties we use are $\mathcal{S}^2 = \check{Q}^{-1}$ and $\mathcal{P}^2 =\mathcal{P}$. Here $\mathcal{P} = (1 + \zeta \check{G}_0)/2$ is the projection onto the '$\zeta$' eigenspace of $\check{G}_0$, and as we show in appendix \ref{sec:alternative_action}, $\mathcal{S} = \mathcal{P}\mathcal{S}_\phi\mathcal{P}$ is the projection of the system 'scattering matrix' onto the same eigenspace, where
\begin{align}
	\mathcal{S}_\phi& = \sqrt{1-T}\eta_3 + c\sqrt{T}\eta_1 - s\sqrt{T}\eta_2 \tau_3.
\end{align}
Let us also give the spectral representation of $G$. By using the general spectral formula Eq. \eqref{eq:spectral_general} from appendix \ref{sec:spectral} with $G$ we obtain
\begin{align}
	G &= \frac{1}{2}\mathcal{S}_{\phi}+ \int_{-\infty}^{\infty} d\epsilon' \frac{\mathcal{A}_{\epsilon'}}{\epsilon'-\check{\epsilon}}
\end{align}
where the contact term comes from $S_\phi = G^R(\infty) + G^A(\infty)$ and the spectral function is
\begin{align}\label{eq:spectral_function}
	\mathcal{A}_{\epsilon} \equiv \frac{1}{2\pi i}(G^R(\epsilon) - G^A(\epsilon))  = \mathcal{A}^{+} \delta(\epsilon - E_A ) + \mathcal{A}^{-} \delta(\epsilon + E_A ) + \mathcal{A}^c_{\epsilon}\rho(\epsilon)
\end{align}
where
\begin{align}
	\mathcal{A}^\pm &\equiv  \pm \frac{\Delta^2 - E_A^2}{2 E_A} \mathcal{P}_{\pm E_A}\mathcal{S}_{\pm E_A},\\
	\mathcal{A}^c_{\epsilon} &\equiv  \frac{\zeta}{2\pi \epsilon} (( \Delta\tau_1-i \epsilon \tau_3 )(\sqrt{1-T}\eta_3 + c\sqrt{T} \eta_1) + i s \sqrt{T}\epsilon \eta_2),\\
	\rho(\epsilon) &\equiv \theta(|\epsilon|-\Delta)\frac{|\epsilon| \sqrt{\epsilon^2-\Delta^2}}{\epsilon^2- E_A^2},
\end{align}
and $\mathcal{P}_{\pm E_A}$ and $\mathcal{S}_{\pm E_A}$ means we evaluate $\mathcal{P}^{R/A}$ and $\mathcal{S}^{R/A}$ at $\epsilon = \pm E_A$ (the expressions are independent of $R/A$ since $\epsilon = \pm E_A$ is bellow the branch cut). From the spectral form in Eq. \eqref{eq:spectral_function} we explicitly see that $G$ contains spectral information about both the bound states and the lead quasiparticles.

The version of the action in Eq. \eqref{eq:keldysh_final} is particularly useful for the $\chi$ expansion, since the $\chi$ dependence is entirely contained in $v$ which is diagonal in time. In particular, it is straight forward to take functional derivatives with respect to $\chi$. By expanding the logarithm we have
\begin{equation}
	\frac{\delta S_J}{\delta \chi_\alpha(t)} = -\frac{i}{2} \tr(\mathcal{G} v' X_{\alpha t}),
\end{equation}
where we defined
\begin{equation}\label{eq:dyson}
	\mathcal{G} \equiv [1- Gv]^{-1}G = G + GvG + GvGvG + \hdots,
\end{equation}
and
\begin{align}
	&v' \equiv \partial_{\check{\chi}}v = \sqrt{T} e^{i \tau_3 \eta_3 (\check{\chi}+\varphi/2)} \tau_3 \eta_2,\\
	& X_{\alpha t} \equiv \frac{\delta \check{\chi}}{\delta \chi_\alpha(t)} = \sigma_\alpha \ket{t}\bra{t}.
\end{align}
Here we use the convention $\sigma_{cl} \equiv \sigma_0$ and $\sigma_{q}\equiv \sigma_1$. To obtain higher order derivatives we use
\begin{equation}
	\frac{\delta \mathcal{G}}{\delta \chi_\alpha(t)} = \mathcal{G} v' X_{\alpha t} \mathcal{G}
\end{equation}
and the product rule, where $X_{\alpha t}$ commutes with $\check{\chi}$ and therefore also $v$.

\subsection{Doublet addition shift}\label{sec:doublet_shift}
\begin{figure}[h]
    \centering
    \includegraphics[width=0.7\textwidth]{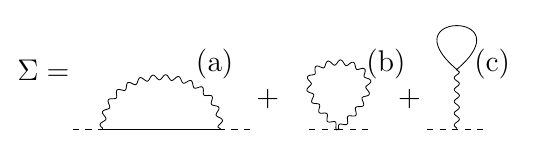}
    \caption{The Feynman diagrams that contribute to the fermion self energy $\Sigma$ at first order in $Z$.}
    \label{fig:sigma}
\end{figure}

In the absence of fluctuations $\mathcal{G}(t;t')\rightarrow G(\epsilon)$ has a pole at $\epsilon = E_A$. However, because $G(\epsilon)$ is projected onto a block in Nambu space given by $\mathcal{P}(\epsilon)$ we have to trace over the Nambu indices before we can obtain the pole from the determinant. Let us define
\begin{equation}
	\bar{G} \equiv \tr_\tau(G) = \check{Q} \mathcal{S}.
\end{equation}
The convention is that subscripts corresponding to Pauli bases on $\tr$  ($\det$) means we take the trace (determinant) with respect to the corresponding matrix label. We do not write out identity matrices and from the context we understand that $\bar{G}$ is not proportional to $\tau_0$ since the Nambu label is traced over. We have
\begin{equation}
	\det_{\eta, \zeta}(\bar{G})=\bar{G}^2 = \check{Q}.
\end{equation}
so the determinant of $\bar{G}$ is a Keldysh matrix with a pole at $\epsilon = E_A$. If we average over environment fluctuations $\langle \mathcal{G}\rangle$ also becomes diagonal in frequency, but with a shifted pole. By expanding $\langle\mathcal{G} \rangle $ like in Eq. \eqref{eq:dyson} we obtain
\begin{equation}\label{eq:dyson_2}
	\langle \mathcal{G} \rangle  = G + G\langle v\rangle G + G\langle vGv \rangle G + \hdots = G + G \Sigma G +  G \Sigma G \Sigma G  +  \hdots
\end{equation}
where
\begin{equation}\label{eq:sigma_original}
	\Sigma =  \langle v\rangle+\langle vGv \rangle + \mathcal{O}(Z^2),
\end{equation}
and we used that $-\langle v\rangle G\langle v \rangle \propto Z^2$. The averages are defined via
\begin{equation} \label{eq:average}
    \langle A \rangle \equiv \frac{\int \mathcal{D}[\chi]e^{iS}A }{\int \mathcal{D}[\chi]e^{iS} },
\end{equation}
where $S = S_J + S_{env}$. In Eq. \eqref{eq:sigma_original} both $G(\epsilon)$ and $\Sigma(\epsilon)$ is projected onto a block in Nambu space given by $\mathcal{P}(\epsilon)$ and before doing the resummation we have to pick out the finite component in Nambu space. By tracing over the Nambu indices in Eq. \eqref{eq:dyson_2} we have
\begin{equation}\label{eq:g_full_bar}
	\tr_{\tau}(\langle\mathcal{G} \rangle )\equiv \langle \bar{\mathcal{G}} \rangle  = \bar{G} + \bar{G} \bar{\Sigma} \bar{G} +  \bar{G} \bar{\Sigma} \bar{G} \bar{\Sigma} \bar{G}  +  \hdots = [\bar{G}^{-1} - \bar{\Sigma}]^{-1} = [\mathcal{S} - \bar{\Sigma}]^{-1},
\end{equation}
where
\begin{equation}\label{eq:bar_sigma}
	\bar{\Sigma} \equiv  \tr_{\tau}(\mathcal{P}\langle v\rangle\mathcal{P}+\mathcal{P}\langle vGv \rangle\mathcal{P}) + \mathcal{O}(Z^2)=\tr_{\tau}(\mathcal{P}v'_0\langle \check{\chi}\rangle+\frac{1}{2}\mathcal{P}v''_0\langle \check{\chi}^2\rangle +\mathcal{P}v'_0\langle \check{\chi} G\check{\chi} \rangle v'_0) + \mathcal{O}(Z^2).
\end{equation}
Here $v^{(n)}_0 \equiv v^{(n)}[\check{\chi}=0]$, so
\begin{align}
	&v'_0 = \sqrt{T} e^{i \tau_3 \eta_3 \phi/2} \tau_3 \eta_2 =
	\sqrt{T} (c \tau_3 \eta_2 + s \eta_1), \\
	&v''_0 = \sqrt{T} e^{i \tau_3 \eta_3 \phi/2} \eta_1 =\sqrt{T} (c \eta_1 - s  \tau_3 \eta_2 ).
\end{align}
To obtain the pole shift we take the square of Eq. \eqref{eq:g_full_bar} to obtain
\begin{equation}
	\langle \bar{\mathcal{G}} \rangle^2  \sim [\mathcal{S}^2 - \{\mathcal{S}, \bar{\Sigma}\}]^{-1} = [\check{Q}^{-1}-  \tr_\eta (\mathcal{S} \bar{\Sigma})]^{-1} = (\check{\epsilon}^2 - \Delta^2)[\check{\epsilon}^2 - E_A^2 - (\check{\epsilon}^2 - \Delta^2)  \tr_\eta (\mathcal{S} \bar{\Sigma})]^{-1} ,
\end{equation}
where in the first equality we neglected $\bar{\Sigma}^2 \sim Z^2$ and in the second we used $\mathcal{S}^2 = \check{Q}^{-1}$ and that both $\tr_\eta(\mathcal{S})=0$ and $\tr_\eta(\bar{\Sigma})=0$ which is straight forward to verify from Eqs. \eqref{eq:S} and \eqref{eq:bar_sigma}. We obtain the pole shift of the retarded Keldysh component evaluated close to $\epsilon = E_A$
\begin{equation}\label{eq:d_odd_original_almost}
	\delta^{\rm odd} = - \frac{\Delta^2 - E_A^2}{2E_A} [\tr_\eta(\mathcal{S}\bar{\Sigma})]^R(E_A).
\end{equation}
As we show in the following by direct calculation of the traces, $\tr_\eta(\mathcal{S}\bar{\Sigma})$ does not have any matrix structure $\propto \zeta$ and hence we can regard the right-hand side as a constant that reduces to the real singlet addition shift at zero temperature. Let us insert Eq. \eqref{eq:bar_sigma} and separate in three contributions
\begin{align}
    \delta^{\rm odd}_{(a)} &\equiv - \frac{\Delta^2 - E_A^2}{2E_A} [\tr_{\eta \tau}(\mathcal{S}\mathcal{P}v'_0\langle \check{\chi} G\check{\chi} \rangle v'_0)]_R(E_A)=- T\frac{\Delta^2 - E_A^2}{2E_A} [ -\check{Q}^{-1} \langle \check{\chi}\check{Q}\check{\chi}\rangle+  \check{E}\langle \check{\chi}\check{Q}\check{E} \check{\chi}\rangle \nonumber\\
	&+  (c^2 - (1-T)s^2)\check{D}\langle\check{\chi}\check{Q}  \check{D} \check{\chi}  \rangle  + T s^2[ 1 + (c^2 - s^2)\check{D}^2]\langle \check{\chi} \check{Q} \check{D}^2 \check{\chi}\rangle   +  T s^2\check{E}\check{D}\langle \check{\chi}\check{Q} \check{E}\check{D} \check{\chi}  \rangle]_R(E_A)\label{eq:d3},\\
	\delta^{\rm odd}_{(b)} &\equiv - \frac{\Delta^2 - E_A^2}{4E_A} [\tr_{\eta \tau}(\mathcal{S}\mathcal{P}v''_0 ) \langle \check{\chi}^2\rangle]_R(E_A)= -T \frac{\Delta^2 - E_A^2}{2E_A} [ (1-s^2 \check{D}^2)      \langle \check{\chi}^2\rangle]_R(E_A),\\
	\delta^{\rm odd}_{(c)} &\equiv - \frac{\Delta^2 - E_A^2}{2E_A} [\tr_{\eta \tau} (\mathcal{S}\mathcal{P}v'_0)\langle \check{\chi}\rangle )]_R(E_A)=-T \frac{\Delta^2 - E_A^2}{2E_A}  [2cs \check{D}^2\langle \check{\chi}\rangle ]_R(E_A)
\end{align}
where $\delta^{\rm odd}= \delta^{\rm odd}_{(a)}+\delta^{\rm odd}_{(c)}+\delta^{\rm odd}_{(c)}$. The three contributions originate from parts of $\Sigma$ that can be illustrated by three Feynman diagrams as in Fig. \ref{fig:sigma}, where the diagram in $(i)$ corresponds to $\delta_{(i)}^{\rm odd}$. Here the wiggly line represents the photon propagator $D$ and the straight line the fermion propagator $G$. To proceed we need to calculate the averages over the phase $\chi_{\alpha}(t)$.

\subsubsection{Phase averages}
To the lowest order in impedance it is sufficient to expand the junction action to first order in $\chi_{\alpha}(t)$ when calculating the phase averages,
\begin{equation}
	S \sim S_{\rm env}-2 \frac{\langle \hat{I}\rangle}{e}\int dt \chi_q(t),
\end{equation}
for $S_{env}$ in Eq. \eqref{eq:environment}. Using this we calculate the gaussian integrals (Eq. \eqref{eq:average}) to obtain
\begin{align}
    &\langle \chi_{\alpha}(t) \rangle = 2 \delta_{\alpha, cl}\frac{\langle \hat{I}\rangle}{e}
	\int dt' D_R(t-t')= - \delta_{\alpha, cl}  L  I_J e \label{Average_1}\\
	&\langle \chi_{\alpha}(t) \chi_\beta(t') \rangle =
	\begin{pmatrix}
		iD_K(t-t') & iD_R(t-t')\\
		iD_A(t-t') & 0
	\end{pmatrix}_{\alpha \beta}\label{Average_2}
\end{align}
where $Z(\omega) \sim -i\omega L$ for small $\omega$, and we use the convention where $cl$ ($q$) if the first (second) matrix index. By using Eq. \eqref{Average_1} we can immediately insert $\langle \check{\chi}\rangle = \langle \chi_{cl}\rangle \sigma_0$ in the expression for $\delta_1^{odd}$ and evaluate the remaining terms at $\epsilon = E_A$ to obtain
\begin{equation}\label{eq:d1_final}
	\delta^{\rm odd}_{(c)} = L  I_J^2
\end{equation}
As we show bellow this combines with another inductive contribution in $\delta^{odd}_3$ to produce the correct prefactor $1/2$. By using Eq \eqref{Average_2} we can calculate the contraction between some fermionic greens functions $\check{G}$ and the fluctuating phase
\begin{align}\label{eq:contracted}
	\langle \check{\chi}(t) \check{G}(t-t') \check{\chi}(t') \rangle  &=
	\sum_{\alpha \beta} \sigma_{\alpha} \check{G}(t-t') \sigma_\beta \langle \chi_{\alpha}(t) \chi_{\beta}(t') \rangle=
	i
	\begin{pmatrix}
		G_R D_K + G_K D_R & G_K D_K + (G_R - G_A) (D_R - D_A)\\
		0 & G_A D_K + G_K D_A
	\end{pmatrix}
\end{align}
where we suppressed the time label and used that $G_R D_A + G_A D_R =0$, which holds also in the case where $G_{R/A}(t-t')$ contains a contact term $\propto \delta(t-t')$ which can be seen in the frequency representation by inserting the spectral representation for $D_{R/A}(\omega)$. The matrix on the right-hand side of Eq. \eqref{eq:contracted} has the form of a fermionic Keldysh matrix and the frequency representation can be written
\begin{equation}\label{eq:contracted_2}
	\langle \check{\chi} \check{G} \check{\chi}\rangle(\epsilon) = \check{M}^{-1}(\epsilon)
	\begin{pmatrix}
		\langle \check{\chi} \check{G} \check{\chi}\rangle_{R}(\epsilon) & 0\\
		0 & \langle \check{\chi} \check{G} \check{\chi}\rangle_{A}(\epsilon)
	\end{pmatrix}
	\check{M}(\epsilon),
\end{equation}
where
\begin{align}
    \langle \check{\chi} \check{G} \check{\chi} \rangle_{R/A}(\varepsilon) &=  \int \frac{d\omega}{2\pi} (G_{R/A}(\varepsilon - \omega)  iD_K(\omega) + G_K(\varepsilon - \omega) iD_{R/A}(\omega)).
\end{align}
We can use the spectral representation (see appendix \ref{sec:spectral}) to write Eq. \eqref{eq:contracted_2} as
\begin{align}\label{eq:spectral_contracted}
	\langle \check{\chi} \check{G} \check{\chi} \rangle(\check{\varepsilon})&=
	\frac{\pi G_Q}{2}(G_R(\infty) + G_A(\infty) )\int_0^\infty \frac{d\omega}{\pi}   \frac{ N_\omega  {\rm Re}(Z(\omega)) }{\omega }\\
	& +\frac{1}{2}\int \frac{d\omega}{ \omega}G_Q{\rm Re}(Z(\omega)) \int \frac{d\varepsilon'}{2\pi i} (G^R(\varepsilon') -G^A(\varepsilon')) \frac{F_{\varepsilon'}+N_{\omega}}{ \varepsilon'  + \omega-\check{\varepsilon} }.
\end{align}
In particular if we let $\check{G}$ be the identity matrix we obtain
\begin{equation}
	\langle \check{\chi}^2\rangle =  \pi G_Q \int_0^\infty \frac{d\omega}{\pi}   N_\omega \frac{ {\rm Re}(Z(\omega)) }{\omega } =  \pi G_Q \langle \hat{\Phi}^2\rangle ,
\end{equation}
where $\hat{\Phi}$ is the junction flux operator. By inserting this in the expression for $\delta_2^{\rm odd}$ we obtain
\begin{align} \label{eq:d2_final}
	\delta^{\rm odd}_{(b)} &= \pi G_Q \langle \hat{\Phi}^2\rangle (1-T) \frac{ \Delta^2 - E_A^2}{2E_A} =  \delta T \partial_T E_A ,
\end{align}
where
\begin{equation}
	\delta T \equiv  -\pi G_Q \langle \hat{\Phi}^2\rangle T(1-T).
\end{equation}
represents a renormalization of the transmission coefficient $T$ from UV photons \cite{kindermannInteractionEffectsCounting2003}. By evaluating everything outside the correlators in the expression Eq. \eqref{eq:d3} for $\delta^{odd}_3$ at $\epsilon = E_A$ we obtain
\begin{equation}\label{eq:d3_2}
	\delta_{3}^{odd}  =\langle \check{\chi} \check{L} \check{\chi}\rangle_R(2E_A)
\end{equation}
where
\begin{equation}
	\check{L} \equiv \frac{1}{2E_A} \frac{ 2(1-T)(\Delta^2 - E_A^2) -TE_A (E_A+  \check{\varepsilon})}{E_A^2 - \check{\varepsilon}^2}\Bigl(\Delta^2 - E_A^2 + \sqrt{\Delta^2 - E_A^2}\sqrt{\Delta^2 - \check{\varepsilon}^2}\Bigr).
\end{equation}
Let us evaluate $L_R(\epsilon) - L_A(\epsilon)$ and insert in Eq. \eqref{eq:spectral_contracted} to obtain
\begin{align}\label{eq:d3_final}
	\delta_{(c)}^{\rm odd}  &=
	\int_0^\infty \frac{d\omega}{\omega}G_Q{\rm Re}(Z(\omega)) \Bigl\{(1-T)\frac{(\Delta^2 - E_A^2)^2}{(2E_A  +\omega)E_A^2} - \frac{(\Delta^2 - E_A^2)(E_A^2 - (1-T)\Delta^2)}{\omega E_A^2}  \\
&+ \frac{\sqrt{\Delta^2 - E_A^2}}{2E_A}\int_\Delta^\infty \frac{d\epsilon}{\pi \epsilon} \rho(\epsilon) \Bigl(\frac{2(1-T)(\Delta^2 - E_A^2)-TE_A^2 + TE_A\epsilon}{ \epsilon  + \omega+E_A }  + \frac{ 2(1-T)(\Delta^2-E_A^2) -  T E_A^2 - TE_A \epsilon }{ \epsilon  + \omega- E_A }\Bigr)    \Bigr\},
\end{align}
where $\Lambda_R(\infty) + \Lambda_A(\infty)=0$ and we let the temperature go to zero. We can symmetrize over negative frequencies and use Kramers-Kronig to write the second term on a more transparent form
\begin{align}
	- \frac{1}{2}\int_{-\infty}^\infty d\omega \frac{G_Q{\rm Re}(Z(\omega)/\omega)}{\omega-0} \frac{(\Delta^2 - E_A^2)(E_A^2 - (1-t)\Delta^2)}{E_A^2} &=
	\frac{1}{2}\lim_{\omega \rightarrow 0}{\rm Im}(\frac{Z(\omega)}{\omega}) \frac{\pi G_Q\Delta^4 T^2 s^2 c^2 }{ E_A^2} = -\frac{1}{2} L I_J^2.
\end{align}
By adding the three contributions to $\delta^{odd}$ from Eqs. \eqref{eq:d1_final}, \eqref{eq:d2_final}, and \eqref{eq:d2_final} we obtain
\begin{equation}
	\delta^{\rm odd} = \frac{1}{2}L I_J^2 + \Lambda,
\end{equation}
where
\begin{align} \label{eq:lambda}
	\Lambda &= \delta T \partial_T E_A +  \int_0^\infty \frac{d\omega}{\pi \omega}{\rm Re} Z(\omega) \int_0^\infty \frac{d\omega'}{\pi }\frac{\omega'}{\omega' +\omega}  {\rm Re}(Y_3(\omega') + \frac{1}{2}Y_2(\omega')  - Y_5(\omega') ),
\end{align}
and
\begin{align}
	{\rm Re} Y_5(\omega)  & \equiv \frac{1}{2} \pi \Delta G_Q T^{3/2} |\sin(\phi/2)|\theta(\omega + E_A - \Delta) \frac{\rho(\omega + E_A)}{\omega } ( 1 +\Delta^2\frac{1-(2-T)\sin^2(\phi/2)}{ E_A (\omega + E_A) }),
\end{align}
while ${\rm Re}Y_2$ and ${\rm Re}Y_3$ are given in Eqs. \eqref{eq:y2} and \eqref{eq:y3}.

\subsection{Singlet addition shift} \label{sec:singlet_shift}
\begin{figure}[h]
    \centering
    \includegraphics[width=0.7\textwidth]{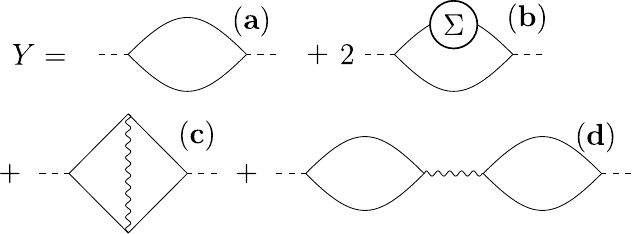}
    \caption{The Feynman diagrams that contribute to the admittance to first order in $Z$.}
    \label{fig:Y}.
\end{figure}

The addition energy of the singlet state with two excited bound states appears as a simple pole in the junction admittance $Y(\omega) \sim Y_3 = i\bar{Y} 2E_A/(\omega+i0 - 2E_A)$, where
\begin{equation}\label{eq:barY}
	\bar{Y} = \pi G_Q(1-T)\frac{(\Delta^2 - E_A^2)^2}{4E_A^4}.
\end{equation}
If the junction is coupled in parallel with an external impedance, the pole in the current correlator is shifted to $\omega = 2E_A + \delta^{even}$. To generate the current response functions in the coupled system we define the functional
\begin{equation}\label{eq:z}
	\mathcal{Z}[\varphi] \equiv \int \mathcal{D}[\chi] e^{i S_{J}[\varphi + \chi] +iS_{env}[\chi] },
\end{equation}
where $\varphi_{\alpha}(t)$ is a source field defined on the Keldysh contour. We note that the normalization $\mathcal{Z}[\varphi_{cl}, \varphi_{q}=0]=1$ is independent of $\varphi_{cl}$ since the right-hand side Eq. \eqref{eq:z} reduces to unitary time evolution which cancels between the two Keldysh branches. In the interacting system the admittance generalizes to
\begin{equation} \label{eq:Y_fluct}
	Y(\omega)=-\frac{ie^2}{2\omega} \Pi_{q, cl}(\omega) ,
\end{equation}
where
\begin{equation}\label{eq:Pi-expansion}
     \Pi_{\alpha \beta}(t-t')\equiv
	 -i\frac{\delta^2  \mathcal{Z} }{\delta \varphi_{\alpha}(t) \delta \varphi_{\beta}(t') }|_{\varphi=0} = i\langle \frac{\delta S_J}{\delta \chi_{\alpha}(t) }\frac{\delta S_J}{\delta \chi_{\beta}(t') } \rangle+  \langle   \frac{\delta^2 S_J}{\delta \chi_{\alpha}(t) \delta \chi_{\beta}(t') } \rangle,
\end{equation}
and $\Pi_{q, cl}(\omega)$ has a pole at the shifted frequency $\omega = 2E_A + \delta^{even}$, where $\delta^{even}\propto Z$. When expanded directly in $Z$ the pole shift has the following analytical structure
\begin{align}
	&\Pi_{q, cl}(\omega) \sim - \bar{\Pi} \frac{2E_A}{\omega - 2E_A - \delta^{\rm even}} =
		 - \bar{\Pi} \frac{2E_A}{\omega - 2E_A } -\bar{\Pi} \frac{2E_A \delta^{\rm even}}{(\omega - 2E_A)^2} + \mathcal{O}(Z^2),\\
	&\bar{\Pi}  = \frac{4E_A }{e^2}\bar{Y}.\label{eq:pi_bar}
\end{align}
Here $\bar{\Pi}$ follows from Eq. \eqref{eq:Y_fluct}, but we also obtain it independently below (see Eq. \eqref{eq:bar_pi_conf}). To first order in $Z$, $\delta^{even}$ is obtained from the prefactor of the contributions to $\Pi_{q, cl}(\omega)$ that give a second order pole at $\omega = 2E_A$. Let us define
\begin{equation}
	\Pi^{(a)}_{\alpha \beta}(t-t') \equiv  \frac{\delta^2 S_J}{\delta \chi_{\alpha}(t) \delta \chi_{\beta}(t') }|_{\chi=0},
\end{equation}
so $\Pi^{(a)}_{q, cl}(\omega) = 2\omega i Y_{\chi =0}(\omega) / e^2$ gives the bare admittance in absence of fluctuations. To first order in $Z$ we can represent the admittance expansion with the diagrams in Fig. \ref{fig:Y}, where the wiggly lines represent the photon propagator $Z \sim  D$ and the straight lines represents the fermion propagator $G$. Here '$\Sigma$' is a placeholder for the two diagrams in Fig. \ref{fig:sigma}, and the factor of '$2$' in front of the diagram in (b) is to include the symmetric diagram where '$\Sigma$' appears on the lower fermion line. The diagram in (a) gives the bare admittance $ \sim \Pi^{(a)}$. Let us define
\begin{equation}
	\delta^{\rm even} \equiv \delta^{\rm even}_{(b)} + \delta^{\rm even}_{(c)}+\delta^{even}_{(d)},
\end{equation}
where $\delta^{\rm even}_{(i)}$ gives the contribution to the pole shift from diagram (i) in Fig. \ref{fig:Y}.

\subsubsection{Diagram (d) contribution}
Let us expand the first term in Eq \eqref{eq:Pi-expansion} to second order in $\chi$ to obtain
\begin{align}
	i\langle \frac{\delta S_J}{\delta \chi_{\alpha}(t) }\frac{\delta S_J}{\delta \chi_{\beta}(t') } \rangle \sim -\sum_{\gamma \delta}\int dt_1 \int dt_2 \Pi_{\alpha \gamma}^0(t-t_1) D_{\gamma \delta}(t_1 - t_2) \Pi_{\delta \beta}^0(t_2-t') \equiv \Pi^{(d)}_{\alpha \beta }(t-t').
\end{align}
Here we used that the disconnected diagrams do not contribute due to the normalization of $\mathcal{Z}$ ($\delta \mathcal{Z} / \delta \varphi_{cl}|_{\varphi=0} =0$). The label (d) signifies that the contribution can be represented by diagram (d) in Fig. \ref{fig:Y}. In the frequency domain we have
\begin{equation}
	\Pi^{(d)}_{q, cl}(\omega) =-(\Pi^{(d)}_{q, cl}(\omega))^2 D^R(\omega) \sim- \bar{\Pi}\frac{ 2E_A (2E_A D^R(2E_A) \bar{\Pi}) }{(\omega - 2E_A)^2}
\end{equation}
And we can read of the contribution to the real pole shift
\begin{equation} \label{eq:delta_d}
	\delta^{\rm even}_{(d)} =  {\rm Re} (D^R(2E_A)) 2E_A \bar{\Pi}   =2E_A \bar{Y} {\rm Im} ( Z(2E_A) ).
\end{equation}

\begin{figure}[h]
    \centering
    \includegraphics[width=0.8\textwidth]{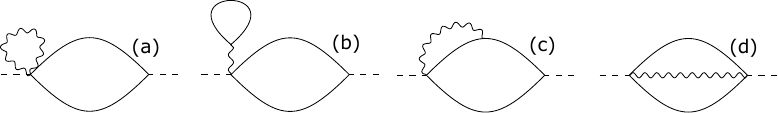}
    \caption{Feynman diagrams that comes from fluctuation in the external vertices. }
    \label{fig:irrelevant}
\end{figure}

Let us next consider the second term in Eq. \eqref{eq:Pi-expansion}
\begin{align}
	\langle   \frac{\delta^2 S_J}{\delta \chi_{\alpha}(t) \delta \chi_{\beta}(t') } \rangle &= \frac{i}{2} \langle \tr (v' X_{\alpha t} \mathcal{G} v' X_{\beta t'} \mathcal{G} )\rangle + \frac{i}{2}\langle \tr (v'' X_{\alpha t} X_{\beta t'}  \mathcal{G}) \rangle \label{eq:general_d2S}\\
 & \sim \frac{i}{2}\langle \tr (v'_0 X_{\alpha t} \mathcal{G} v'_0 X_{\beta t'} \mathcal{G})\rangle \label{eq:simple_d2S}\\
	&= \frac{i}{2}\langle \tr (v'_0 X_{\alpha t} (G + GvG + GvGvG) v'_0 X_{\beta t'} (G + GvG + GvGvG) )\rangle + \mathcal{O}(Z^2)\\
	&= \Pi^{(b)}_{\alpha \beta}(t-t') + \Pi^{(c)}_{\alpha \beta}(t-t') + \mathcal{O}(Z^2) ,
\end{align}
where we defined
\begin{align}
	\Pi^{(b)}_{\alpha \beta}(t-t') &\equiv \frac{i}{2}\tr (v'_0 X_{\alpha t} (G + G\Sigma G) v'_0 X_{\beta t'}  (G + G\Sigma G) ),\label{eq:pib}\\
	\Pi^{(c)}_{\alpha \beta}(t-t')&\equiv  \frac{i}{2}\langle \tr (v'_0 X_{\alpha t}  (GvG) v'_0 X_{\beta t'} (GvG) )\rangle, \label{eq:pic}
\end{align}
which gives the contribution to the admittance from diagram (b) and (c) in Fig. \ref{fig:Y}. In going from Eq. \eqref{eq:general_d2S} to Eq. \eqref{eq:simple_d2S} we neglected the diamagnetic contribution $\propto \delta(t-t')$ which clearly does not contribute to the admittance pole. We also neglected fluctuations in the external vertices and fixed $v' \rightarrow v'_0$. This step requires some explanation. In Fig. \ref{fig:irrelevant} we illustrate the diagrams that result from fluctuations in the external vertices. All of these can be seen as limits of diagram (b) in Fig. \ref{fig:Y}, where one or both fermion propagators next to '$\Sigma$' are gone. This is also the simple reason why they do not contribute to the second order pole, since there are simply not enough fermion propagators to produce a higher order pole at $\omega = 2E_A$. It is not difficult to calculate the diagram contributions exlicity, but here we instead sketch the mechanism that produces the second order pole at $\omega = 2E_A$ to show it does not apply to the diagrams in Fig. \ref{fig:irrelevant}.

The fermions lines have poles which split into a principle value part and a delta contribution
\begin{equation}
	G^R(\epsilon) \sim {\rm P.V}. \frac{1}{\epsilon \pm E_A}  - i\pi\delta(\epsilon \pm E_A ).
\end{equation}
In the bare admittance (represented by (a) in Fig. \ref{fig:Y}) the pole at $\omega=2E_A$ comes about only after the integral over the internal frequency is fixed by the $\delta(\epsilon \pm E_A)$ contribution of one fermion line, which leaves us with the pole at $\omega = 2E_A$ in the other. To produce a higher order pole we require three fermion lines in the 'inner loop', where one fixes the integral and the two others give a second order pole. This immediately implies that the diagrams in (a), (b), and (d) of Fig. \ref{fig:irrelevant} do not produce a higher order pole. As for diagram (c) in Fig. \ref{fig:irrelevant}, the third fermion propagator is a part of a loop that requires an additional frequency integral to contract the fermion propagator with the external impedance. This essentially 'washes out' the resonant structure of the fermion line, and is why also this diagram fails to produce a higher order pole at $\omega = 2E_A$.

\subsubsection{Diagram (b) contribution}\label{sec:odd_contribution}
To obtain the contribution from $\Pi^{(b)}$ to the pole shift our strategy is to replace $\Sigma$ with $\tr_{\tau, \eta}\{\mathcal{S} \bar{\Sigma} \}\equiv \Sigma_Q$, which we used to obtain $\delta^{odd}$ and has no matrix structure in $\tau, \eta, \xi$ (see Sec. \ref{sec:doublet_shift}). The Keldysh matrix $\Sigma_Q$ appears as a self energy correction $\check{Q} \rightarrow \check{
\mathcal{Q}}$ in $Q$. Then we calculate the pole like contribution to the admittance $\sim \Pi^{(b)}$ using $\mathcal{Q}$.

Let us first note that $\Sigma$ appears between two projection operators ($G \propto \mathcal{P}$) and we can therefore replace $\Sigma \rightarrow  \tr_{\tau}\{\Sigma\} \equiv  \bar{\Sigma}$. To proceed we write

\begin{align}
	\Pi^{{(b)}}_{\alpha \beta}(t-t') &\equiv \frac{i}{2}\tr (v'_0 X_{\alpha t} G(1 + \{\bar{\Sigma}, G\}) v'_0 X_{\beta t'}  G(1 + \{\bar{\Sigma}, G\})- \frac{i}{2}\tr (v'_0 X_{\alpha t} \mathcal{P}\check{Q } (\mathcal{S} + \bar{\Sigma})v'_0 X_{\beta t'}  \mathcal{P}\check{Q}  (\mathcal{S} + \bar{\Sigma})),
\end{align}
where we used $G^2 = Q$. The second term gives no contribution to the second order pole since it only contains two factors of $\check{Q}(\epsilon) \sim (\check{\epsilon} \pm E_A)^{-1}$. Here $\bar{\Sigma}$ is a traceless matrix in $\eta_i$ so we can write
\begin{align}
	G(1+ \{\bar{\Sigma}, G\}) &= \mathcal{P} \mathcal{S}( \check{Q}+ \check{Q}^2\tr_{\eta}(\mathcal{S}\bar{\Sigma} )) = \mathcal{P} \mathcal{S}\check{\mathcal{Q}} + \mathcal{O}(Z^2)
\end{align}
where we defined
\begin{equation}
	\check{\mathcal{Q}} = \frac{1}{Q^{-1} - \tr_\eta(\bar{\Sigma} \mathcal{S}) } = \frac{\Delta^2 - \check{\varepsilon}^2}{E_A^2+ 2E_A \Sigma_Q - \check{\varepsilon}^2} = \frac{\Delta^2 - \check{\varepsilon}^2}{(E_A+\Sigma_Q)^2  - \check{\varepsilon}^2} + \mathcal{O}(Z^2),
\end{equation}
and
\begin{equation}
	\Sigma_Q \equiv - \frac{\Delta^2 - \check{\varepsilon}^2}{2E_A}\tr_{\eta}(\bar{\Sigma} \mathcal{S}).
\end{equation}
As we showed in Sec. \ref{sec:doublet_shift}, $\Sigma_Q$ only has nontrivial matrix structure in Keldysh and frequency, and $\delta^{odd} = \Sigma_Q^R(E_A)$ (see Eq. \eqref{eq:d_odd_original_almost}). Let us define
\begin{align}
	\tilde{G} & \equiv \mathcal{P} \mathcal{S}\check{\mathcal{Q}},\\
	\tilde{\mathcal{A}}_{\epsilon} &= \frac{1}{2\pi i}(\tilde{G}^R(\epsilon) - \tilde{G}^A(\epsilon)).
\end{align}
By using the spectral representation (see Eq. \eqref{eq:spectral_general}) we can separate out the Keldysh structure in $\Pi^1$
\begin{align}
	\Pi^{(b)}_{\alpha \beta}(t-t') &= \frac{i}{2}\int_{-\infty}^{\infty} d\epsilon_1 \int_{-\infty}^{\infty} d\epsilon_2 \tr (v'_0 X_{\alpha t}\tilde{\mathcal{A}}_{\epsilon_1}\frac{1}{\epsilon_1-\check{\epsilon}} v'_0 X_{\beta t'}  \tilde{\mathcal{A}}_{\epsilon_2}\frac{1}{\epsilon_2-\check{\epsilon}}),\\
	&= \frac{i}{2}\int_{-\infty}^{\infty} d\epsilon_1 \int_{-\infty}^{\infty} d\epsilon_2 \tr_{\zeta \eta \tau} (v'_0 \tilde{\mathcal{A}}_{\epsilon_1} v'_0  \tilde{\mathcal{A}}_{\epsilon_2}) \tr_{\sigma}(\sigma_{\alpha} \bra{t}\frac{1}{\epsilon_1-\check{\epsilon}}  \ket{t'}\sigma_{\beta}\bra{t'} \frac{1}{\epsilon_2-\check{\epsilon}} \ket{t}),
\end{align}
where we neglected the frequency independent contact term that does not contribute to the frequency pole in $\Pi^{(b)}$. Performing the Keldysh trace results in
\begin{align}
	\Pi^{(b)}_{q, cl}(\omega) &= \frac{1}{2}\int_{-\infty}^{\infty} d\epsilon_1 \int_{-\infty}^{\infty} d\epsilon_2 (F_{\epsilon_1} - F_{\epsilon_2})\frac{-\tr_{\zeta \eta \tau} (v'_0 \tilde{\mathcal{A}}_{\epsilon_1} v'_0  \tilde{\mathcal{A}}_{\epsilon_2})}{\omega +i0 +\epsilon_2- \epsilon_1 }.
\end{align}
The pole in $\Pi_{q, cl}(\omega)$ comes from the part of the spectral function that has a delta peak close to $\varepsilon = \pm E_A$. So
\begin{align}
	\tilde{\mathcal{A}}_\epsilon &= \frac{1}{2\pi i}	(\mathcal{Q}^R \mathcal{P}^R\mathcal{S}^R - \mathcal{Q}^A \mathcal{P}^A\mathcal{S}^A ) \sim \frac{1}{2}	\frac{\mathcal{Q}^R - \mathcal{Q}^A}{2\pi i}(\mathcal{P}^R\mathcal{S}^R +  \mathcal{P}^A\mathcal{S}^A )\\
	& = \mathcal{A}^{+} \delta(\varepsilon - E_A - \delta^{\rm odd}) + \mathcal{A}^{-} \delta(\varepsilon + E_A + \delta^{\rm odd}),
\end{align}
where we used Eq. \eqref{eq:d_odd_original_almost} and consider zero temperature so $\delta^{odd}$ is real. By using this we have
\begin{align}\label{eq:pi1_pole}
	\Pi^{(b)}_{q, cl}(\omega) &\sim \frac{-\tr_{\zeta \eta \tau} (v'_0 \mathcal{A}^+ v'_0  \mathcal{A}^-)}{\omega +i0 -2E_A - 2\delta^{\rm odd}},
\end{align}
and we can read off the contribution to the pole shift
\begin{equation}\label{eq:delta_b}
	\delta^{\rm even}_{(b)} =2\delta^{\rm odd}.
\end{equation}
From Eq. \eqref{eq:pi1_pole} we also calculate
\begin{equation}\label{eq:bar_pi_conf}
	\bar{\Pi} = \frac{1}{2E_A} \tr(v_0' \mathcal{A}^+ v_0' \mathcal{A}^-) = (1-T)\frac{(\Delta^2 - E_A^2)^2}{E_A^3}
\end{equation}
in agreement with Eq. \eqref{eq:pi_bar}.

\subsubsection{Diagram (c) contriution}
To obtain the contribution to $\delta^{even}$ from $\Pi^{(c)}$ we proceed by direct calculation of $\Pi^{(c)}$. To second order in $\check{\chi}$ we have
\begin{align}
    \Pi^{(c)}_{\alpha \beta}(t-t') &= \frac{i}{2}\langle \tr (v'_0 X_{\alpha t}  G v'_0 \check{\chi} G v'_0  X_{\beta t'}  G v'_0 \check{\chi} G )\rangle\\
    &=-\frac{1}{2}\sum_{\gamma \delta}\int dt_1 \int dt_2 D_{\gamma \delta}(t_1 - t_2)  \tr ( X_{\alpha t}  v'_0 G X_{\gamma t_1} v'_0  G X_{\beta t'} v'_0  G X_{\delta t_2} v'_0 G )
\end{align}
By using the spectral representation we separate out the Keldysh structure
\begin{align}
    \Pi^{(c)}_{\alpha \beta}(t-t') &=  \int  d\epsilon_1 \hdots  d\epsilon_4  \tr_{\zeta, \eta, \tau}(v_0'\mathcal{A}_{\varepsilon_1} v_0' \mathcal{A}_{\varepsilon_2 } v_0' \mathcal{A}_{\varepsilon_3}v_0' \mathcal{A}_{\varepsilon_4}) W^{\{\epsilon_i\}}_{\alpha, \beta}(t-t'),
\end{align}
where
\begin{align}
	W^{\{\epsilon_i\}}_{\alpha, \beta}(t-t') &\equiv -\frac{1}{2}\sum_{\gamma \delta}\int dt_1 \int dt_2 D_{\gamma \delta}(t_1 - t_2)  \\
	&\times \tr_{\sigma} ( \sigma_\alpha \bra{t} \frac{1}{\epsilon_1 - \check{\epsilon}}  \ket{t_1} \sigma_{\gamma}  \bra{t_1}\frac{1}{\epsilon_2 - \check{\epsilon}} \ket{t'}\sigma_{\beta}\bra{t'} \frac{1}{\epsilon_3 - \check{\epsilon}}  \ket{t_2}\sigma_{\delta} \bra{t_2} \frac{1}{\epsilon_4 - \check{\epsilon}} \ket{t}).
\end{align}
Performing the Keldysh trace results in
\begin{align}
	&{\rm Re}W^{\{\epsilon_i\}}_{q, cl}(\omega)= \mathrm{P.V.} \frac{1}{(\varepsilon_4 -\varepsilon_1- \omega)(\varepsilon_3  - \varepsilon_2 - \omega )}\\
	&\times \int \frac{d\nu}{2\pi} {\rm Im}(D^{R}_{\nu}) \Bigl(\frac{2 N_{\nu}(F_{\varepsilon_3} -F_{\varepsilon_4})+2(1-F_{\varepsilon_4}F_{\varepsilon_3})}{\varepsilon_4 - \varepsilon_3 - \nu} - \frac{2 N_{\nu} F_{\varepsilon_3} +1-F_{\varepsilon_1} F_{\varepsilon_3} }{\varepsilon_1 - \varepsilon_3  + \omega - \nu}-\frac{2 N_{\nu} F_{\varepsilon_4} + 1-F_{\varepsilon_2} F_{\varepsilon_4} }{\varepsilon_2 - \varepsilon_4 + \omega - \nu}  \Bigr).
\end{align}
To simplify we also symmetrized over $\epsilon_i \leftrightarrow - \epsilon_i$. It is not true that $\mathcal{A}_{\epsilon}=-\mathcal{A}_{-\epsilon}$, however we can define $\kappa \equiv \zeta_1 \tau_3$, such that  $\kappa \mathcal{A}_{\epsilon}\kappa =-\mathcal{A}_{-\epsilon}$, and using that $\kappa^2 =1$ and $\kappa$ commutes with $v_0'=\sqrt{T}(c\tau_3\eta_2 + s\eta_1)$ we have
\begin{align}
	\tr_{\zeta, \eta, \tau}(v_0'\mathcal{A}_{\varepsilon_1} v_0' \mathcal{A}_{\varepsilon_2 } v_0' \mathcal{A}_{\varepsilon_3}v_0' \mathcal{A}_{\varepsilon_4}) &= \tr_{\zeta, \eta, \tau}(v_0'(\kappa \mathcal{A}_{\varepsilon_1}\kappa) v_0' (\kappa \mathcal{A}_{\varepsilon_2}\kappa) v_0' (\kappa \mathcal{A}_{\varepsilon_3}\kappa)v_0'(\kappa \mathcal{A}_{\varepsilon_4}\kappa))\\
	& =(-1)^2\tr_{\zeta, \eta, \tau}(v_0'\mathcal{A}_{\varepsilon_1} v_0' \mathcal{A}_{\varepsilon_2 } v_0' \mathcal{A}_{\varepsilon_3}v_0' \mathcal{A}_{\varepsilon_4})
\end{align}
The second order pole in $\Pi^{(c)}$ comes from the part of the spectral function that fixes $\epsilon_4 = \epsilon_3=E_A$, and $\epsilon_2 = \epsilon_1 = -E_A$. This results in
\begin{align}
    {\rm Re}\Pi^2_{q, cl}(\omega) &\sim  -\frac{e^2 L}{2 } \frac{\tr_{\zeta \eta \tau}(v_0'\mathcal{A}^+ v_0' \mathcal{A}^+ v_0' \mathcal{A}^-v_0' \mathcal{A}^-) }{(\omega-2E_A)^2}.
\end{align}
After some algebra we have
\begin{align}
    \mathcal{A}^{\pm} v_0' \mathcal{A}^{\pm} =   \mp 2 \partial_\phi E_A  \mathcal{A}^+,
\end{align}
so
\begin{align}
    {\rm Re}(\Pi^2_{q, cl}(\omega)) &\sim   -L I_J^2 \frac{\tr_{\zeta \eta \tau}(v_0'\mathcal{A}^+ v_0' \mathcal{A}^-) }{(\omega-2E_A)^2} =L I_J^2 \bar{\Pi}\frac{2E_A}{(\omega-2E_A)^2},
\end{align}
and we can read off the contribution
\begin{equation}\label{eq:delta_c}
	\delta^{\rm even}_{(c)} = -L I_J^2
\end{equation}
\subsubsection{Full pole shift}
By combining the contributions from Eqs. \eqref{eq:delta_d}, \eqref{eq:delta_b}, and \eqref{eq:delta_c} we obtain
\begin{equation}
	\delta^{\rm even} = 2E_A \bar{Y}{\rm Im}(Z(2E_A))  + 2\delta^{\rm odd} - L I_J^2 = 2E_A \bar{Y}{\rm Im}Z(2E_A)+ 2\Lambda.
\end{equation}

\section{Energy shifts from the quasiparticle Hamiltonian} \label{sec:Hamiltonian}
In this section we present an alternative derivation of the ground state shift $\Delta E_g$, and the addition shifts $\delta^{odd}, \delta^{even}$ by using the quasiparticle Hamiltonian and doing second order perturbation theory to determine the level shifts of the states with $0, 1,$ or $2$ excited quasiparticles.

The full system Hamiltonian is
\begin{equation}
	\hat{H} = \hat{H}_{\rm env} + \hat{H}_J.
\end{equation}
The environment Hamiltonian $H_{env}$ can be expressed using the photon ladder operators $b_k, b_k^\dagger$
\begin{equation}
	\hat{H}_{\rm env} = \sum_k \omega_k b_k^\dagger b_k,
\end{equation}
where $[b_k, b_{k'}^\dagger] = \delta_{k, k'}$ and the photon vacuum state $\ket{\Omega}$ is defined by $b_k \ket{\Omega} = 0$ for all $k$. The junction Hamiltonian is coupled through the environment through a parametric dependence on the flux
\begin{equation}
	\hat{H}_J(\Phi_0 + \hat{\Phi}) = \hat{H}^{(0)}_J + \hat{H}^{(1)}_J \hat{\Phi} + \frac{1}{2} \hat{H}^{(2)}_J \Hat{\Phi}^2 + \mathcal{O}(\hat{\Phi}^3),
\end{equation}
where $\hat{H}^{(n)}_J  \equiv \partial^n_{\Phi}\hat{H}(\Phi_0) $ and the flux operator $\hat{\Phi}$ is hermitian and linear in the photon ladder operators
\begin{equation}
	\hat{\Phi} \equiv \sum_k (\Phi_k b_k +\Phi_k^* b^\dagger_k).
\end{equation}
Let us define the non-interacting Hamiltonian
\begin{equation}
	\hat{H}_0 = \hat{H}_{\rm env} + \hat{H}_J^{(0)},
\end{equation}
and consider the eigenstate $\ket{\psi_i} = \ket{\Omega} \ket{E_i}$ of $\hat{H}_0$ which is the combination of the photon vacuum (defined by $b_k \ket{\Omega}=0$) and the eigenstate $\ket{E_i}$ of the non-interacting junction Hamiltonian
\begin{equation}
	\hat{H}_J^{(0)}\ket{E_i} = E_i \ket{E_i}.
\end{equation}
We would like to obtain the perturbative shift $\Delta E_i$ in the energy $E_i$ of $\ket{\psi_i}$ due to environmental coupling in the case where $\ket{E_i}$ is the state with $0, 1$, or $2$ excited bound junction particles. The general formula for the shift can be obtained from second order perturbation theory and reads
\begin{align}\label{eq:DE0}
    \Delta E_i  = \sum_k |\Phi_k|^2 \Bigl( \frac{1}{2}\bra{E_i} \hat{H}_J^{(2)} \ket{E_i}- \sum_{n} \frac{|\bra{E_n}\hat{H}^{(1)}_J\ket{E_i}|^2}{E_n + \omega_k -E_i} \Bigr),
\end{align}
where the $n$ sum runs over all eigenstates $\ket{E_n}$ of $H_J^{(0)}$.
To proceed we relate the quantities in Eq. \eqref{eq:DE0} to the circuit impedance and junction admittance.

\subsection{Level shifts in terms of the circuit impedance}
Let us define the ground state flux correlator
\begin{align}
	\chi^{\Phi \Phi}(t-t') = -i \theta(t-t') \bra{\Omega} [\hat{\Phi}(t),\hat{\Phi}(t')] \ket{\Omega}
\end{align}
Where time evolution is with respect to $\hat{H}_{env}$. In the frequency domain we have
\begin{equation}
	\chi^{\Phi \Phi}(\omega) =  \sum_{k} |\Phi_k|^2\Bigl( \frac{1}{\omega +i0 - \omega_k} -\frac{1 }{\omega +i0 +\omega_k}\Bigr),
\end{equation}
and is related to the impedance as
\begin{equation}
	Z(\omega) = i \omega \chi^{\Phi \Phi}(\omega).
\end{equation}
For $\omega > 0$ we have
\begin{equation}
	{\rm Re} Z(\omega)  =  \pi\omega \sum_k |\Phi_k|^2 \delta(\omega - \omega_k),
\end{equation}
which we insert in Eq. \eqref{eq:DE0} to obtain

\begin{align}\label{eq:DE1}
    \Delta E_i  = \int_0^\infty \frac{d\omega}{\pi \omega} {\rm Re} Z(\omega)  \Bigl( \frac{1}{2}\bra{E_i} H_J^{(2)} \ket{E_i}- \sum_{n} \frac{|\bra{E_n}\hat{H}^{(1)}\ket{E_i}|^2}{E_n  -E_i + \omega} \Bigr).
\end{align}

\subsection{Level shifts expressed with the Junction admittance}
If the junction is prepared in some state $\ket{E_i}$ and coupled to a small classical flux source $\delta \Phi(t)$ which we have
\begin{equation}
	\langle \hat{I}(t)\rangle = \langle \hat{I} \rangle_i + \int_{-\infty}^\infty dt \chi^{II}_i(t-t') \delta \Phi(t')
\end{equation}
where
\begin{align}
	\chi^{II}_i(t-t') &= \delta(t-t') \langle \partial_{\Phi}\hat{I}\rangle_i  -i \theta(t-t')\langle \comm{\hat{I}(t)}{\hat{I}(t')}\rangle_i.
\end{align}
Here $\hat{I} = H_{J}^{(1)}$, $\partial_{\Phi}\hat{I} = H_{J}^{(2)}$, the time evolution is with respect to $H_{J}^{(0)}$, and the averages are taken with respect to the state $\ket{E_i}$. In the frequency domain we have
\begin{align}\label{eq:chiII}
	\chi^{II}_i(\omega) &= \langle H_{J}^{(2)} \rangle_i  +\sum_{n}|\bra{E_i}\hat{H}^{(1)} \ket{E_n}|^2    (\frac{1}{\omega+i0+E_n - E_i}-\frac{1}{\omega +i0 - E_n + E_i}),
\end{align}
which is related to the junction admittance
\begin{equation}
	Y^{(i)}(\omega) = \frac{1}{i\omega} \chi^{II}_i(\omega)
\end{equation}
For $\omega > 0$ we have

\begin{align}
	{\rm Re} Y^{(e)}(\omega) &= \frac{\pi}{\omega} \sum_{n}|\bra{E_e}\hat{H}^{(1)} \ket{E_n}|^2 \delta(\omega -E_n + E_e), \\
	{\rm Re} Y^{(o)}(\omega) &= \frac{\pi}{\omega} \sum_{n}|\bra{E_e}\hat{H}^{(1)} \ket{E_n}|^2 \delta(\omega -E_n + E_o), \\
	{\rm Re} Y^{(x)}(\omega) &= \frac{\pi}{\omega} \sum_{n \neq e}|\bra{E_x}\hat{H}^{(1)} \ket{E_n}|^2 \delta(\omega -E_n + E_x) - \frac{\pi}{\omega}|\bra{E_x}\hat{H}^{(1)} \ket{E_n}|^2 \delta(\omega -E_x + E_e).
\end{align}
Where we used that $\hat{H}^{(1)}$ does not couple states of different parity. By combining this with Eq. \eqref{eq:DE1} we obtain the shift in the even ground state $\Delta E_e$, odd ground state $\Delta E_o$, and excited singlet state $\Delta E_x$,
\begin{align}
    \Delta E_e  &= -\frac{\langle \hat{\Phi}^2\rangle}{2 L_\infty^{(e)}} -\int_0^\infty \frac{d\omega}{\pi \omega} {\rm Re} Z(\omega)  \int_0^\infty \frac{d\omega'}{\pi}\Bigl( {\rm Re} Y^{(e)}(\omega')  \frac{\omega'}{ \omega +\omega'} + \frac{|\langle \hat{H}_J^{(1)}\rangle_e|^2}{ \omega}\Bigr), \label{eq:dee} \\
    \Delta E_o  &= -\frac{\langle \hat{\Phi}^2\rangle}{2 L_\infty^{(o)}} -\int_0^\infty \frac{d\omega}{\pi \omega} {\rm Re} Z(\omega)  \Bigl(\int_0^\infty \frac{d\omega'}{\pi} {\rm Re} Y^{(o)}(\omega')  \frac{\omega'}{ \omega +\omega'} + \frac{|\langle \hat{H}_J^{(1)}\rangle_o|^2}{ \omega} \Bigr), \label{eq:deo}\\
    \Delta E_x  &= -\frac{\langle \hat{\Phi}^2\rangle}{2 L_\infty^{(x)}} -\int_0^\infty \frac{d\omega}{\pi \omega} {\rm Re} Z(\omega) \Bigl( \int_0^\infty \frac{d\omega'}{\pi} {\rm Re} Y^{(x)}(\omega')  \frac{\omega'}{ \omega +\omega'}  +\frac{|\bra{E_e}\hat{H}_J^{(1)}\ket{E_x}|^2}{ \omega - 2E_A} - \frac{|\bra{E_e}\hat{H}_J^{(1)}\ket{E_x}|^2}{ -\omega - 2E_A} + \frac{|\langle \hat{H}_J^{(1)}\rangle_x|^2}{ \omega} \Bigr)\label{eq:dex}
\end{align}
where,  $E_x - E_e = 2E_A$ and we defined
\begin{align}
	\frac{1}{L_\infty^{(i)}} \equiv \lim_{\omega \rightarrow \infty}  \omega {\rm Im} Y^{(i)}(\omega), \quad \quad \quad
	\langle \hat{\Phi}^2\rangle  \equiv \int_0^\infty \frac{d\omega}{\pi \omega} {\rm Re} Z(\omega).
\end{align}
Previously we have used $\Delta E_g$ to denote the ground state shift, but here we prefer $\Delta E_e$ to distinguish it from the odd ground state. By using \cite{kosFrequencydependentAdmittanceShort2013b} ${\rm Re} Y_3(\omega) = 2E_A \bar{Y} \pi \delta(\omega - 2E_A)$, we can read of the matrix element
\begin{equation}
	|\bra{E_e}\hat{H}^{(1)}\ket{E_x}|^2  = (2E_A)^2 \bar{Y}
\end{equation}
where $\bar{Y}$ was defined in Eq. \eqref{eq:barY}. The diagonal matrix elements $ \langle \hat{H}_J^{(1)}\rangle_i$ can be obtained by differentiating $\langle \hat{H}_J \rangle_{e, x} =\mp E_A$ and $\langle \hat{H}_J \rangle_{o} =0$ with repect to the flux to obtain
\begin{align}
	&\langle \hat{H}_J^{(1)} \rangle_e = I_J, \\
	&\langle \hat{H}_J^{(1)} \rangle_o = 0,\\
    & \langle \hat{H}_J^{(1)} \rangle_x = -I_J.
\end{align}
The terms $\propto$ $\langle \hat{\Phi}^2\rangle$ gives a log-divergent contributions from photons with energy $\omega \gg \Delta$. It is known~\cite{kindermannInteractionEffectsCounting2003} that the effect of high frequency photons can be included in a renormalization of the transmission coefficient $T\rightarrow T + \delta T$, where $\delta T = -\pi G_Q \langle \hat{\Phi}^2 \rangle T (1-T)$. This requires that
\begin{align}
	&-\langle \hat{\Phi}^2\rangle /(2 L_\infty^{(e)}) = -\delta T \partial_{T}E_A, \\
	&-\langle \hat{\Phi}^2\rangle/(2 L_\infty^{(o)}) = 0,\\
	& -\langle \hat{\Phi}^2\rangle /(2 L_\infty^{(x)})  = \delta T \partial_{T}E_A.
\end{align}
Finally by using the Kramer-Kronig relations to rewrite the last terms in Eqs. \eqref{eq:dee}, \eqref{eq:deo}, and \eqref{eq:dex} we obtain
\begin{align}
    \Delta E_e  &= -\delta T \partial_{T}E_A - \frac{1}{2}LI_J^2 -\int_0^\infty \frac{d\omega}{\pi \omega} {\rm Re} Z(\omega)  \int_0^\infty \frac{d\omega'}{\pi} {\rm Re} Y^{(e)}(\omega')  \frac{\omega'}{ \omega +\omega'}, \label{eq:deo1} \\
    \Delta E_o  &= -\int_0^\infty \frac{d\omega}{\pi \omega} {\rm Re} Z(\omega)  \int_0^\infty \frac{d\omega'}{\pi} {\rm Re} Y^{(o)}(\omega')  \frac{\omega'}{ \omega +\omega'}  , \label{eq:dee1}\\
    \Delta E_x  &= \delta T \partial_{T}E_A - \frac{1}{2}LI_J^2+2E_A \bar{Y} {\rm Im} Z(2E_A)  -\int_0^\infty \frac{d\omega}{\pi \omega} {\rm Re} Z(\omega) \int_0^\infty \frac{d\omega'}{\pi} {\rm Re} Y^{(x)}(\omega')  \frac{\omega'}{ \omega +\omega'} .\label{eq:dex1}
\end{align}
Finally we obtain $\delta^{\rm odd}$ and $\delta^{\rm even}$ from the differences
\begin{align}
	\delta^{\rm odd} &\equiv \Delta E_o - \Delta E_e=  \frac{1}{2}LI_J^2 + \Lambda\\
	\delta^{\rm even} &\equiv \Delta E_x - \Delta E_e=  2E_A \bar{Y} {\rm Im} Z(2E_A) +2 \Lambda,
\end{align}
where
\begin{equation}
    \Lambda = \delta T \partial_{T}E_A -\int_0^\infty \frac{d\omega}{\pi \omega} {\rm Re} Z(\omega) \int_0^\infty \frac{d\omega'}{\pi} ({\rm Re} Y^{(o)}(\omega')  - {\rm Re} Y^{(e)}(\omega'))\frac{\omega'}{ \omega +\omega'}
\end{equation}
is the same as in Eq. \eqref{eq:lambda}, and we used~\cite{kosFrequencydependentAdmittanceShort2013b}
\begin{equation}
	{\rm Re} Y^{(x)}(\omega')  - {\rm Re} Y^{(e)}(\omega') = 2({\rm Re} Y^{(o)}(\omega')  - {\rm Re} Y^{(e)}(\omega')).
\end{equation}

\appendix

\section{Derivation of the alternative representation of the junction action}\label{sec:alternative_action}
In this appendix we show the equivalence between the representation of the junction action $S_J$ in Eqs. \eqref{eq:keldysh_lead_version} and \eqref{eq:keldysh_final}. Let us introduce $\eta_i$ as a Pauli basis for a 'lead structure' and collect the lead greens functions into a common matrix
\begin{equation}
    g \equiv
    \begin{pmatrix}
        \check{G}_{\phi+2\chi} & 0 \\
        0 & \check{G}_{-\phi-2\check{\chi}}
    \end{pmatrix}
    =
    e^{-i \tau_3\eta_3(\phi+2\check{\chi}(t))/4 } \check{G}_{0}e^{i \tau_3\eta_3(\phi+2\check{\chi}(t))/4 },
\end{equation}
We also define a 'scattering matrix'
\begin{equation}
	\mathcal{S}_0 = \sqrt{1-T}\eta_3 + \sqrt{T}\eta_1.
\end{equation}
By using $(\check{G}_{0})^2=1$ it is straight forward to show that
\begin{equation}
	\Bigl(\frac{1}{2}\{\mathcal{S}_0,g\}\Bigr)^2 =
	\begin{pmatrix}
		1-\frac{T}{4}(\check{G}_{\phi+2\check{\chi}} - \check{G}_{-\phi-2\check{\chi}})^2 & 0\\
		0 & 1-\frac{T}{4}(\check{G}_{\phi+2\check{\chi}} - \check{G}_{-\phi-2\check{\chi}})^2
	\end{pmatrix},
\end{equation}
where we wrote out the lead structure $\propto \eta_0$. Therefore,
\begin{equation}\label{eq:claim}
	\tr_\eta  \ln\Bigl(\frac{1}{2}\{\mathcal{S}_0,g\}\Bigr) =\frac{1}{2} \tr_\eta \ln\Bigl(\Bigl(\frac{1}{2}\{\mathcal{S}_0,g\}\Bigr)^2\Bigr) =\ln(1-\frac{T}{4}(\check{G}_{\phi+2\check{\chi}} - \check{G}_{-\phi-2\check{\chi}})^2)
\end{equation}
where $\tr_\eta$ only trace over the lead index. Now if we include the lead index in the trace we can rewrite Eq. \eqref{eq:keldysh_lead_version} as
\begin{equation}
	S_J = -\frac{i}{2}\tr\ln(\frac{1}{2}\{\mathcal{S}_0,g\}).
\end{equation}
Through a unitary transformation that does not affect the trace we can transfer the phase dependence from the greens functions to the scattering matrix
\begin{equation}
	S_J = -\frac{i}{2}\tr\ln(\frac{1}{2}\{\mathcal{S}_{\phi + 2\check{\check{\chi}}},\check{G}_0\}),
\end{equation}
where
\begin{equation}
	\mathcal{S}_{\phi + 2\check{\chi}} \equiv e^{ i\tau_3 \eta_3 (\phi+2\check{\chi})/4} \mathcal{S}_0 e^{ -i\tau_3 \eta_3 (\phi+2\check{\chi})/4}
\end{equation}
To proceed let us next define the 'interaction vertex' (see Eq \eqref{eq:v})
\begin{equation}
	v \equiv \mathcal{S}_{\phi} -\mathcal{S}_{\phi + \check{\chi}} = -\sqrt{T}(e^{i \tau_3 \eta_3 \check{\chi}} -1)e^{i \tau_3 \eta_3 \varphi/2}\eta_1
\end{equation}
and write
\begin{align}
	S_J &= -\frac{i}{2}\tr\ln(\frac{1}{2}\{\mathcal{S}_{\phi},\check{G}_0\} - \frac{1}{2}\{v,\check{G}_0\}) =  -\frac{i}{2}\tr\ln(1 - \frac{1}{2}\{M^{-1} v,\check{G}_0\}),
\end{align}
where
\begin{equation}
	M = \frac{1}{2}\{\mathcal{S}_{\phi},\check{G}_0\},
\end{equation}
and we used that $\{\mathcal{S}_{\phi},\check{G}_0\}$ and therefore also its inverse commutes with $\check{G}_0$. We also used that $\tr\ln(\frac{1}{2}\{\mathcal{S}_{\phi},\check{G}_0\})=0$ since the Keldysh branches cancel without a quantum field. Because $(\check{G}_0)^2=1$ we can define the projection operators onto the $\pm1$ eigenvalues of $\check{G}_0$,
\begin{equation}
	\mathcal{P}_{\pm} \equiv \frac{1}{2}(1 \pm \check{G}_0).
\end{equation}
By using $\mathcal{P}_{-}+\mathcal{P}_{+} =1 $, $(\mathcal{P}_{\pm})^2 = \mathcal{P}_{\pm}$ and $\mathcal{P}_{-}\mathcal{P}_{+}=0$ we obtain
\begin{align}
	S_J = -\frac{i}{2}\tr\ln(1 - \mathcal{P}_+ M^{-1} v)-\frac{i}{2}\tr\ln(1 + \mathcal{P}_- M^{-1} v ).
\end{align}
By introducing another matrix label for the $\pm$ projected eigenvalues we can write the junction action as in Eq. \eqref{eq:keldysh_final}
\begin{align}
	S_J &= -\frac{i}{2}\tr\ln(1 - G v),
\end{align}
where
\begin{equation}
	G \equiv
	\begin{pmatrix}
		\mathcal{P}_+ M^{-1} & 0 \\
		0 & -\mathcal{P}_- M^{-1}
	\end{pmatrix},
\end{equation}
and the trace now includes the additional matrix label which runs over the eigenspaces of $\check{G}_0$.

\subsection{Explicit expression for $G$}
In terms of the projection operators
\begin{equation}
	M =  \mathcal{P}_+ \mathcal{S}_\phi \mathcal{P}_+ - \mathcal{P}_- \mathcal{S}_\phi  \mathcal{P}_-,
\end{equation}
where
\begin{align}\label{eq:Sphi}
	\mathcal{S}_\phi& = \sqrt{1-T}\eta_3 + \cos(\varphi/2)\sqrt{T}\eta_1 - \sin(\varphi/2)\sqrt{T}\eta_2 \tau_3
\end{align}
By using
\begin{equation}
	\mathcal{P}_\pm \tau_3 \mathcal{P}_{\pm} = \pm \frac{-i \check{\varepsilon}}{\sqrt{\Delta^2 - \check{\varepsilon}^2}} \mathcal{P}_{\pm}
\end{equation}
we get
\begin{align}
	\mathcal{P}_\pm \mathcal{S}_\phi\mathcal{P}_\pm &=(\sqrt{1-T}\eta_3 + \cos(\varphi/2)\sqrt{T}\eta_1 \pm \sin(\varphi/2)\sqrt{T}\eta_2 \frac{i \check{\varepsilon}}{\sqrt{\Delta^2 - \check{\varepsilon}^2}}) \mathcal{P}_{\pm}
\end{align}
and
\begin{equation}
    M^2 =(\mathcal{P}_+ \mathcal{S}_\phi \mathcal{P}_+)^2 + (\mathcal{P}_- \mathcal{S}[0] \mathcal{P}_-)^2 =  \frac{E_A^2 - \check{\varepsilon}^2}{\Delta^2 - \check{\varepsilon}^2}.
\end{equation}
Therefore,
\begin{equation}
	M^{-1} = M \check{Q}
\end{equation}
where $\check{Q}$ was defined in Eq. \eqref{eq:Q_matrix}. Using this we have
\begin{equation}
	G \equiv
	\begin{pmatrix}
		\mathcal{P}_+ M^{-1} & 0 \\
		0 & -\mathcal{P}_- M^{-1}
	\end{pmatrix}
	=
	\begin{pmatrix}
		\mathcal{P}_+ \mathcal{S}_\phi \mathcal{P}_+ & 0 \\
		0 & \mathcal{P}_- \mathcal{S}_\phi \mathcal{P}_-
	\end{pmatrix}
	\check{Q}
\end{equation}
Let us introduce $\zeta_i$ as a Pauli basis that acts on the space of $\pm1$ projected eigenvalues, but note that we only require one matrix $\zeta_3 \equiv \zeta$. We have
\begin{equation}
    \begin{pmatrix}
    	\mathcal{P}_+ \mathcal{S}_\phi \mathcal{P}_+ & 0 \\
    	0 & \mathcal{P}_-\mathcal{S}_\phi\mathcal{P}_-
    \end{pmatrix}
    \equiv \mathcal{P} \mathcal{S}_\phi \mathcal{P} = \mathcal{S} \mathcal{P}.
\end{equation}
Where we defined
\begin{equation}
	\mathcal{S} \equiv \sqrt{1-T}\eta_3 + \cos(\varphi/2)\sqrt{T}\eta_1 + \sin(\varphi/2)\sqrt{T}\eta_2 \frac{i\zeta \check{\epsilon}}{\sqrt{\Delta^2 - \check{\epsilon}^2}}
\end{equation}
which contains the lead structure of $G$ and
\begin{equation}
	\mathcal{P} \equiv \frac{1}{2}(1 + \zeta \check{G}_0) = \frac{1}{2}(1 + \frac{\Delta \zeta}{\sqrt{\Delta^2 - \check{\epsilon}^2}}\tau_1 + \frac{-i\check{\epsilon} \zeta }{\sqrt{\Delta^2 - \check{\epsilon}^2}} \tau_3)
\end{equation}
which contains the Nambu structure of $G$. Hence, we can write $G$ on the same form as in Eq. \eqref{eq:G},
\begin{equation}
	G = \check{Q} \mathcal{P} \mathcal{S}.
\end{equation}

\section{The spectral representation of equilibrium Keldysh greens functions}\label{sec:spectral}
In this section we derive the spectral representation which we frequently use when working with Keldysh matrices. Many of Keldysh matrix functions we use derive from $\check{G}_0$ which is normalized $\check{G}_0^2 = 1$ even at infinity. We therefore have to consider functions that do not vanish at infinity.

Let us consider a function $f^R(\epsilon)$, which can be extended to an analytical function in the upper half plane, and has a well-defined limit at infinity $\lim_{|z|\rightarrow \infty}f^R(z)$ which it approaches algebraically. Now we have
\begin{align} \label{eq:lambda_R}
	f^R(\epsilon) &= f^R_{\infty} + \int_{-\infty}^{\infty} \frac{d\epsilon'}{2\pi i} \frac{f^R(\epsilon') - f^R_\infty}{\epsilon'-\epsilon -i0}
\end{align}
which can be seen by extending the integral around the semicircle at infinity in the upper half plane. Let us similarly consider a function $f^A(z)$ which can be extended to an analytical function in the lower half plane and algebraically approach some well defined $f^A_\infty$ at large $|z|$. Now
\begin{align}\label{eq:lambda_A}
	0 &=  \int_{-\infty}^{\infty} \frac{d\epsilon'}{2\pi i}\frac{f^A(\epsilon') - f^A_\infty}{\epsilon'-\epsilon -i0}
\end{align}
By subtracting Eqs. \eqref{eq:lambda_R} and \eqref{eq:lambda_A} we have

\begin{align}
	f^R(\epsilon) &= f^R_{\infty} + \int_{-\infty}^{\infty} \frac{d\epsilon'}{2\pi i}  \frac{1}{\epsilon'-\epsilon -i0}(f^R(\epsilon') - f^A(\epsilon') - f^R_\infty + f^A_\infty)\\
	&= \frac{1}{2}(f^R_{\infty} + f^A_\infty ) + \int_{-\infty}^{\infty} \frac{d\epsilon'}{2\pi i}  \frac{f^R(\epsilon') - f^A(\epsilon')}{\epsilon'-\epsilon -i0} \label{eq:spectral_R}
\end{align}
Where we assume the prescription of symmetric upper and lower limits on the integral in the second line (which is improper if $f^R_\infty \neq f^A_\infty$). Equation \eqref{eq:spectral_R} holds also for matrices as long as they satisfy our assumptions at infinity componentwise. We can derive an analogous expression of $f^A(\epsilon)$
\begin{align}
	f^A(\epsilon) &= \frac{1}{2}(f^R_{\infty} + f^A_\infty ) + \int_{-\infty}^{\infty} \frac{d\epsilon'}{2\pi i}  \frac{f^R(\epsilon') - f^A(\epsilon')}{\epsilon'-\epsilon +i0} \label{eq:spectral_A}
\end{align}
and by combining Eqs. \eqref{eq:spectral_R} and \eqref{eq:spectral_A} we have
\begin{align}
	\check{f}=
	\frac{1}{2}(f^R_{\infty} + f^A_\infty ) + \int_{-\infty}^{\infty} \frac{d\epsilon'}{2\pi i}  \frac{f^R(\epsilon') - f^A(\epsilon')}{\epsilon'-\check{\epsilon}}\label{eq:spectral_general},
\end{align}
where
\begin{equation}
    \check{f}(\epsilon) \equiv
	\check{M}^{-1}(\epsilon)
	\begin{pmatrix}
		f^R(\epsilon) & 0\\
		0 & f^A(\epsilon)
	\end{pmatrix}
	\check{M}(\epsilon).
\end{equation}

\bibliography{Supplement_energy_shift}